\documentclass[10pt,a4paper,final]{iopart}

\usepackage{graphicx}
\usepackage{amsthm}
\usepackage{iopams}
\usepackage{xcolor}
\usepackage{stackrel}
\usepackage{fullpage}
\usepackage{cite}
\usepackage{bm}
\usepackage{graphicx}
\graphicspath{{figures/}}
\usepackage{SIunits}
\usepackage{hyphenat}
\usepackage{multirow}
\usepackage{breqn}
\usepackage{array}
\usepackage{mdframed}
\usepackage{scrextend}
\usepackage[bookmarks=false]{hyperref}
\usepackage{amsfonts}
\usepackage{comment}
\usepackage[format=hang,font=small,labelfont=bf,textfont=sf,figurename=Fig.]{caption}[2013/02/03]
\usepackage[margin=0.2cm]{subcaption}
\usepackage[capitalise]{cleveref}
\crefname{section}{Sec.}{Secs.}
\Crefname{section}{Section}{Sections}

\usepackage{braket}

\let\square\relax

\renewcommand{\sectionautorefname}{Sec.}

\renewcommand\ket[1]{{|{#1}\rangle}}

\begin{document}

\title{Asymmetric twin-field quantum key distribution}

\author{Federico Grasselli$^{1}$} 
\ead{federico.grasselli@hhu.de}
\author{\'Alvaro Navarrete$^{2}$}
\ead{anavarrete@com.uvigo.es}
\author{Marcos Curty$^{2}$}
\address{$^1$ Institut f\"ur Theoretische Physik III, Heinrich-Heine-Universit\"at D\"usseldorf, Universit\" atsstra{\ss}e 1, D-40225 D\"usseldorf, Germany}
\address{$^2$ EI Telecomunicaci\'on, Department of Signal Theory and Communications, University of Vigo, Vigo E-36310, Spain}

\begin{abstract}
Twin-Field (TF) quantum key distribution (QKD) is a major candidate to be the new benchmark for far-distance QKD implementations, since its secret key rate can overcome the repeaterless bound by means of a simple interferometric measurement. Many variants of the original protocol have been recently proven to be secure. Here, we focus on the TF-QKD type protocol proposed by Curty et al~[preprint arXiv:1807.07667], which can provide a high secret key rate and whose practical feasibility has been demonstrated in various recent experiments. The security of this protocol relies on the estimation of certain detection probabilities (yields) through the decoy-state technique. Analytical bounds on the relevant yields have been recently derived assuming that both parties use the same set of decoy intensities, thus providing sub-optimal key rates in asymmetric-loss scenarios. Here we derive new analytical bounds when the parties use either three or four independent decoy intensity settings each. With the new bounds we optimize the protocol's performance in asymmetric-loss scenarios and show that the protocol is robust against uncorrelated intensity fluctuations affecting the parties' lasers.
\end{abstract}

\maketitle

\section{Introduction}
\label{sec:intro}
Quantum Key Distribution (QKD)~\cite{bennett1984,ekert1991,scarani2009,lo2014} allows two separated parties (typically called Alice and Bob) to generate identical bit strings with information-theoretic security. Due to the loss in the quantum channel connecting the parties, the performance of point-to-point QKD generally decreases with the distance, being unpractical for far-distance applications. Nonetheless, there have been remarkable efforts towards improving its range of applicability, such as the recent QKD experiments performed over 421 km of optical fiber~\cite{boaron2018secure} and over 1000 km of free space in satellite-to-ground links~\cite{liao2017satellite,takenaka2017satellite}. However, even for the most outstanding far-distance experiments, the secret key rate turns out to be probably too low for commercial purposes. In fact, it has been proven that there exist fundamental limits on the secret key rate that can be extracted from such point-to-point configurations. These limits say that the secret key rate scales linearly with the transmittance of the quantum channel linking the parties, or in other words, that it decreases exponentially with the channel length~\cite{takeoka2014fundamental,pirandola2017fundamental}. 

Quantum repeaters~\cite{briegel1998,duan2001long,sangouard2011} and measurement-device-independent QKD (MDI-QKD) protocols with either quantum memories~\cite{abruzzo2014measurement,panayi2014memory} or with quantum non-demolition measurements~\cite{azuma2015all} are possible theoretical solutions to overcome these limits. Unfortunately, in practice they require a technology that seems to be far from available in the near future. A more realistic solution was proposed recently by Lucamarini \textit{et al.}~\cite{lucamarini2018overcoming}. They devised an MDI-QKD type protocol -- called twin-field QKD (TF-QKD) -- in which the untrusted central node performs a single-photon interference measurement on the two incoming pulses, causing the key rate to scale with the square-root of the channel transmittance by using simple optical devices. Since the original proposal, several variants of the TF-QKD protocol were proven to be secure~\cite{wang2018twin,curty2018simple,ma2018phase,cui2019twin,tamaki2018information,lin2018simple} and some of them were experimentally implemented~\cite{zhong2019proof,minder2019experimental,liu2019experimental,wang2019beating}.

Here we focus on the TF-QKD scheme proposed in~\cite{curty2018simple}. In this protocol, Alice and Bob use the decoy-state technique to upper bound the detection probabilities associated to various photon-number states (called yields), which are subsequently used to obtain a bound on the phase error rate. Importantly, and in contrast to other solutions~\cite{wang2018twin,ma2018phase,tamaki2018information,lin2018simple} which use a post-selection step based on the matching of a global phase, the scheme in~\cite{curty2018simple} pre-selects the value of the global phase and thus it can provide a higher secret key rate. Moreover, the practical feasibility of this scheme has been recently demonstrated in~\cite{zhong2019proof,minder2019experimental,wang2019beating}. A complete analysis of the symmetric scenario where both users analytically estimate the yields using the \textit{same} intensity settings was performed recently in~\cite{grasselli2019practical}. However, using the same set of intensities is an optimal strategy only when the quantum channels connecting the users to the central node have approximately the same transmittance. Thus, the bounds derived in~\cite{grasselli2019practical} are not suitable for several real-world situations in optical networks where the distances between the users and the central node can be notoriously different. Furthermore, assuming that the parties employ exactly the same intensities is problematic even when the losses are symmetric. This is due to the fact that, typically, neither Alice nor Bob can ensure that their lasers emit pulses with a perfectly locked intensity. Instead, their intensities are typically fluctuating randomly and independently from the other party. For these reasons, the derivation of yields bounds based on \textit{asymmetric} decoy intensities is crucial for the protocol's security in the presence of intensity fluctuations and for addressing asymmetric-loss scenarios. A similar analysis has been recently carried out for another type of TF-QKD protocol in~\cite{zhou2019asymmetric}.

In this paper, we address this problem by analysing the performance of the TF-QKD scheme proposed in~\cite{curty2018simple} in the presence of asymmetric losses and independent laser intensity fluctuations. For this, we derive analytical bounds on the yields when Alice and Bob use asymmetric intensity settings. In particular, we consider the practical cases where each of Alice and Bob uses three and four decoy intensity settings, which are the most efficient solutions for covering long distances. In doing so, we show that the protocol can tolerate highly-asymmetric loss scenarios and is quite robust against intensity fluctuations, thus demonstrating its practicality for realistic network configurations.

The paper is organized as follows. In \sectionautorefname~\ref{sec:TFQKD} we summarize the TF-QKD protocol introduced in~\cite{curty2018simple}. Then, in \sectionautorefname~\ref{sec:Sym} we analyze the performance of the aforementioned protocol under the assumption that Alice and Bob use the same signal and decoy intensities. In \sectionautorefname~\ref{yields-bounds-2decoys} we derive analytical bounds on the yields when the parties are allowed to use independent decoy intensity settings. With the derived bounds, we investigate the protocol's performance in \sectionautorefname~\ref{simulations} when using independent signal and decoy intensities and in the presence of uncorrelated intensity fluctuations affecting the users' lasers. Finally, in \sectionautorefname~\ref{sec:conclusion} we present our conclusions. The paper includes also a few Appendixes with additional calculations.

\section{TF-QKD}
\label{sec:TFQKD}
\subsection{Protocol description}
In this section we briefly summarize the considered TF-QKD protocol~\cite{curty2018simple}. As shown in Fig.~\ref{Fig_SchemeTFQKD}, it consists in both Alice and Bob sending optical pulses through a quantum channel to an untrusted third party, Charles, who is in charge of performing joint measurements on the incoming pulses and announcing the results. The protocol is composed of the following seven steps:
\begin{figure}[ht]
	\begin{center}
		\includegraphics[width=0.7\columnwidth]{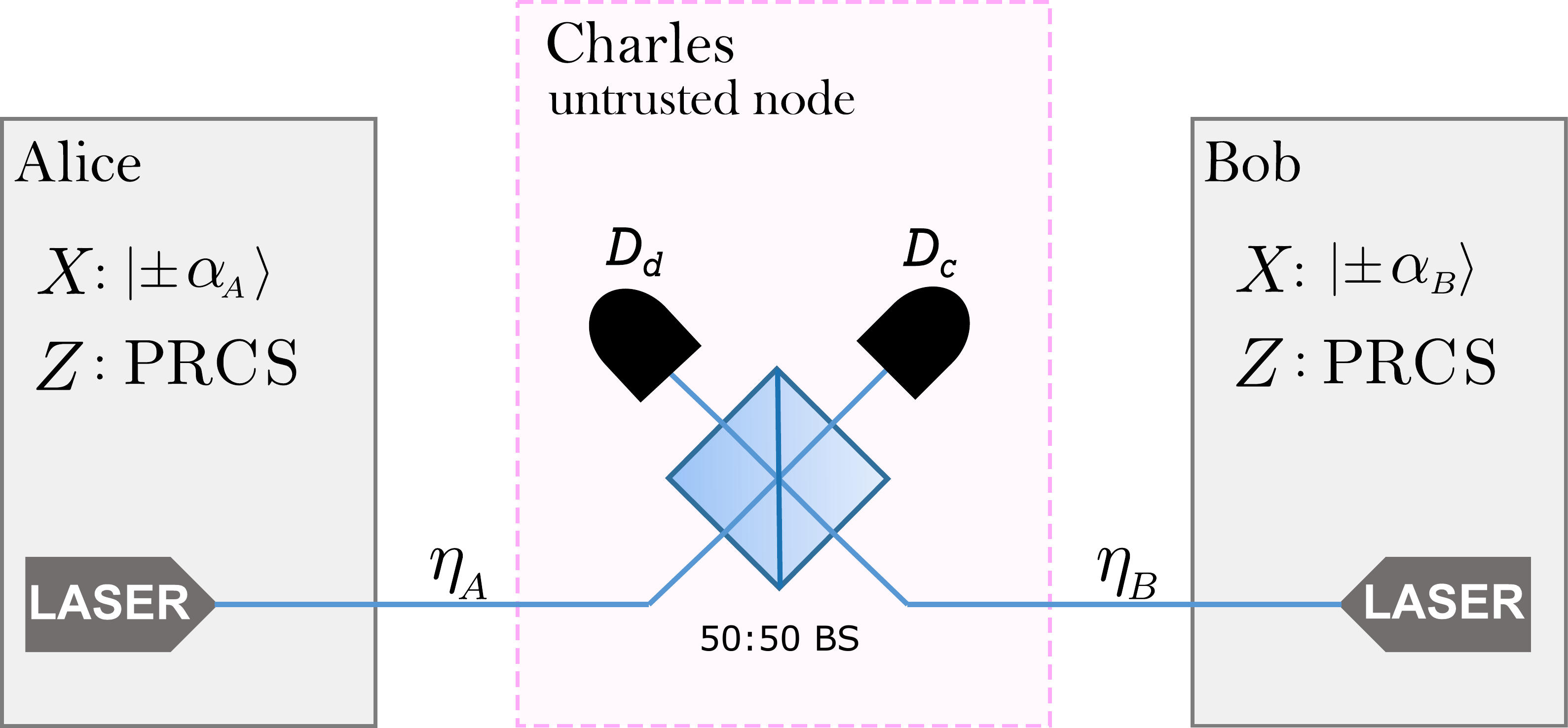}
	\end{center}
	\caption{Scheme of the TF-QKD protocol proposed in~\cite{curty2018simple}. After selecting the $X$ or $Z$ basis randomly, Alice and Bob send optical pulses through a quantum channel to an intermediate node controlled by the untrusted party Charles. In an honest implementation, Charles makes the incoming pulses interfere in a 50:50 beam splitter (BS) and publicly announces which of the two threshold detectors placed at the beam splitter output ports clicks. For the $X$ basis, Alice and Bob send coherent states $\ket{\pm\alpha_A}$ and $\ket{\pm\alpha_B}$, respectively. For the $Z$ basis, they send phase-randomized coherent states (PRCS) whose intensities are chosen at random from predefined sets.}\label{Fig_SchemeTFQKD}
\end{figure}
\begin{itemize}
	\item[(i)] Alice (Bob) chooses the $X$ basis with probability $p_x^A$ ($p_x^B$) and the $Z$ basis with probability $p_z^A=1-p_x^A$ ($p_z^B=1-p_x^B$). For the $X$ basis, Alice (Bob) prepares an optical pulse in a coherent state $\ket{(-1)^{b_A}\alpha_A}$ ($\ket{(-1)^{b_B}\alpha_B}$), with $b_A$ ($b_B$) being a randomly chosen bit and $\alpha_A,\alpha_B \in\mathbb{R}$, for simplicity. For the $Z$ basis, Alice (Bob) prepares an optical pulse in a phase-randomized coherent state (PRCS) $\rho_{\mu_k}$ ($\rho_{\nu_l}$) whose intensity $\mu_k$ ($\nu_l$) is chosen from a set $\mathcal{S}_A=\{\mu_{k}\}_k$ ($\mathcal{S}_B=\{\nu_{l}\}_l$) with probability $p_{k}$ ($p_{l}$).
	\item[(ii)] Both Alice and Bob send their pulses to an intermediate untrusted node, Charles, through optical channels with transmittances $\eta_A$ and $\eta_B$, respectively, in a synchronized manner.
	\item[(iii)] Charles interferes the incoming pulses in a 50:50 beam splitter, followed by two threshold detectors associated with the constructive (detector $D_c$) and destructive (detector $D_d$) interference, respectively.
	\item[(iv)] Charles announces the measurement outcomes $k_c$ and $k_d$ of the detectors $D_c$ and $D_d$, respectively, with $k_c=1$ ($k_d=1$) corresponding to a click event and $k_c=0$ ($k_d=0$) corresponding to a no-click event.
	\item[(v)] Alice and Bob reveal a small fraction of the bits $b_A$ ($b_B$) collected from those events when both parties chose the $X$ basis and Charles reported a click only in one detector ($k_c+k_d=1$) to estimate the bit error rate. Their raw keys consist on the remaining undisclosed bits. Also, Bob flips all the bits $b_B$ collected when the click occurred in $D_d$.
	\item[(vi)] Alice and Bob publicly announce the intensities used in all the events when both chose the $Z$ basis, and they use that information to estimate the phase error rate.
	\item[(vii)] Alice and Bob apply error correction and privacy amplification techniques to their raw keys to distill two identical secret keys.
\end{itemize}

\subsection{Secret key rate}
The asymptotic secret key rate of the protocol described above is lower bounded by~\cite{curty2018simple}
\begin{equation}\label{Eq_Rate}
	R\geq \max\{R_{X}^{\Omega_c},0\} + \max\{R_{X}^{\Omega_d},0\},
\end{equation}
where $R_{X}^{\Omega}$ is a lower bound on the secret key rate that Alice and Bob can obtain from the event $\Omega\in\{\Omega_c,\Omega_d\}$, being $\Omega_c\equiv (k_c=1\wedge k_d=0)$ and $\Omega_d\equiv (k_c=0\wedge k_d=1)$. This lower bound is given by 
\begin{eqnarray}\label{Eq_RateO}
R_{X}^{\Omega}&=&p_x^Ap_x^Bp_{X}(\Omega)[1-h_2(e_{Z,\Omega}^{\text{upp}})-fh_2(e_{X,\Omega})],
\end{eqnarray}
where $p_{X}(\Omega)=\frac{1}{4}\sum_{b_Ab_B} p_{X}(\Omega|b_A,b_B)$ is the conditional probability that the event $\Omega$ occurs given that Alice and Bob select the $X$ basis, $e_{Z,\Omega}^{\text{upp}}$ is an upper bound on the phase error rate, $e_{X,\Omega}$ is the bit error rate, $f$ is the reconciliation efficiency of the error correction process and \mbox{$h_2(x)=-x\log_2(x)-(1-x)\log_2(1-x)$} is the binary entropy function. Note that in the asymptotic scenario, which is the scenario we consider in this work, we assume for simplicity that $p_x^A=p_x^B\approx1$. The upper bound on the phase error rate, $e_{Z,\Omega}^{\text{upp}}$, is given by~\cite{curty2018simple}:
\begin{equation}\label{Eq_error_z}
e_{Z,\Omega}^{\text{upp}}\times p_{X}(\Omega)=\left(\sum_{n,m\in2\mathbb{N}^0}^{\infty}c_{n,m}\sqrt{Y_{nm}^{\Omega}}\right)^2+\left(\sum_{n,m\in2\mathbb{N}^0+1}^{\infty}c_{n,m}\sqrt{Y_{nm}^{\Omega}}\right)^2,
\end{equation}
where $Y_{nm}^{\Omega}\equiv p_{ZZ}(\Omega|n,m)$ is the conditional probability of the event $\Omega$ given that Alice and Bob sent $n$ and $m$ photons, respectively, $c_{n,m}=e^{-\frac{\alpha_A^2+\alpha_B^2}{2}}\frac{\alpha_A^n\alpha_B^m}{\sqrt{n!m!}}$ and $\mathbb{N}^0$ denotes the set of non-negative integers. The yields $Y_{nm}^{\Omega}$ are not experimentally observed but can be estimated through the decoy-state method~\cite{hwang2003,lo2005,wang2005a} (see \autoref{yields-bounds-2decoys}). The bit error rate is given by
\begin{eqnarray}\label{Eq_error_x1}
e_{X,\Omega_c}&=&p_{X}(b_A\neq b_B|\Omega_c)=\frac{1}{4}\sum_{b_A\neq b_B}\frac{p_{X}(\Omega_c|b_A,b_B)}{p_{X}(\Omega_c)},\\
e_{X,\Omega_d}&=&p_{X}(b_A = b_B|\Omega_d)=\frac{1}{4}\sum_{b_A= b_B}\frac{p_{X}(\Omega_d|b_A,b_B)}{p_{X}(\Omega_d)}.\label{Eq_error_x2}
\end{eqnarray}
The values of the bit error rate $e_{X,\Omega}$ and of the probability $p_{X}(\Omega)$ for a typical channel model are given in \ref{AppendixChannelModel}. These are the values we use in our simulations.

\section{Symmetric intensities}
\label{sec:Sym}
When analyzing QKD protocols based on a central-node architecture, it is common to consider the symmetric scenario where the transmittances of the channels Alice-Charles and Bob-Charles are equal. This is, however, an unrealistic assumption. In a practical scenario, the loss introduced by the quantum channel Alice-Charles could significantly differ from the loss in the channel Bob-Charles. In this case, the yields bounds obtained by using the decoy-state technique with the same intensity settings for Alice and Bob are not optimal anymore, i.e. they are looser than those obtained when the channel losses are instead symmetric. 

\begin{figure}
	\begin{subfigure}{0.5\textwidth}
		\centering
		\includegraphics[width=\textwidth]{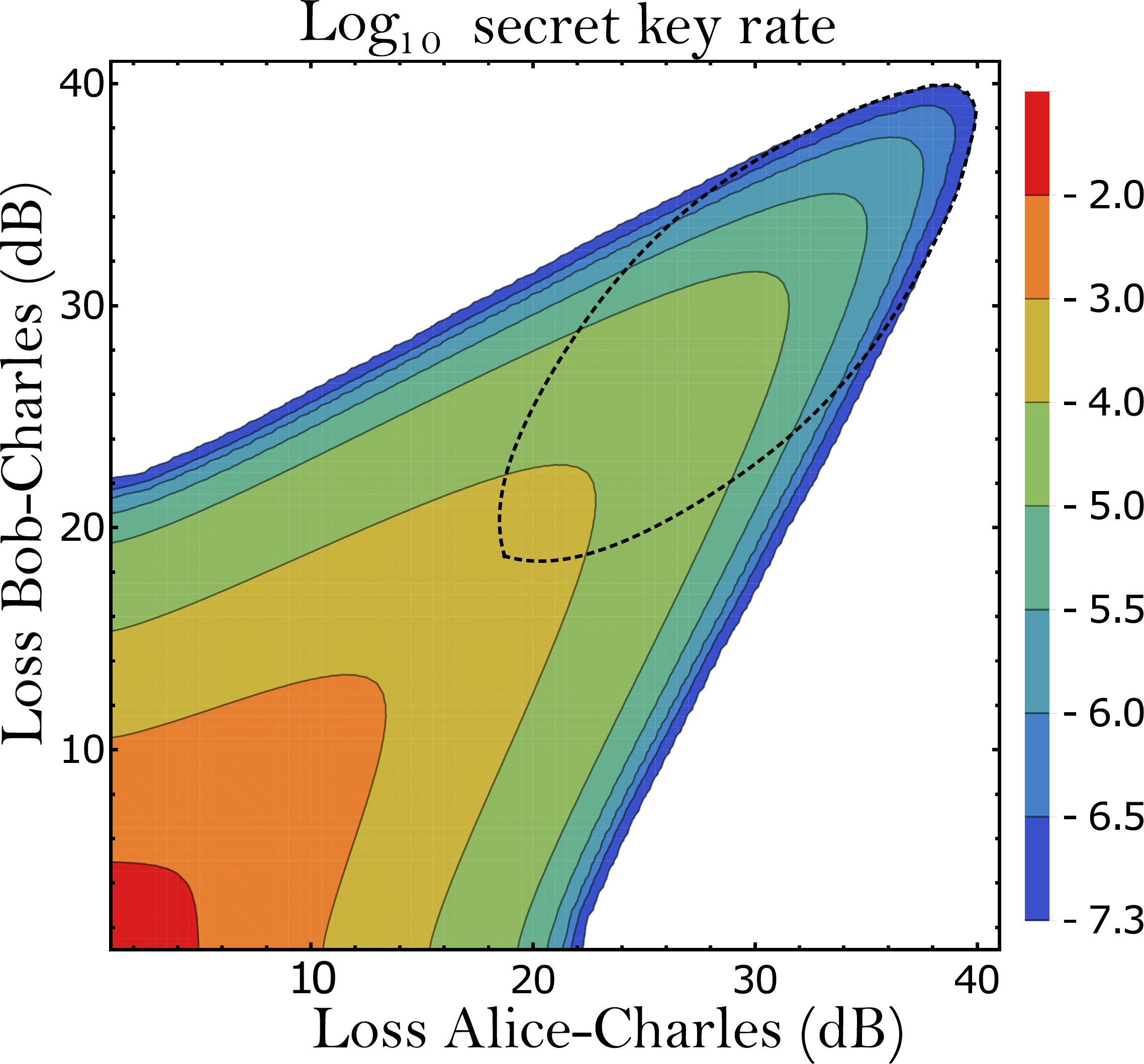}
		\caption{}\label{Fig_SymDecoy3TFQKD}
	\end{subfigure}
	\begin{subfigure}{0.5\textwidth}
		\centering
		\includegraphics[width=\textwidth]{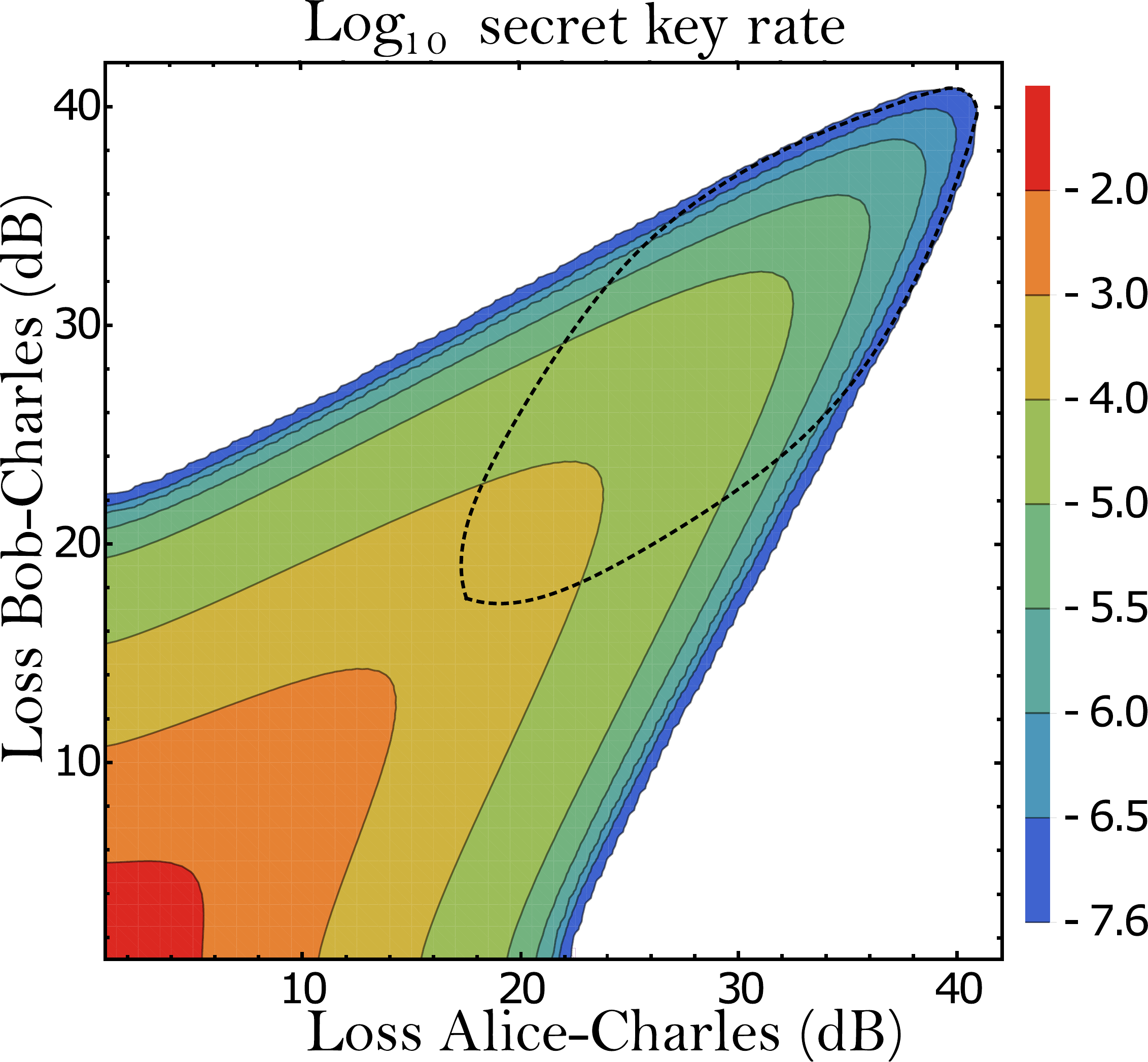}
		\caption{}\label{Fig_SymDecoy4TFQKD}
	\end{subfigure}
	\caption{Contour lines for the secret key rate of the TF-QKD protocol introduced in~\cite{curty2018simple} as a function of the channel loss in Alice's and Bob's sides, assuming that both Alice and Bob use the same signal and decoy intensities. Specifically, in (a) and (b) they use three and four decoy intensities each, respectively. The black dashed line encloses the loss region where the key rate overcomes the repeaterless bound \cite{pirandola2017fundamental}. We note that for several combinations of losses, it might be beneficial for the parties to increase the loss in one of the channels in order to make them more symmetric, thus maximizing the key rate. The experimental parameters used in the simulations are given in Table~\ref{tab:experimental-parameters}.}\label{Fig_SymDecoyTFQKD}
\end{figure}

Indeed, as already shown in MDI-QKD~\cite{wang2018asymmetric,xu2013,xu2014}, if Alice and Bob use the same intensity settings, they might be in a situation where it is convenient for them to symmetrize the channels losses by increasing the loss in one of the channels, in order to enhance the key rate. In doing so, the intensities of the pulses arriving at the central node are now of similar magnitude, which results in an improvement of the key rate. The same happens in the TF-QKD scheme introduced in~\cite{curty2018simple}. This is clear from Fig.~\ref{Fig_SymDecoyTFQKD}, where we plot the secret key rate assuming that Alice and Bob use the same set of three and four decoy intensities. The plots are obtained by using the analytical yields bounds for the symmetric-intensities scenario derived in~\cite{grasselli2019practical}. The experimental parameters used for the simulations are given in Table~\ref{tab:experimental-parameters} and the corresponding channel model is given in \ref{AppendixChannelModel}.
\begin{table}
	\caption{Experimental parameters used in the simulations. See \ref{AppendixChannelModel} for the definitions.}\label{tab:experimental-parameters}
	\renewcommand*{\arraystretch}{1.15}
	\begin{center}
		\begin{tabular}[t]{lcc}
			\hline\hline
			Dark count probability & $p_d$ & $10^{-7}$ \\
			Total polarization misalignment & $\theta$ &  2\% \\
			Phase mismatch & $\phi$ &  2\% \\
			\hline\hline
		\end{tabular}
	\end{center}
\end{table}

In Fig.~\ref{Fig_SymDecoyTFQKD}, the key rate is optimized over the signal intensity $\alpha_A^2=\alpha_B^2$ and over the strongest decoy intensity (assumed to be equal for the two parties), while the other decoy intensities are fixed to the same values for both parties. As a matter of fact, after having observed that in the asymptotic scenario the optimal values of the weaker decoy intensities tend to be as small as possible regardless of the losses in the two channels, we fixed them to reasonably low values in the key rate optimization. More precisely: the weakest and the second-to-the-weakest decoy intensities of Alice and Bob in the three- and four-decoy case are fixed to $10^{-5}$ and $10^{-4}$, while the third-to-the-weakest decoy intensity in the four-decoy case is fixed to $10^{-3}$. The resulting key rate basically reproduces the rate one would obtain when optimizing even on the weaker decoy intensities~\cite{grasselli2019practical}. The disadvantage of using symmetric signal and decoy intensity settings is clear in both the three- and the four-decoy case, where increasing the loss in one of the channels can lead to an increase of the key rate in asymmetric-loss scenarios.

Furthermore, as already mentioned in the introduction, assuming that Alice and Bob are using exactly the same intensities is not realistic in most experimental implementations~\cite{minder2019experimental,liu2019experimental,wang2019beating} due to the intensity fluctuations on the transmitters' lasers. The effect of intensity fluctuations was already considered in~\cite{grasselli2019practical} under the assumption that the fluctuations are correlated among the two parties, which is satisfied in the experiment reported in~\cite{zhong2019proof}, but does not hold in general.

\section{Asymmetric intensities}  \label{yields-bounds-2decoys}
In order to enhance the protocol's performance in the presence of asymmetric losses and to investigate uncorrelated intensity fluctuations, in this work we derive analytical upper bounds on the yields for the three- and four-decoy scenarios with \textit{independent} intensity settings for Alice and Bob. We note that the use of three or four decoy intensity settings is already enough to obtain a secret key rate close to the one that could be achieved with infinite decoy intensity settings~\cite{curty2018simple,grasselli2019practical}. The derivation of these bounds is presented in \ref{3intensities} and \ref{4intensities}. However, for illustration purposes, we present in this section the resulting upper bounds for the three-decoy case.

According to the TF-QKD protocol \cite{curty2018simple} summarized in Sec.~\ref{sec:TFQKD} --when both parties choose the $Z$ basis-- Alice prepares a PRCS whose intensity belongs to the set $\{\mu_0,\mu_1,\mu_2\}$, with $\mu_0>\mu_1>\mu_2$. Analogously, Bob prepares a state whose intensity is instead drawn from the set $\{\nu_0,\nu_1,\nu_2\}$, with $\nu_0>\nu_1>\nu_2$. The key assumption of the decoy-state method is that the yields are independent of the chosen intensities and are thus subjected to the following four equality constraints:
\begin{equation}
\tilde{Q}^{k,l} \equiv e^{\mu_k + \nu_l} Q^{k,l} =\sum_{n,m=0}^{\infty} \frac{Y_{nm}}{n!m!} {\mu_k}^n {\nu_l}^m \quad k,l \in \{0,1,2\} \,\,, \label{constr-3decoys}
\end{equation}
where $Q^{k,l}$ is the gain in the $Z$ basis given that Alice and Bob choose intensities $\mu_k$ and $\nu_l$, respectively. Note that we omit here and in what follows, for readability, the dependency of the variables with $\Omega$. Being probabilities, the yields are additionally subjected to the inequality constraints:
\begin{equation}
0 \leq Y_{nm} \leq 1 \quad \forall\, n,m  \label{ineq-constr} \,\,.
\end{equation}
By properly combining the constraints (\ref{constr-3decoys}) with a procedure similar to Gaussian elimination, we can obtain analytical upper bounds on the yields $Y_{00},Y_{11},Y_{22},Y_{02},Y_{04},Y_{20},Y_{40},Y_{13}$ and $Y_{31}$. The other yields are trivially upper bounded by 1. The upper bounds are then inserted in the expression for the phase error rate (\ref{Eq_error_z}), enabling us to obtain a fully analytical expression of the asymptotic secret key rate (\ref{Eq_Rate}). In what follows, we present the resulting upper bounds on the aforementioned yields (we refer the reader to~\ref{3intensities} for their derivation). For this, let's consider the most general combination of the nine constraints (\ref{constr-3decoys}):
\begin{eqnarray}
G_{uv}=\sum_{i,j=0}^{2} c_{i,j} \tilde{Q}^{i,j}. \label{eq:Guv}
\end{eqnarray}
For simplicity, in~(\ref{eq:Guv}) and also below, we omit the explicit dependence of the coefficients $c_{i,j}$ with the value of $u$ and $v$. Then, we can obtain an upper bound on the yield $Y_{uv}$ by appropriately choosing the coefficients $c_{i,j}$ that appear in~(\ref{eq:Guv}).

\subsection{Upper bound on $Y_{00}$}
An upper bound on the yield $Y_{00}$ is given by
\begin{eqnarray}
Y^U_{00}=\frac{\mu_1 \mu_2 \nu_1 \nu_2 G_{00}}{(\mu_0 -\mu_1)(\mu_0 - \mu_2)(\nu_0 -\nu_1)(\nu_0 - \nu_2)}  \label{Y00-upperbound-3decoys} \,\,.
\end{eqnarray}
where $G_{00}$ is given by~(\ref{eq:Guv}) by fixing the $c_{i,j}$ coefficients to those given in~(\ref{cij-00}).

\subsection{Upper bound on $Y_{11}$}
An upper bound on the yield $Y_{11}$ is given by
\begin{eqnarray}
\fl &Y^U_{11}=\frac{G_{11} (\mu_1+\mu_2)(\nu_1+\nu_2)}{(\mu_0 -\mu_1)(\mu_0 -\mu_2)(\nu_0 -\nu_1)(\nu_0 -\nu_2)} +\frac{Y_{13}^U}{6} (\nu_1 \nu_2 + \nu_0\nu_1+ \nu_2 \nu_0) + \frac{Y_{31}^U}{6} (\mu_1 \mu_2 + \mu_0 \mu_1 +\mu_2\mu_0) \nonumber\\
\fl &-\frac{(e^{\nu_1}-\nu_1- \frac{\nu_1^2}{2}-\frac{\nu_1^3}{6})(\nu_0^2 - \nu_2^2)+(e^{\nu_2}-\nu_2- \frac{\nu_2^2}{2}-\frac{\nu_2^3}{6})(\nu_1^2 - \nu_0^2) + (e^{\nu_0}-\nu_0- \frac{\nu_0^2}{2}-\frac{\nu_0^3}{6})(\nu_2^2 - \nu_1^2)}{(\nu_1-\nu_2)(\nu_0 - \nu_1)(\nu_0 - \nu_2)} \nonumber\\
\fl &-\frac{(e^{\mu_1}-\mu_1 -\frac{\mu_1^2}{2}-\frac{\mu_1^3}{6})(\mu_0^2 - \mu_2^2)+(e^{\mu_2}-\mu_2 -\frac{\mu_2^2}{2}-\frac{\mu_2^3}{6})(\mu_1^2 - \mu_0^2) + (e^{\mu_0}-\mu_0 -\frac{\mu_0^2}{2}-\frac{\mu_0^3}{6})(\mu_2^2 - \mu_1^2)}{(\mu_1-\mu_2)(\mu_0 - \mu_1)(\mu_0 - \mu_2)}\,\,.\nonumber\\
\fl  \label{Y11-upperbound-3decoys}
\end{eqnarray}
where $G_{11}$ is given by~(\ref{eq:Guv}) by fixing the $c_{i,j}$ coefficients to those given in~(\ref{cij-11}), and where the upper bounds $Y_{13}^{U}$ and $Y_{13}^{U}$ are provided below.

\subsection{Upper bound on $Y_{22}$}
An upper bound on the yield $Y_{22}$ is given by
\begin{eqnarray}
Y^U_{22}=\frac{4 G_{22}}{(\mu_0 -\mu_1)(\mu_0 -\mu_2)(\nu_0 -\nu_1)(\nu_0 -\nu_2)}  \,\,, \label{Y22-upperbound-3decoys}
\end{eqnarray}
where $G_{22}$ is given by~(\ref{eq:Guv}) by fixing the $c_{i,j}$ coefficients to those given in~(\ref{cij-22}).

\subsection{Upper bounds on $Y_{02}$ and $Y_{04}$}
The upper bounds on the yields $Y_{02}$ and $Y_{04}$ are given by, respectively,
\begin{eqnarray}
Y^U_{02}=\frac{2 G_{02} \mu_1 \mu_2}{(\mu_0 -\mu_1)(\mu_0 - \mu_2)(\nu_0 -\nu_1)(\nu_0 - \nu_2)} \label{Y02-upperbound-3decoys} \,\,,
\end{eqnarray}
and
\begin{eqnarray}
\fl Y^U_{04}=\frac{24 G_{02} \mu_1 \mu_2}{(\mu_0 -\mu_1)(\mu_0 - \mu_2)(\nu_0 -\nu_1)(\nu_0 - \nu_2)(\nu_2^2+\nu_1^2+\nu_0^2+\nu_0 \nu_1 +\nu_0\nu_2 + \nu_1\nu_2)} \label{Y04-upperbound-3decoys} \,\,,
\end{eqnarray}
where $G_{02}$ is given by~(\ref{eq:Guv}) by fixing the $c_{i,j}$ coefficients to those given in~(\ref{cij-02}).

\subsection{Upper bounds on $Y_{20}$ and $Y_{40}$}
The upper bounds on the yields $Y_{20}$ and $Y_{40}$ are given by, respectively,
\begin{eqnarray}
Y^U_{20}=\frac{2 G_{20} \nu_1 \nu_2}{(\mu_0 -\mu_1)(\mu_0 - \mu_2)(\nu_0 -\nu_1)(\nu_0 - \nu_2)} \label{Y20-upperbound-3decoys} \,\,,
\end{eqnarray}
and
\begin{eqnarray}
\fl Y^U_{40}=\frac{24 G_{20} \nu_1 \nu_2}{(\mu_0 -\mu_1)(\mu_0 - \mu_2)(\nu_0 -\nu_1)(\nu_0 - \nu_2)(\mu_2^2+\mu_1^2+\mu_0^2+\mu_0 \mu_1 +\mu_0\mu_2 + \mu_1\mu_2)} \label{Y40-upperbound-3decoys} \,\,,
\end{eqnarray}
where $G_{20}$ is given by~(\ref{eq:Guv}) by fixing the $c_{i,j}$ coefficients to those given in~(\ref{cij-20}).

\subsection{Upper bound on $Y_{13}$}
An upper bound on the yield $Y_{13}$ is given by
\begin{eqnarray}
\fl &Y^U_{13}=\frac{-6(\mu_1+\mu_2)G_{13}}{(\mu_0 -\mu_1)(\mu_0 - \mu_2)(\nu_0 -\nu_1)(\nu_0 - \nu_2)(\nu_0+\nu_1+\nu_2)}  \nonumber\\
\fl &+ \frac{6\left[e^{\nu_2}(\nu_1 -\nu_0) +e^{\nu_1}(\nu_0 - \nu_2) + e^{\nu_0} (\nu_2 - \nu_1)\right]}{(\mu_0 -\mu_1)(\mu_0 - \mu_2)(\mu_1 - \mu_2)(\nu_0 -\nu_1)(\nu_0 -\nu_2)(\nu_1-\nu_2)(\nu_0+\nu_1+\nu_2)} \nonumber\\
\fl &\times \left[e^{\mu_2}(\mu^2_1 -\mu^2_0) +e^{\mu_1}(\mu^2_0 - \mu^2_2) + e^{\mu_0} (\mu^2_2 - \mu^2_1) -(\mu_0 -\mu_1)(\mu_0 - \mu_2)(\mu_1 - \mu_2)\right]\label{Y13-upperbound-3decoys} \,\,,
\end{eqnarray}
where $G_{13}$ is given by~(\ref{eq:Guv}) by fixing the $c_{i,j}$ coefficients to those given in~(\ref{cij-13}).

\subsection{Upper bound on $Y_{31}$}
An upper bound on the yield $Y_{31}$ is given by
\begin{eqnarray}
\fl &Y^U_{31}=\frac{-6(\nu_1+\nu_2)G_{31}}{(\mu_0 -\mu_1)(\mu_0 - \mu_2)(\nu_0 -\nu_1)(\nu_0 - \nu_2)(\mu_0+\mu_1+\mu_2)}  \nonumber\\
\fl &+ \frac{6\left[e^{\mu_2}(\mu_1 -\mu_0) +e^{\mu_1}(\mu_0 - \mu_2) + e^{\mu_0} (\mu_2 - \mu_1)\right]}{(\mu_0 -\mu_1)(\mu_0 - \mu_2)(\mu_1 - \mu_2)(\nu_0 -\nu_1)(\nu_0 -\nu_2)(\nu_1-\nu_2)(\mu_0+\mu_1+\mu_2)} \nonumber\\
\fl &\times \left[e^{\nu_2}(\nu^2_1 -\nu^2_0) +e^{\nu_1}(\nu^2_0 - \nu^2_2) + e^{\nu_0} (\nu^2_2 - \nu^2_1) -(\nu_0 -\nu_1)(\nu_0 - \nu_2)(\nu_1 - \nu_2)\right]\label{Y31-upperbound-3decoys} \,\,,
\end{eqnarray}
where $G_{31}$ is given by~(\ref{eq:Guv}) by fixing the $c_{i,j}$ coefficients to those given in~(\ref{cij-31}).

\section{Simulations}  \label{simulations}
\begin{figure}
	\begin{subfigure}{0.5\textwidth}
	\centering
	\includegraphics[width=\textwidth]{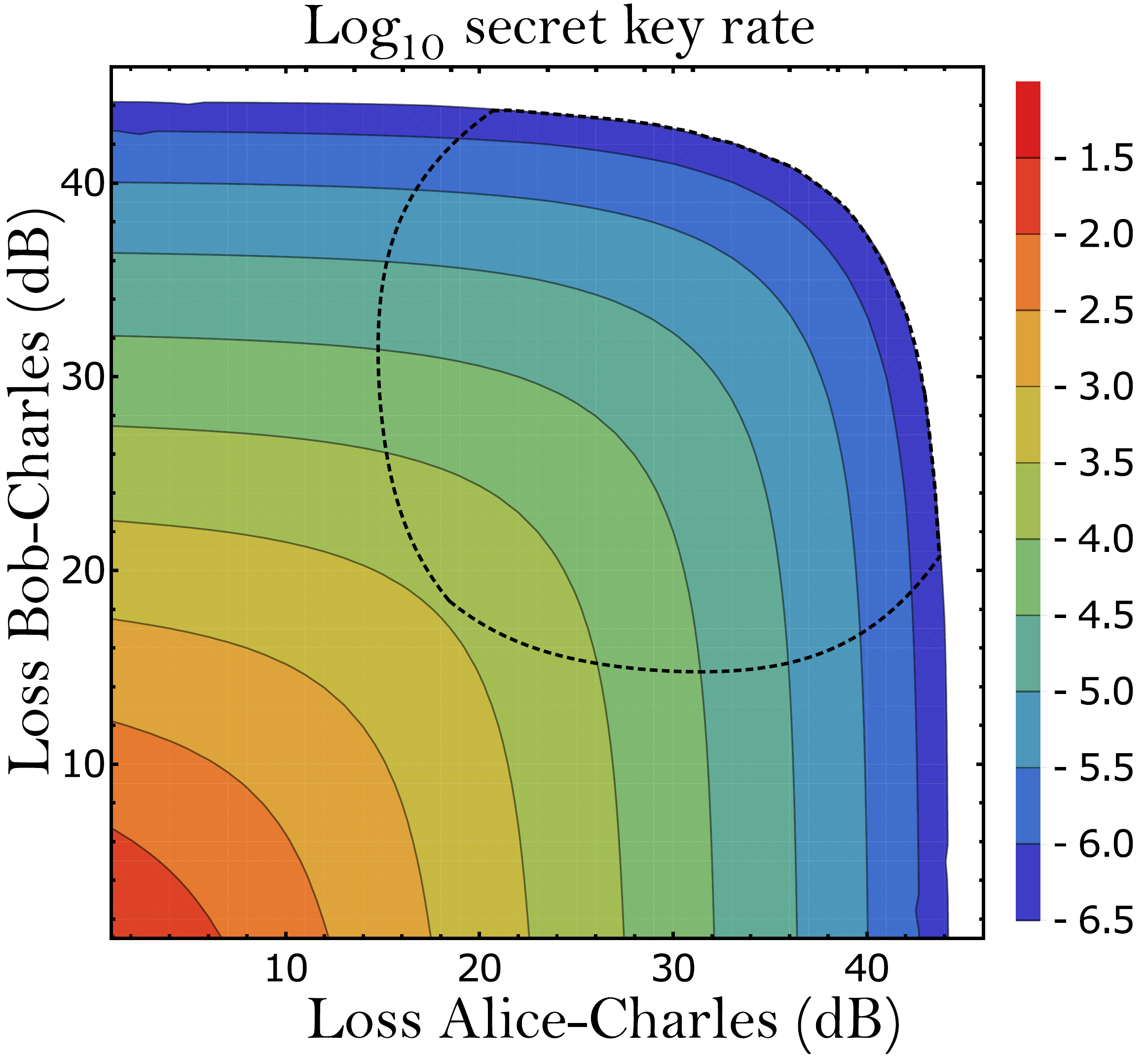}
	\caption{}\label{Fig_AsymDecoyTFQKD3dec}
	\end{subfigure}
	\begin{subfigure}{0.5\textwidth}
	\centering
	\includegraphics[width=\textwidth]{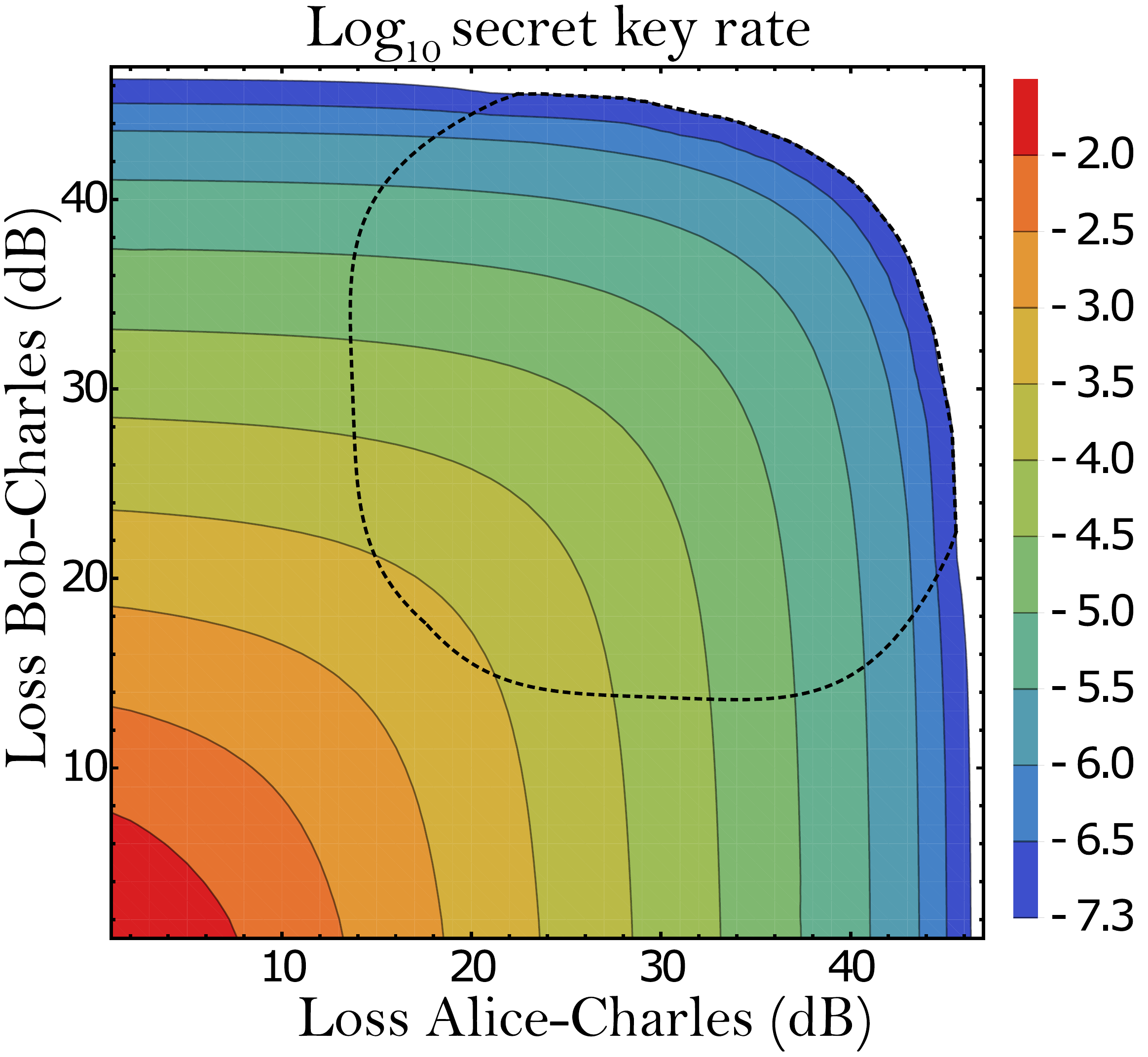}
	\caption{}\label{Fig_AsymDecoyTFQKD4dec}
	\end{subfigure}	
	\caption{Contour lines for the secret key rate of the TF-QKD protocol introduced in~\cite{curty2018simple} as a function of the channel loss in Alice's and Bob's sides assuming that both Alice and Bob use independent signal and decoy intensities. In (a) and (b), the parties use three and four decoy intensities each, respectively. The key rate is never enhanced by increasing the loss in one of the two quantum channels, in contrast to the previous scenario (Figs~\ref{Fig_SymDecoy3TFQKD} and \ref{Fig_SymDecoy4TFQKD}). The black dashed line encloses the loss region where the key rate overcomes the repeaterless bound \cite{pirandola2017fundamental}. The experimental parameters used in the simulations are given in Table~\ref{tab:experimental-parameters}.}\label{Fig_AsymDecoyTFQKDdec}
\end{figure}

In order to obtain the optimal secret key rate in the asymptotic-key regime, one needs to optimize it over the $X$ basis intensities $\alpha_A^2$ and $\alpha_B^2$, and over six or eight decoy intensities, depending on the number of decoys used by Alice and Bob. The key rate depends on the decoy intensities through the yields bounds derived in \ref{3intensities} and \ref{4intensities}. For instance, for the three-decoy case analyzed in the previous section, we have that the vector of parameters to be optimized is $\vec{p}=(\alpha_A,\alpha_B,\mu_0,\mu_1,\mu_2,\nu_0,\nu_1,\nu_2)$. In order to fairly compare the simulation results with those of the symmetric scenario (Fig.~\ref{Fig_SymDecoyTFQKD}), we use the same experimental parameters given by Table~\ref{tab:experimental-parameters} and we again fix the weaker decoy intensities to the same symmetric values for Alice and Bob, namely: $\mu_1=\nu_1=10^{-4}$ and $\mu_2=\nu_2=10^{-5}$ for the three-decoy case, $\mu_0=\nu_0=10^{-3}$, $\mu_1=\nu_1=10^{-4}$ and $\mu_2=\nu_2=10^{-5}$ for the four-decoy case. Thus the key rate is actually optimized over $\vec{p}=(\alpha_A,\alpha_B,\mu_0,\nu_0)$ in the three-decoy case, and over $\vec{p}=(\alpha_A,\alpha_B,\mu_3,\nu_3)$ in the four-decoy case. As explained in~\ref{4intensities}, note that in the four-decoy case, for convenience of our notation, $\mu_3$ and $\nu_3$ denote the strongest decoy intensities, \textit{i.e.}, we use the ordering $\mu_3>\mu_0>\mu_1>\mu_2$ and $\nu_3>\nu_0>\nu_1>\nu_2$. Although having fixed the weaker decoy intensities to the same values for both parties might seem restrictive in the asymmetric-loss scenario considered here, indeed it is not. As a matter of fact, we observed that the optimal values of the weaker decoy intensities (i.e. $\mu_1,\mu_2$ and $\nu_1,\nu_2$ in the three-decoy case, $\mu_0,\mu_1,\mu_2$ and $\nu_0,\nu_1,\nu_2$ in the four-decoy case) tend to be as low as possible, independently of the losses in Alice and Bob's channels. We thus fixed them to symmetric low values that are reasonable from an experimental point of view \cite{zhong2019proof,minder2019experimental}.

Fixing these parameters reduces the computation complexity of the simulations, which is important since, in contrast to the MDI-QKD scenario~\cite{xu2014a}, the key rate is not, in general, a convex function of $\vec{p}$ (see \ref{AppendixNoConvex}). This means that it is not possible to safely use time-efficient optimization methods, such as, for instance, the coordinate descent algorithm~\cite{boyd2004convex}. Our optimization is thus carried out by using the built-in global optimization algorithms of Wolfram Mathematica 11.0~\cite{Mathematica}.

In Fig.~\ref{Fig_AsymDecoyTFQKDdec} we plot the asymptotic secret key rate as a function of the loss when the parties employ independent signal and decoy intensities, and each party uses either three (Fig~\ref{Fig_AsymDecoyTFQKD3dec}) or four (Fig~\ref{Fig_AsymDecoyTFQKD4dec}) decoy intensities. In both plots we observe that the improvement given by the use of independent intensities in the asymmetric-loss regions is significant. That is, introducing extra losses in one of the channels does not enhance the key rate any longer, in contrast to Figs~\ref{Fig_SymDecoy3TFQKD} and \ref{Fig_SymDecoy4TFQKD}, where the intensities are instead symmetric for the two parties.

\begin{figure}
	\begin{subfigure}{0.5\textwidth}
		\begin{center}
			\includegraphics[width=\textwidth]{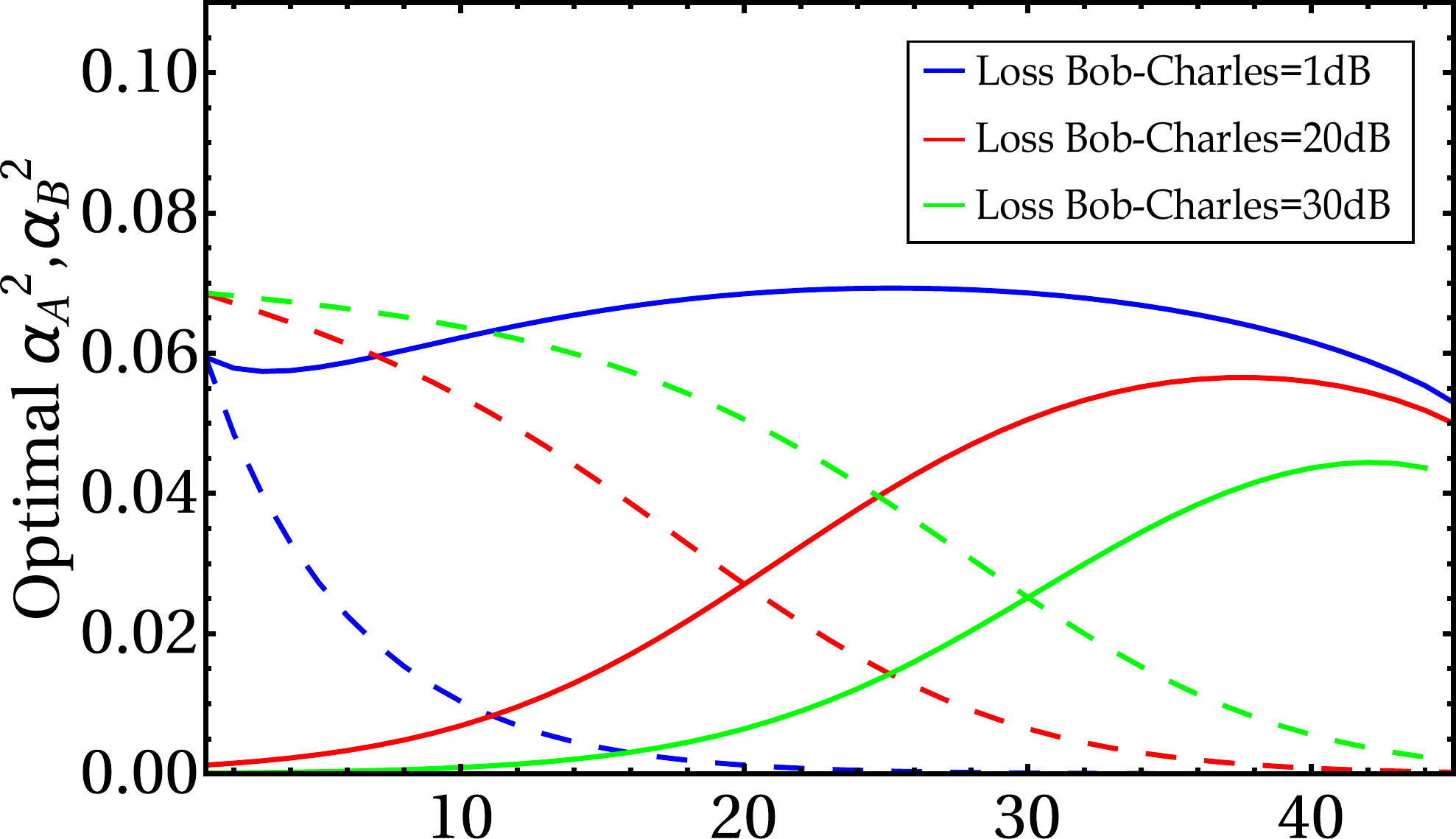}
		\end{center}
		\begin{center}
			\includegraphics[width=\textwidth]{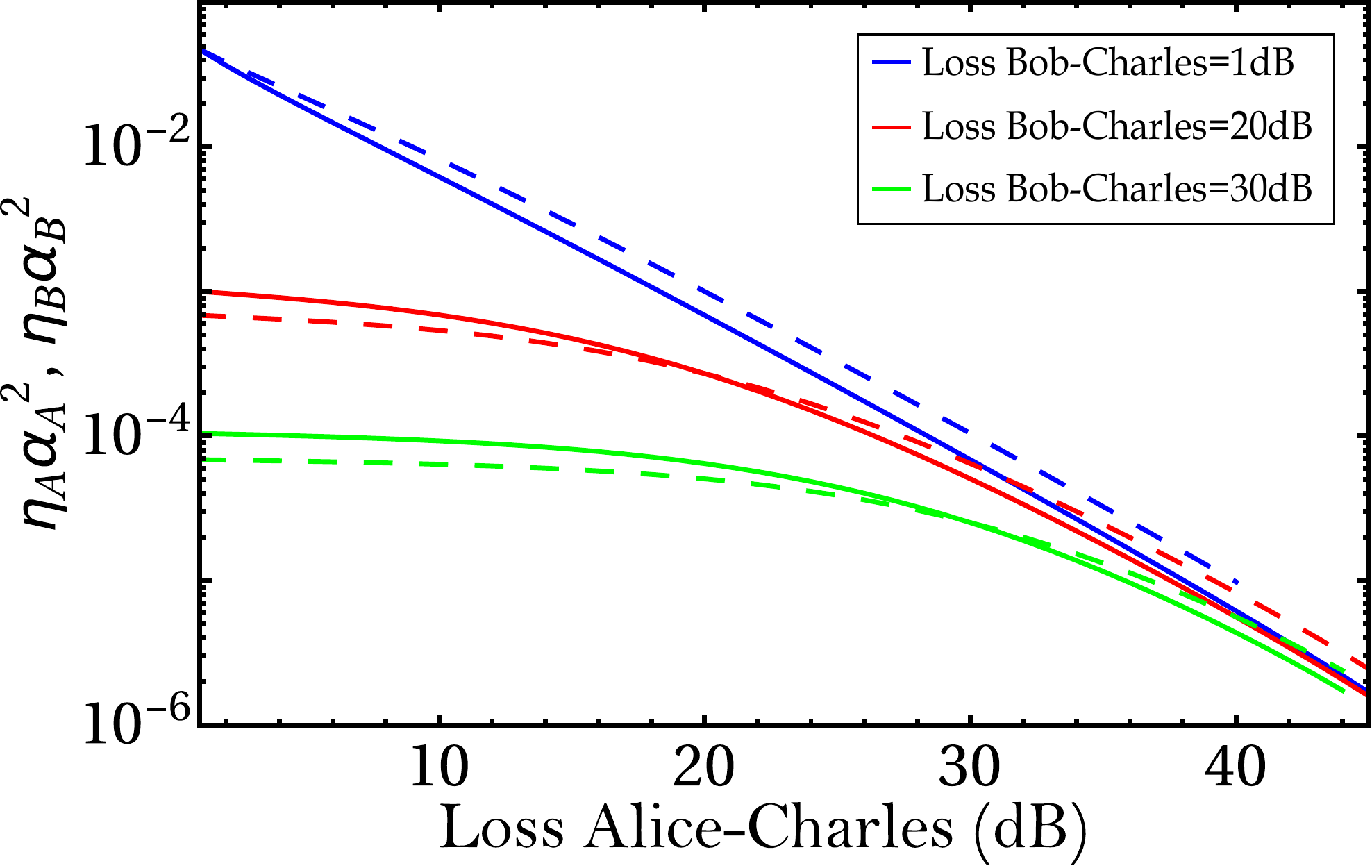}
		\end{center}
		\caption{}
	\end{subfigure}
	\begin{subfigure}{0.5\textwidth}
		\begin{center}
			\includegraphics[width=\textwidth]{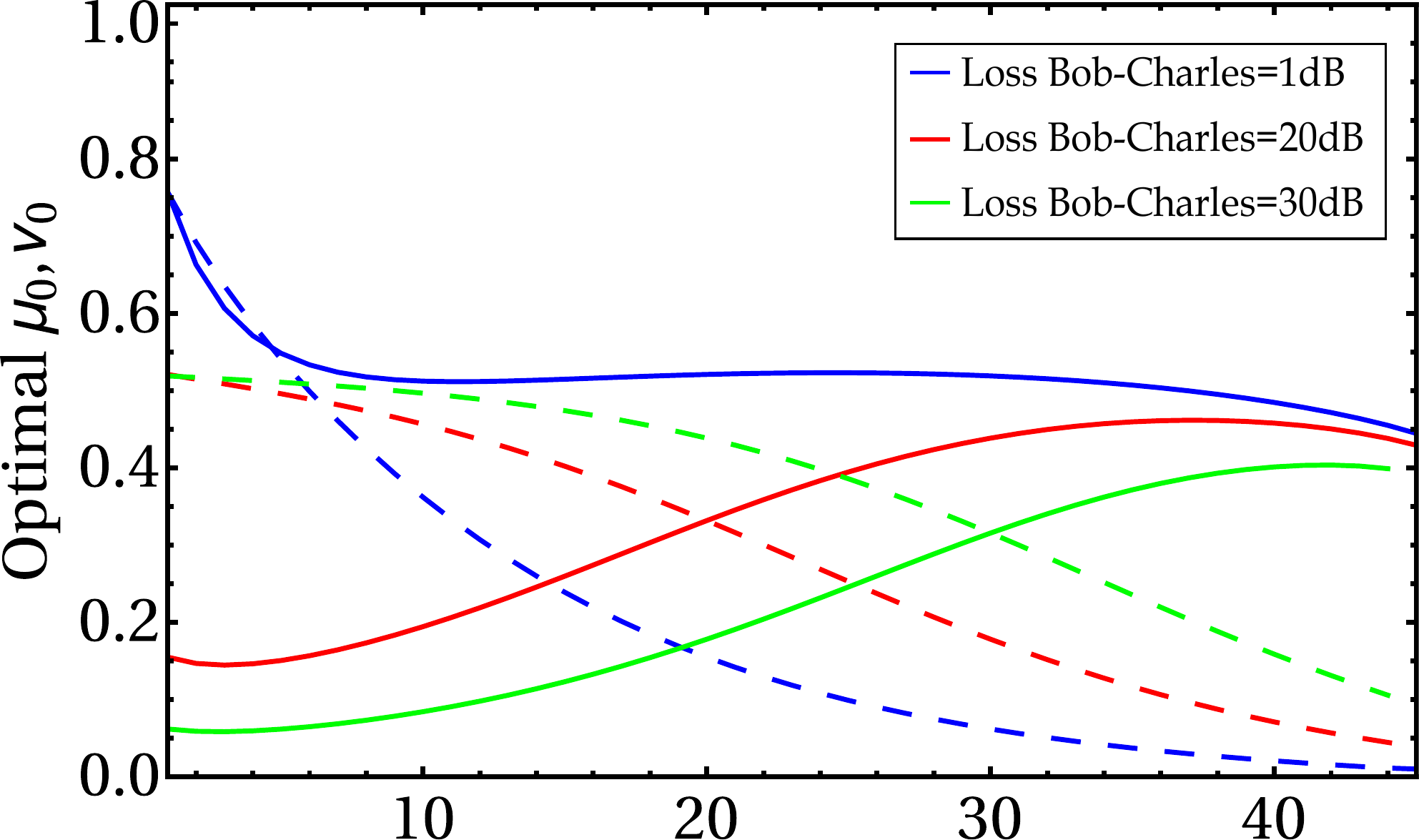}
		\end{center}
		\begin{center}
			\includegraphics[width=\textwidth]{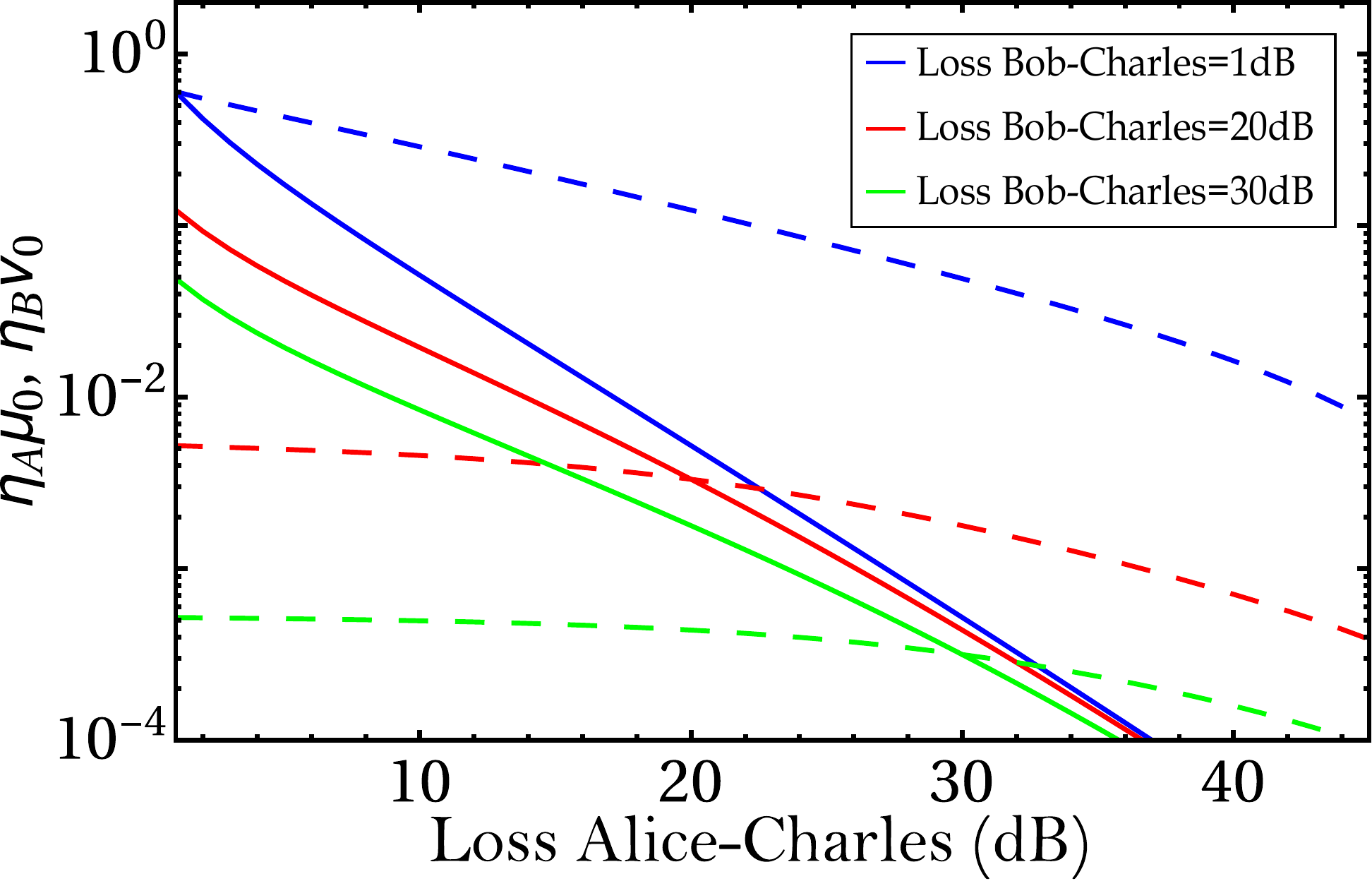}
		\end{center}
		\caption{}
	\end{subfigure}
	\caption{(a) Optimal values of the signal intensities ($\alpha_A^2$ and $\alpha_B^2$) and the arriving signal intensities ($\alpha_A^2\eta_A$ and $\alpha_B^2\eta_B$) both for Alice (solid lines) and Bob (dashed lines). (b) Optimal values of the strongest decoy intensities ($\mu_0$ and $\nu_0$) and the arriving strongest decoy intensities ($\mu_0\eta_A$ and $\nu_0\eta_B$) both for Alice (solid lines) and Bob (dashed lines). All the figures are plotted as a function of the loss in the channel Alice-Charles for three different values of the loss in the channel Bob-Charles. The corresponding optimized key rate is given in Fig~\ref{Fig_AsymDecoyTFQKD3dec}, where each party has independently three decoy intensities. We observe that it is optimal for the parties to prepare the intensities of their pulses such that the signals arriving to Charles have similar intensities, especially for the $X$-basis rounds. The experimental parameters used for the simulations are given in Table~\ref{tab:experimental-parameters}.}\label{optimal_intensities-3dec}
\end{figure}

The simulations suggest that in order to get a high key rate, it is important that the intensities of the pulses arriving at the central node are of similar magnitude (but not exactly the same), so that a \textit{cleaner} interference occurs. This is clear from Fig.~\ref{optimal_intensities-3dec}, where we plot the optimal signal and decoy intensities in the three-decoy scenario, as a function of the loss in the channel Alice-Charles and for fixed losses in the channel Bob-Charles. We note that the optimal signal intensities $\alpha_A^2$ of Alice (solid lines) become greater than the correspondent ones $\alpha_B^2$ of Bob (dashed lines) as soon as the loss in Alice's side is greater than in Bob's side. The same happens for the decoy intensities ($\mu_0$ of Alice and $\nu_0$ of Bob) over which the key rate is optimized. Besides, when the losses at Alice's and Bob's sides are equal (symmetric scenario), the optimal values of both the signal and decoy intensities coincide for Alice and Bob, as expected. Moreover, the bottom plots in Fig.~\ref{optimal_intensities-3dec} show that the signal intensities \textit{arriving} at the untrusted node, i.e. $\eta_A \alpha_A^2$ and $\eta_B \alpha_B^2$, are very similar to each other, while this is less pronounced in the case of the arriving decoy intensities. For completeness, the analogous figure for the four-decoy case is shown in \ref{OptimalIntensities}.

\begin{figure}
	\begin{subfigure}{0.5\textwidth}
		\begin{center}
			\includegraphics[width=\textwidth]{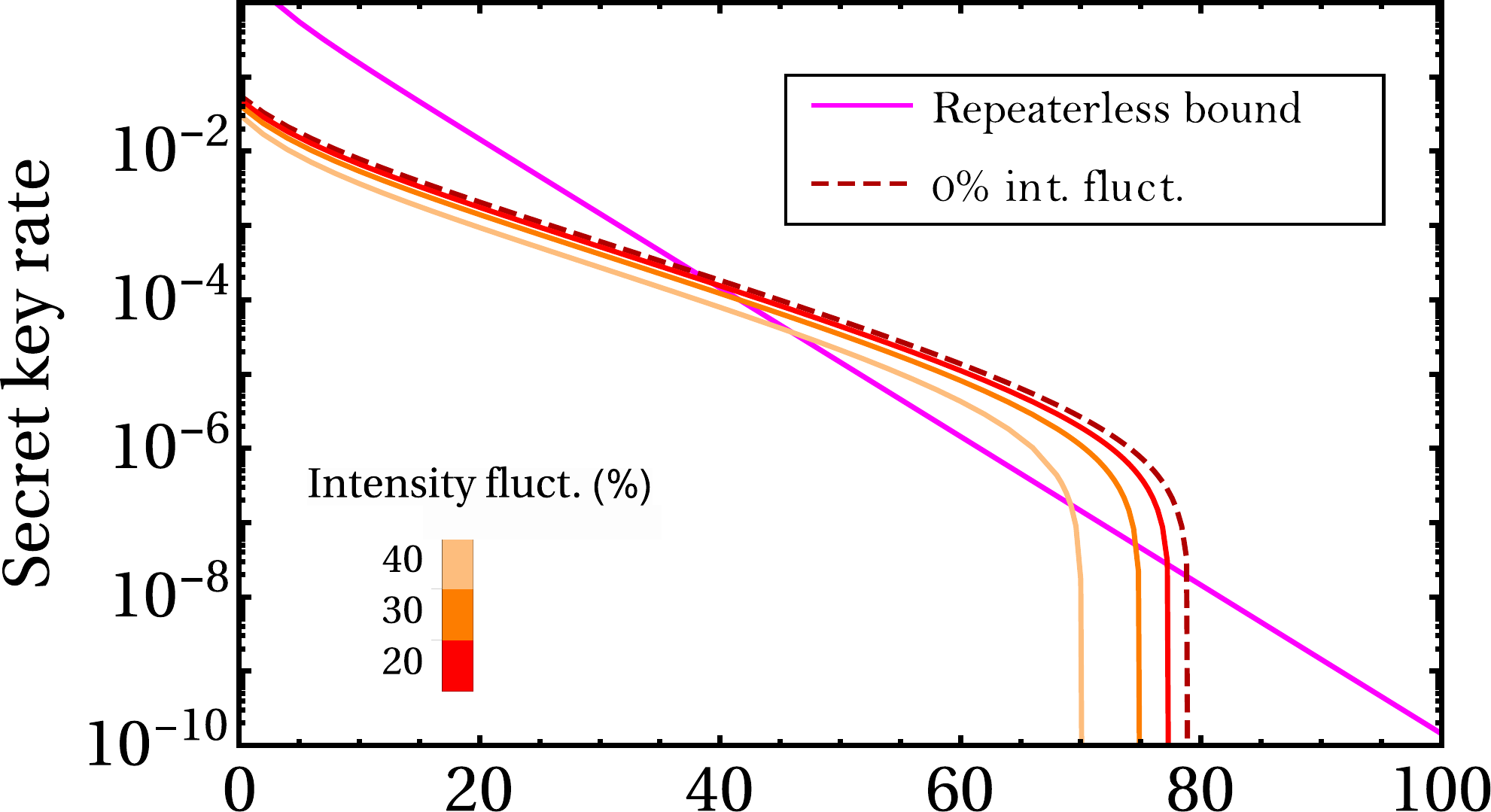}
		\end{center}
		\begin{center}
			\includegraphics[width=\textwidth]{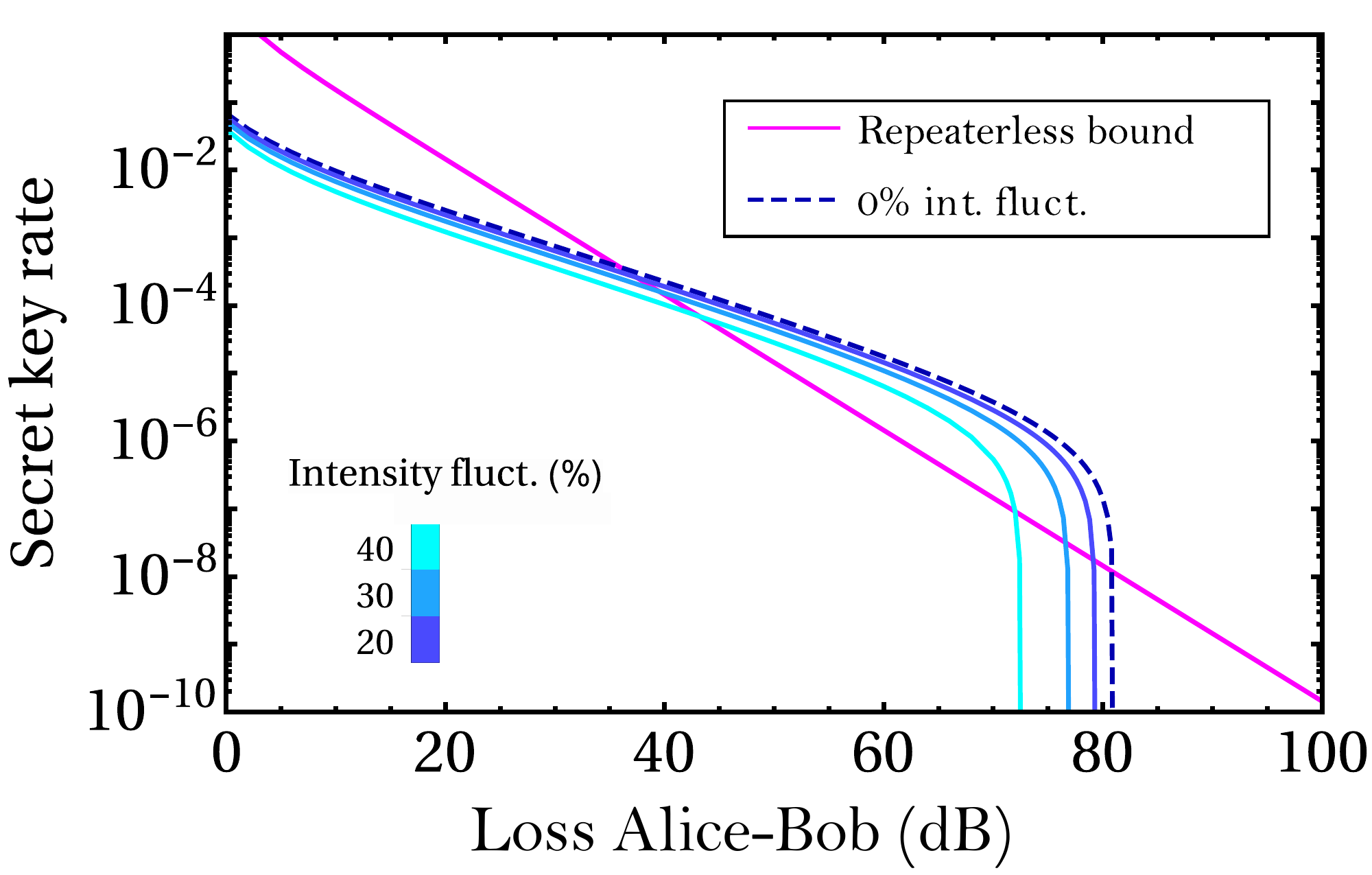}
		\end{center}
		\caption{}\label{Fig_IntFluct}
	\end{subfigure}
	\begin{subfigure}{0.5\textwidth}
		\begin{center}
			\includegraphics[width=\textwidth]{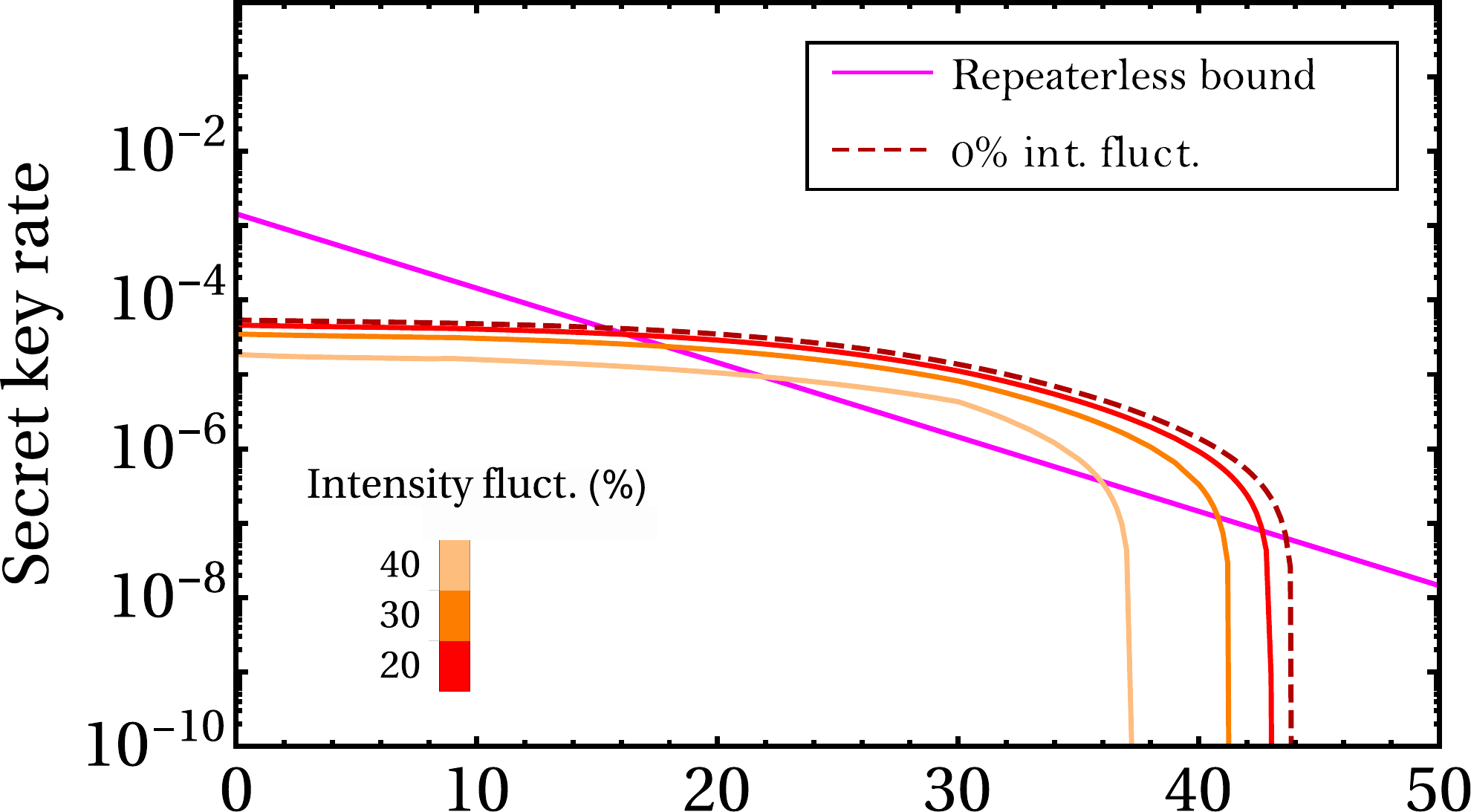}
		\end{center}
		\begin{center}
			\includegraphics[width=\textwidth]{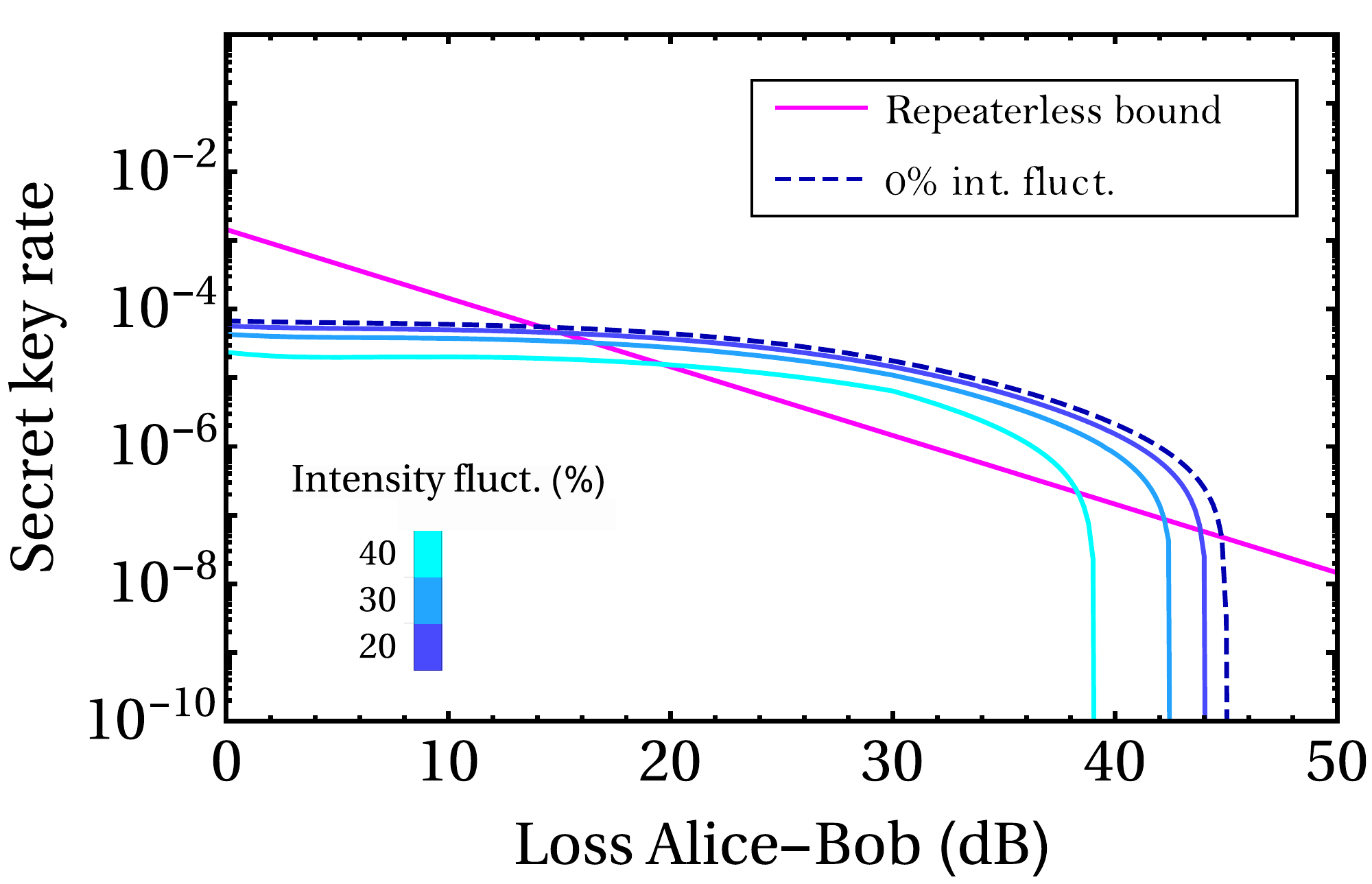}
		\end{center}
		\caption{}\label{Fig_IntFluctAsym}
	\end{subfigure}
	\caption{Comparison between the secret key rate with optimal signal and decoys intensities (dashed lines) with the secret key rates affected by increasing intensity fluctuations (solid lines): 20\%, 30\% and 40\% (brighter colors; right to left). The number of decoy intensity settings are three (red lines) and four (blue lines). In (a) the losses at Alice's and Bob's sides are equal (symmetric-loss scenario) while in (b) the loss in the channel Bob-Charles is fixed to $30\deci\bel$. We assume that the fluctuations affect each decoy intensity and each signal intensity of both parties in a independent way, i.e. the fluctuations are uncorrelated. The plots show that the TF-QKD protocol is quite robust against intensity fluctuations. The weaker decoy intensities have been fixed to the following values: $\mu_2=\nu_2=10^{-3}$ and $\mu_1=\nu_1=10^{-2}$ for three decoy intensity settings, and $\mu_2=\nu_2=10^{-3}$, $\mu_1=\nu_1=10^{-2}$ and $\mu_0=\nu_0=10^{-1}$ for four decoy intensity settings. The other experimental parameters are given in Table~\ref{tab:experimental-parameters}.}
\end{figure}

Apart from the general improvement in the secret key rate that the yields bounds derived in this work entail in asymmetric-loss scenarios, the bounds also allow to incorporate uncorrelated intensity fluctuations into the model, guaranteeing security in more realistic conditions. Fig.~\ref{Fig_IntFluct} illustrates how taking into account the possible intensity fluctuations at the transmitters' lasers affects the key rate. In particular, the dashed lines are obtained by optimizing the key rate over the signal and decoy intensities (we consider three and four decoy intensity settings), for symmetric losses in the two quantum channels. We then apply uncorrelated fluctuations of fixed magnitudes on all the signal and decoy intensities of both parties and take the worst-case key rate (solid lines), i.e. the one minimized by letting each intensity independently fluctuate in its fluctuation range (centered on its optimal value). This effect has already been analyzed in~\cite{grasselli2019practical}, where the fluctuations are, however, assumed to be perfectly correlated among the two users. This is a quite restrictive assumption, which only occurs in practice in certain experimental implementations based on the use of only one laser~\cite{zhong2019proof}, but does not hold in general when two lasers are employed~\cite{minder2019experimental,liu2019experimental,wang2019beating}, even in a scenario with symmetric losses. In order to directly compare the effect of \textit{uncorrelated} fluctuations with the results in~\cite{grasselli2019practical}, we fixed the weaker decoy intensities to exactly the same values used in the intensity fluctuations plots of~\cite{grasselli2019practical}, that is: $\mu_2=\nu_2=10^{-3}$ and $\mu_1=\nu_1=10^{-2}$ for three decoy intensity settings, and $\mu_2=\nu_2=10^{-3}$, $\mu_1=\nu_1=10^{-2}$ and $\mu_0=\nu_0=10^{-1}$ for four decoy intensity settings. The figures suggest that the protocol is quite robust against intensity fluctuations even when the fluctuations are uncorrelated among the two parties. In fact, the maximal tolerable loss in the overall Alice-Bob channel for both the three and four-decoy scenarios decreases less than 2$\deci\bel$ for a 20\% fluctuation of the signal and decoy intensities. Remarkably, even with a fluctuation magnitude of 40\% the decrease is still below 10$\deci\bel$. Similar conclusions hold for the asymmetric scenario shown in Fig.~\ref{Fig_IntFluctAsym}, where the loss in the channel Bob-Charles is fixed to 30 $\deci\bel$.
\begin{figure}[htb!]
	\begin{center}
		\includegraphics[width=0.6\textwidth]{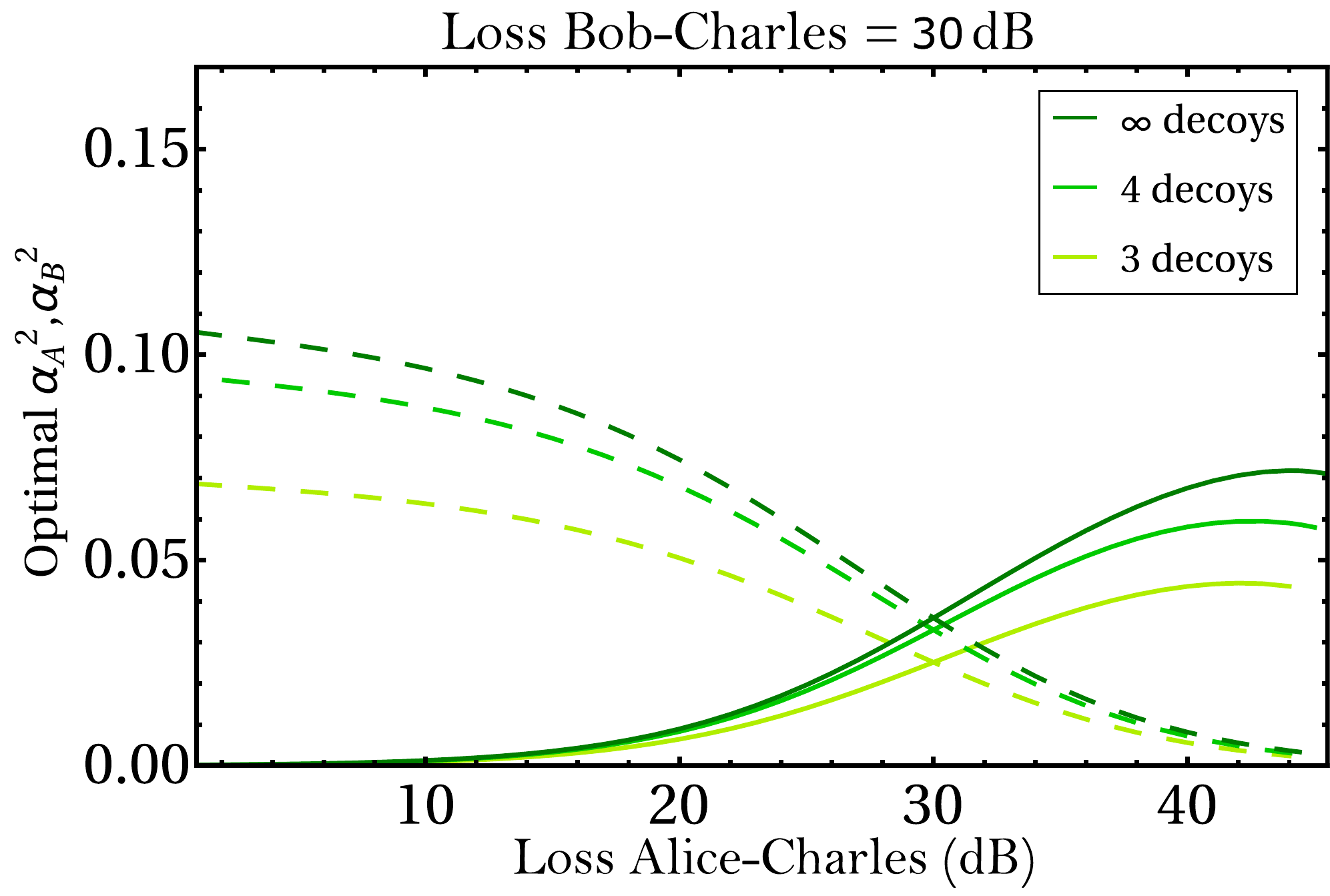}
	\end{center}
	\caption{Optimal values of the signal intensities $\alpha_A^2$ and $\alpha_B^2$ (solid and dashed lines, respectively) as a function of the loss in the channel Alice-Charles for different levels of accuracy in the estimation of the yields, i.e. assuming that the parties have at their disposal three, four and infinite decoy intensity settings (bottom to top). The loss in the channel Bob-Charles is fixed to $30\deci\bel$. We observe that the optimal values of the signal intensities used to prepare states in the $X$ basis increase with the number of decoys used in the $Z$ basis. For the infinite decoy case, we assume that Alice and Bob can estimate the yields precisely and we used the theoretical values of the yields, which are given in \ref{YieldValues}. The experimental parameters used for the simulations are given in Table~\ref{tab:experimental-parameters}.}\label{Fig_OptSignalInt_LossBob30dB}
\end{figure}

Finally, it is also interesting to observe how the optimal values for the signal intensities in the $X$ basis depend on the estimation of the yields in the $Z$ basis. Fig.~\ref{Fig_OptSignalInt_LossBob30dB} shows the variation of the optimal $\alpha_A^2$ and $\alpha_B^2$ ($X$ basis) as a function of the loss in the channel Alice-Charles (the loss in the channel Bob-Charles is fixed to 30$\deci\bel$) for three different levels of accuracy in the estimation of the yields ($Z$ basis).
One can see that, when the yields' estimation is not so tight, the $X$ basis intensities $\alpha_A^2$ and $\alpha_B^2$ tend to be small in order to reduce the weights $c_{n,m}$ of the yields appearing in~(\ref{Eq_error_z}) and compensate the yields' loose upper bounds. By increasing the number of decoys in the $Z$ basis and thus the tightness of the yields' bounds as well as the number of relevant yields which are non-trivially upper bounded, the optimal values of the signal intensities in the $X$ basis also increase, showing that the optimal signal intensities in the $X$ basis depend on the number of decoy states used in the Z basis.


\section{Conclusion}
\label{sec:conclusion}
In this paper we have investigated the performance of the TF-QKD protocol proposed in~\cite{curty2018simple} under the realistic condition of asymmetric losses in the  quantum channels linking Alice and Bob to the intermediate node. For this, we have derived analytical bounds on the relevant yields that appear in the phase error rate expression when the parties use either three or four decoy intensity settings. In contrast to previous results~\cite{grasselli2019practical}, the bounds derived here are valid in the general scenario of independent intensity settings for the two parties, thus optimizing the protocol's performance in the presence of asymmetric losses in the two quantum channels. The simulations show a significant improvement on the secret key rate when using independent signal and decoy intensity settings in several asymmetric-loss scenarios. In particular, the secret key rate is never enhanced by adding fiber in one of the channels in order to symmetrize their losses. Furthermore, we have demonstrated the robustness of the protocol against uncorrelated intensity fluctuations on the transmitters' lasers. These results clearly indicate the suitability of employing the considered TF-QKD protocol in practical QKD networks.

\section{Acknowledgments}
This work was supported by the Spanish Ministry of Economy and Competitiveness (MINECO), the Fondo Europeo de Desarrollo Regional (FEDER) through grant TEC2017-88243-R, and the European Union's Horizon 2020 research and innovation programme under the Marie Sklodowska-Curie grant agreement No 675662. AN gratefully acknowledges support from a FPU scholarship from the Spanish Ministry of Education.

\appendix
\section{Asymmetric channel model}
\label{ChannelModel}\label{AppendixChannelModel}
Here, we present the expected values of the quantities required to calculate the lower bound on the secure key rate given by (\ref{Eq_Rate}), in the case of a typical channel model. The loss between Alice (Bob) and Charles is modeled with a beamsplitter of transmittance $\eta_A$ ($\eta_B$). In order to model the phase and polarization misalignments, let $\phi=\delta\pi$ be a phase shift at Bob's side for some parameter $\delta$ and let $\theta_A$ ($\theta_B$) be the polarization shift angle at Alice's (Bob's) side. Finally, let $p_d$ be the dark-count probability of Charles' detectors, which we assume to be the same for both detectors. Let's define for convenience 
\begin{eqnarray}
\hspace{1cm}\gamma&=&\frac{\eta_A\alpha_A^2 + \eta_B\alpha_B^2}{2},\\
\chi(\phi,\theta)&=&\alpha_A\alpha_B\sqrt{\eta_A\eta_B}\cos(\phi)\cos(\theta) \,\,,
\end{eqnarray}
where $\theta=\theta_A -\theta_B$. Then it can be shown that the bit error rate $e_{X,\Omega}$ and the probability $p_X(\Omega)$ are given by
\begin{eqnarray}
e_{X,\Omega}=\frac{e^{-\chi(\phi,\theta)}-(1-p_d)e^{-\gamma}}{e^{-\chi(\phi,\theta)}+e^{\chi(\phi,\theta)}-2(1-p_d)e^{-\gamma}},
\end{eqnarray}
and
\begin{eqnarray}
	p_X(\Omega)&=&\frac{1}{2}(1-p_d)(e^{-\chi(\phi,\theta)}+e^{\chi(\phi,\theta)})e^{-\gamma}-(1-p_d)^2e^{-2\gamma}.
\end{eqnarray}
Finally, the observed gains $Q^{k,l}$ used by Alice and Bob to calculate the upper bounds on the yields $Y_{nm}$ are just the probabilities that the event $\Omega$ occurred when Alice and Bob chose intensities $\mu_k$ and $\nu_l$ for their PRCS. For this channel model it turns out that the gains read:
\begin{eqnarray}
	\fl Q^{k,l}=(1-p_d)[e^{-(\mu_k\eta_A+\nu_l\eta_B)/2} I_0(\sqrt{\mu_k\nu_l\eta_A\eta_B}\cos(\theta))-(1-p_d)e^{-(\mu_k\eta_A+\nu_l\eta_B)}],
\end{eqnarray}
where $I_0(x)$ is the modified Bessel function of the first kind. Note that due to the balanced redistribution of the incoming photons in the central beam splitter, all the quantities presented here are actually independent of which detector clicked, i.e. they read the same for $\Omega=\Omega_c,\Omega_d$.

In the simulations in the main text we assume that both the total polarization misalignment and phase mismatched are 2\%, that is, we select $\theta=2\arcsin(\sqrt{0.02})$ and $\delta=0.02$.

\section{Theoretical values for the yields}
\label{YieldValues}
In order to check the quality of the analytical bounds on the yields, it is useful to compare them with their theoretical values, i.e. the values directly inferred from the channel model and that Alice and Bob would estimate when using an infinite number of decoy intensities. This is used, for instance, in Fig.~\ref{Fig_OptSignalInt_LossBob30dB}. The theoretical values of the yields $Y_{nm}$, according to the channel model presented in \ref{AppendixChannelModel}, are given by
\begin{eqnarray}
\fl & Y_{nm}=\sum_{k=0}^{n}C_{n,k}^A\sum_{t=0}^{m}C_{m,t}^B\sum_{i=0}^{k}\binom{k}{i}\sum_{j=0}^{t}\binom{t}{j}\sum_{p=\max(0,i+j-t)}^{\min(k,i+j)}  \binom{k}{p}\binom{t}{i+j-p}\tan(\theta_A)^{i+p}\tan(\theta_B)^{i+2j-p}\nonumber\\
\fl &\times (k+t-i-j)!(i+j)!-(1-\eta_A)^n(1-\eta_B)^m,
\end{eqnarray}
where the coefficients $C_{n,k}^A$ and $C_{m,t}^B$ are given by
\begin{eqnarray} 
C_{n,k}^A=\frac{1}{k!2^k}\binom{n}{k}\eta_A^k(1-\eta_A)^{n-k}\cos(\theta_A)^{2k},\nonumber\\
C_{m,t}^B=\frac{1}{t!2^t}\binom{m}{t}\eta_B^t(1-\eta_B)^{m-t}\cos(\theta_B)^{2t}.\nonumber
\end{eqnarray}
Note that the values of the yields are independent of the event $\Omega$.
\section{Non-convexity of the secret key rate with respect to $\vec{p}$}
\label{AppendixNoConvex}

\begin{figure}
	\begin{subfigure}{0.5\textwidth}
		\centering
		\includegraphics[width=\textwidth]{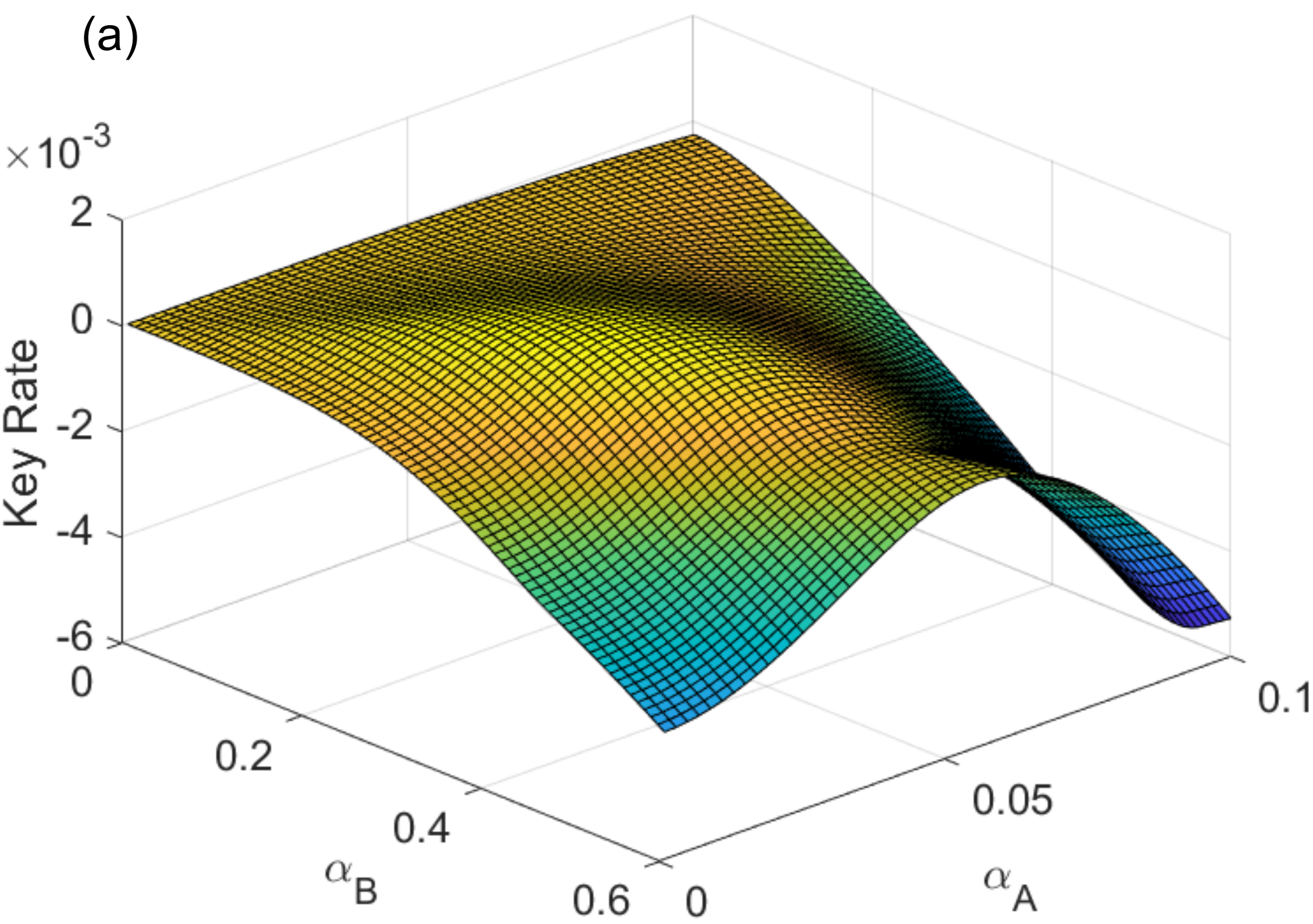}
	\end{subfigure}
	\begin{subfigure}{0.5\textwidth}
		\centering
		\includegraphics[width=\textwidth]{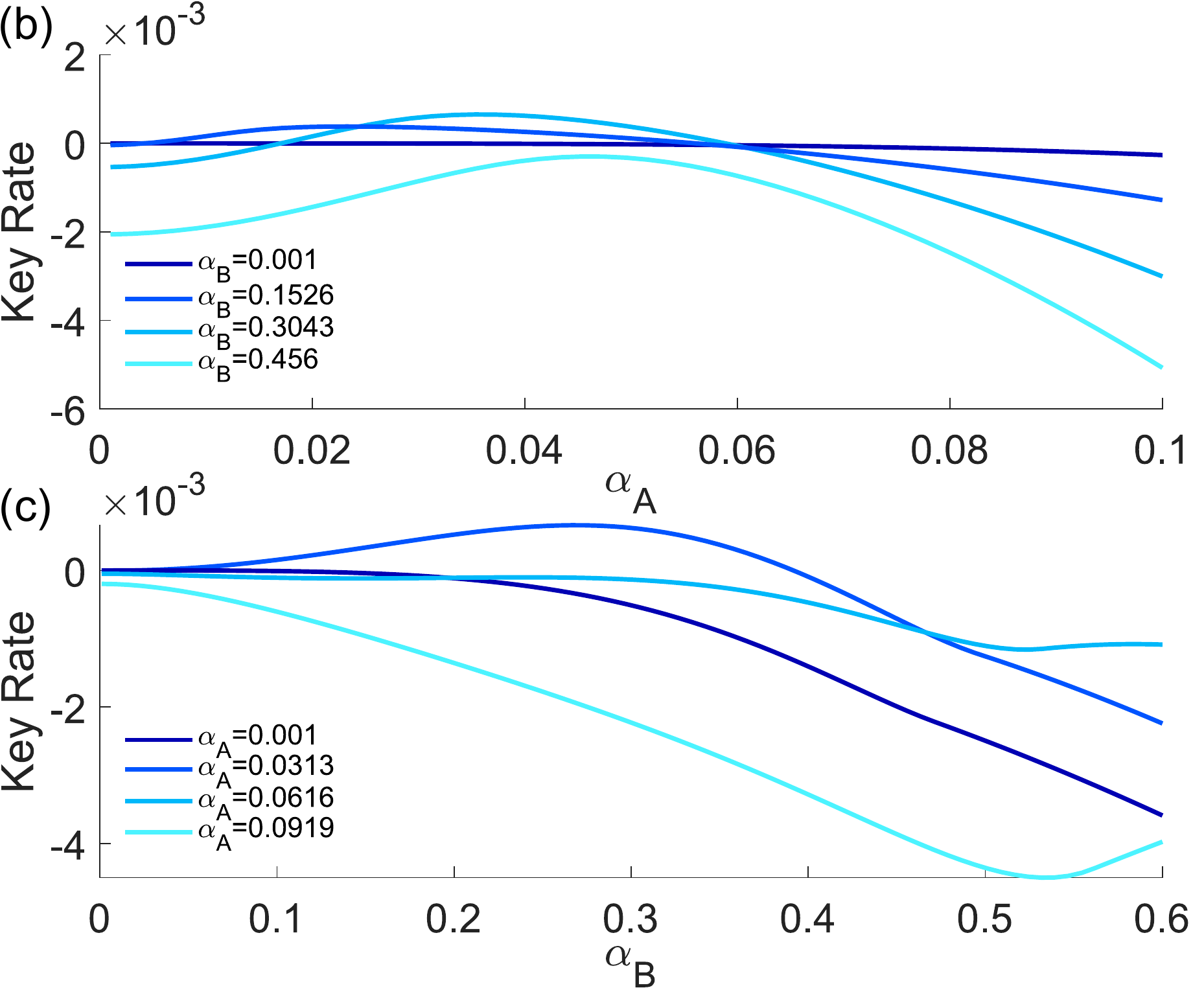}
	\end{subfigure}
	\caption{(a) Lower bound on the secret key rate as a function of $\alpha_A$ and $\alpha_B$. Also, we show in (b) and (c) some specific slices of (a). Here we considered the $\infty$-decoy scenario, and the losses in Alice's and Bob's channels are 20$\deci\bel$ and 0$\deci\bel$, respectively. It is easy to note that the secret key rate function is clearly not convex.}\label{Fig_Conv2}
\end{figure}

As one can notice from Eqs. (\ref{Eq_Rate})-(\ref{Eq_error_x2}), the dependence of the key rate $R$ with its parameters is far from trivial. Here we numerically analyze the convexity of the key rate function $R(\vec{p})$, being $\vec{p}$ the vector of parameters to optimize by the users. It is well-known that this property is noticeably useful since convex functions permit to use efficient optimization methods, which are very important when the length of $\vec{p}$ increases. Unfortunately, it turns out that the key rate function is not convex in general, as shown in Fig.~\ref{Fig_Conv2}, therefore making many efficient optimization algorithms work poorly.

For instance, if we consider the coordinate descent algorithm~\cite{boyd2004convex}, it is clear from the plots that it would not reach the optimal value if the starting point is any corner of the $\alpha_A$-$\alpha_B$ plane and the first variable to optimize is $\alpha_B$. Note that starting from a corner basically means that, in the first step, the algorithm have to maximize the darkest or the lightest line in Fig.~\ref{Fig_Conv2} (c), being both maximized when $\alpha_B$ is minimal. This means that, in the next step, the algorithm has always to optimize the darkest line in Fig.~\ref{Fig_Conv2} (b), which again has its maximum when $\alpha_A$ is minimal. In Fig.~\ref{Fig_Conv2}, for simplicity, we assume that Alice and Bob can estimate the yields precisely. That is, we assume they use an infinite number of decoy intensities.

\section{Optimal signal and decoy intensities for the four-decoy case}
\label{OptimalIntensities}
\begin{figure} 
	\begin{subfigure}{0.5\textwidth}
		\begin{center}
			\includegraphics[width=\textwidth]{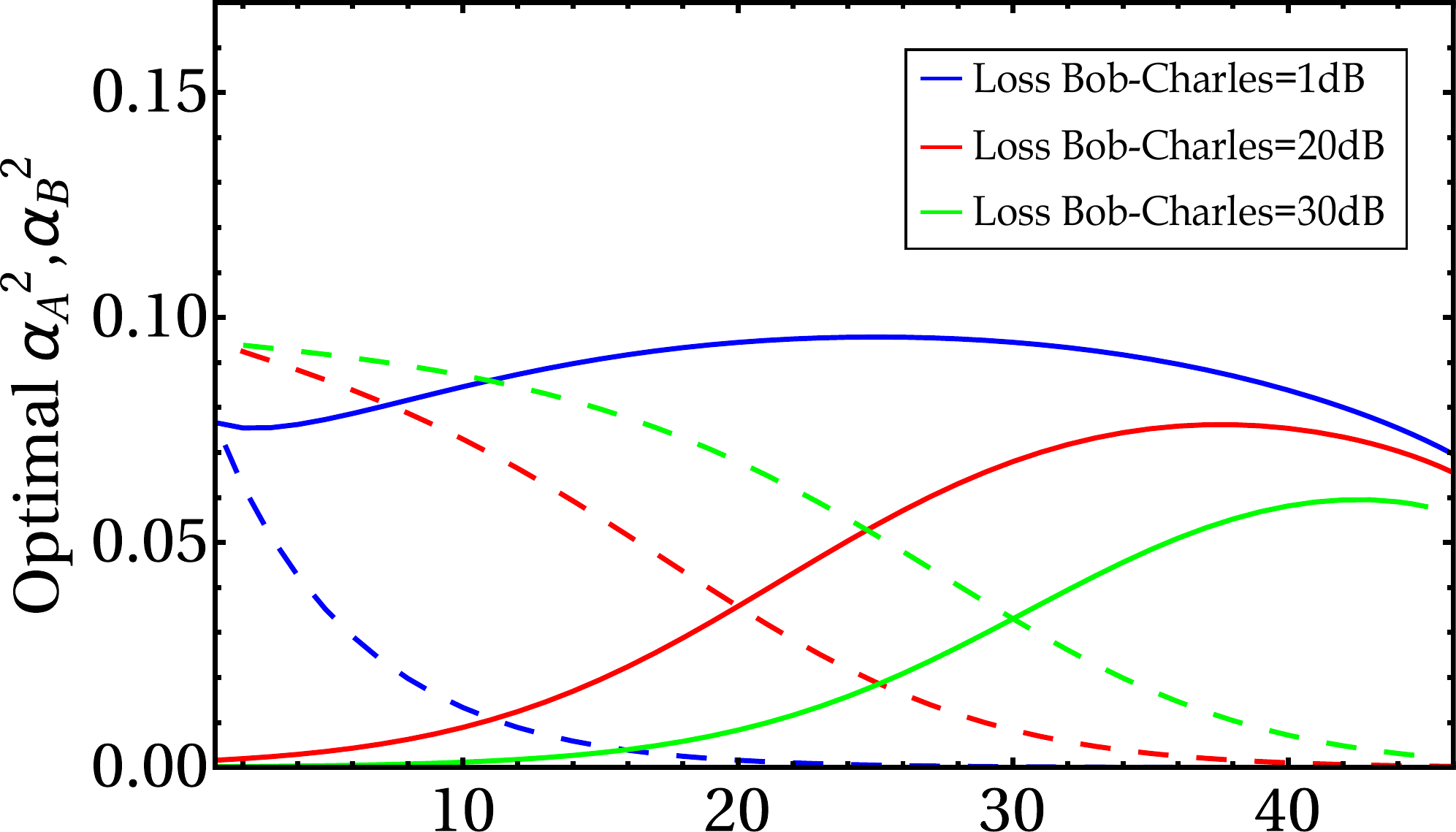}
		\end{center}
		\begin{center}
			\includegraphics[width=\textwidth]{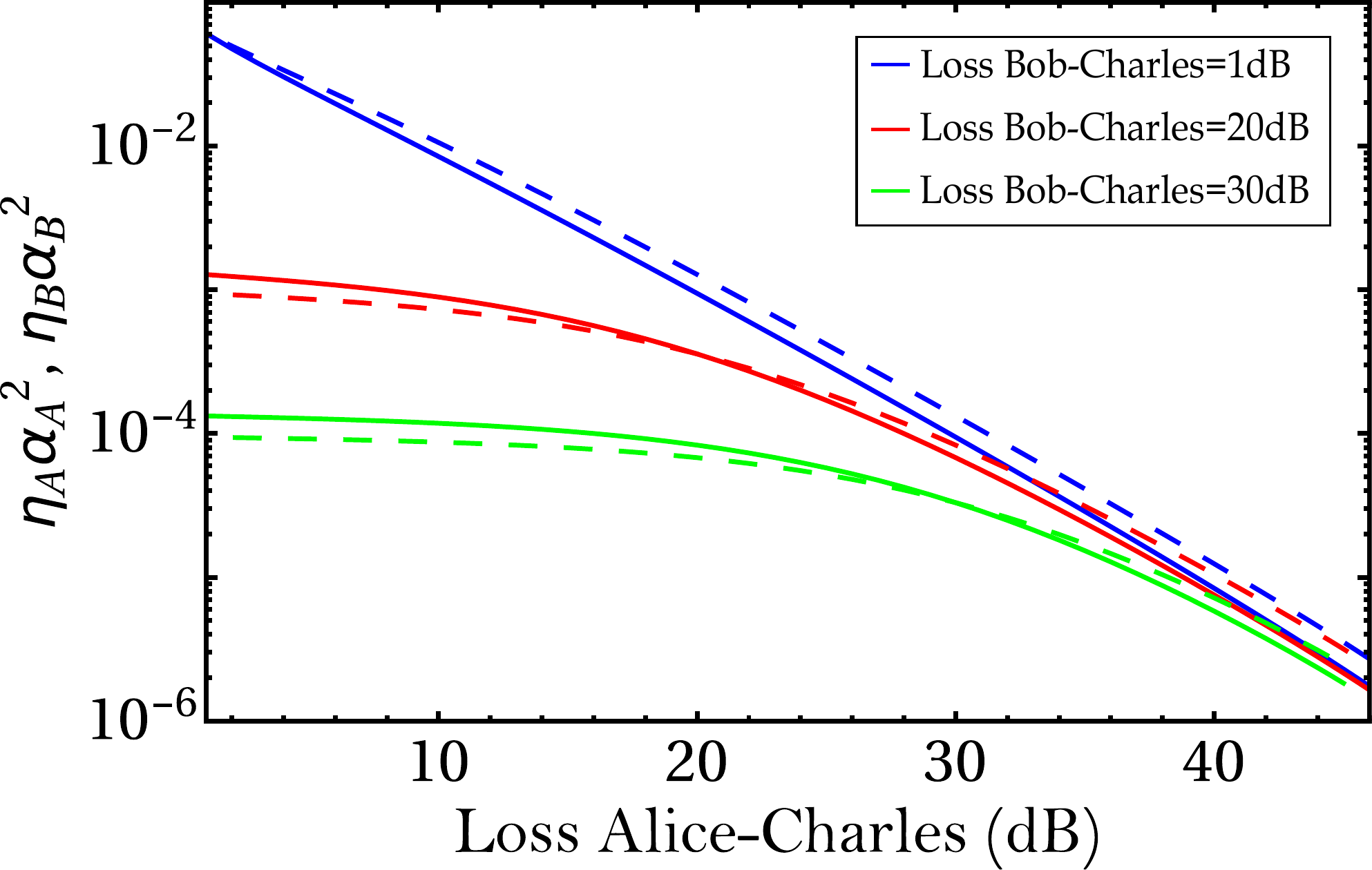}
		\end{center}
	\caption{}
	\end{subfigure}
	\begin{subfigure}{0.5\textwidth}
		\begin{center}
			\includegraphics[width=\textwidth]{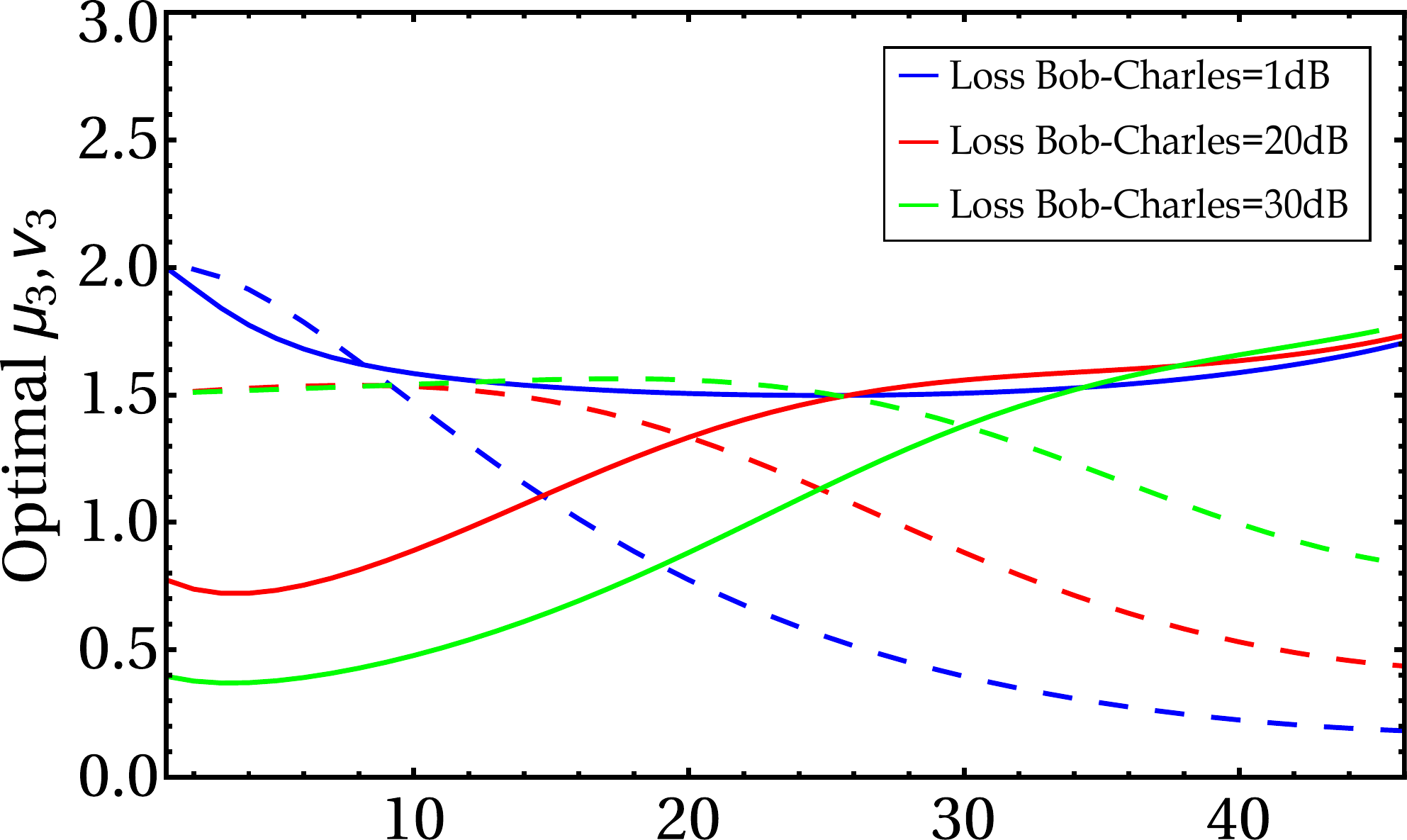}
		\end{center}
		\begin{center}
			\includegraphics[width=\textwidth]{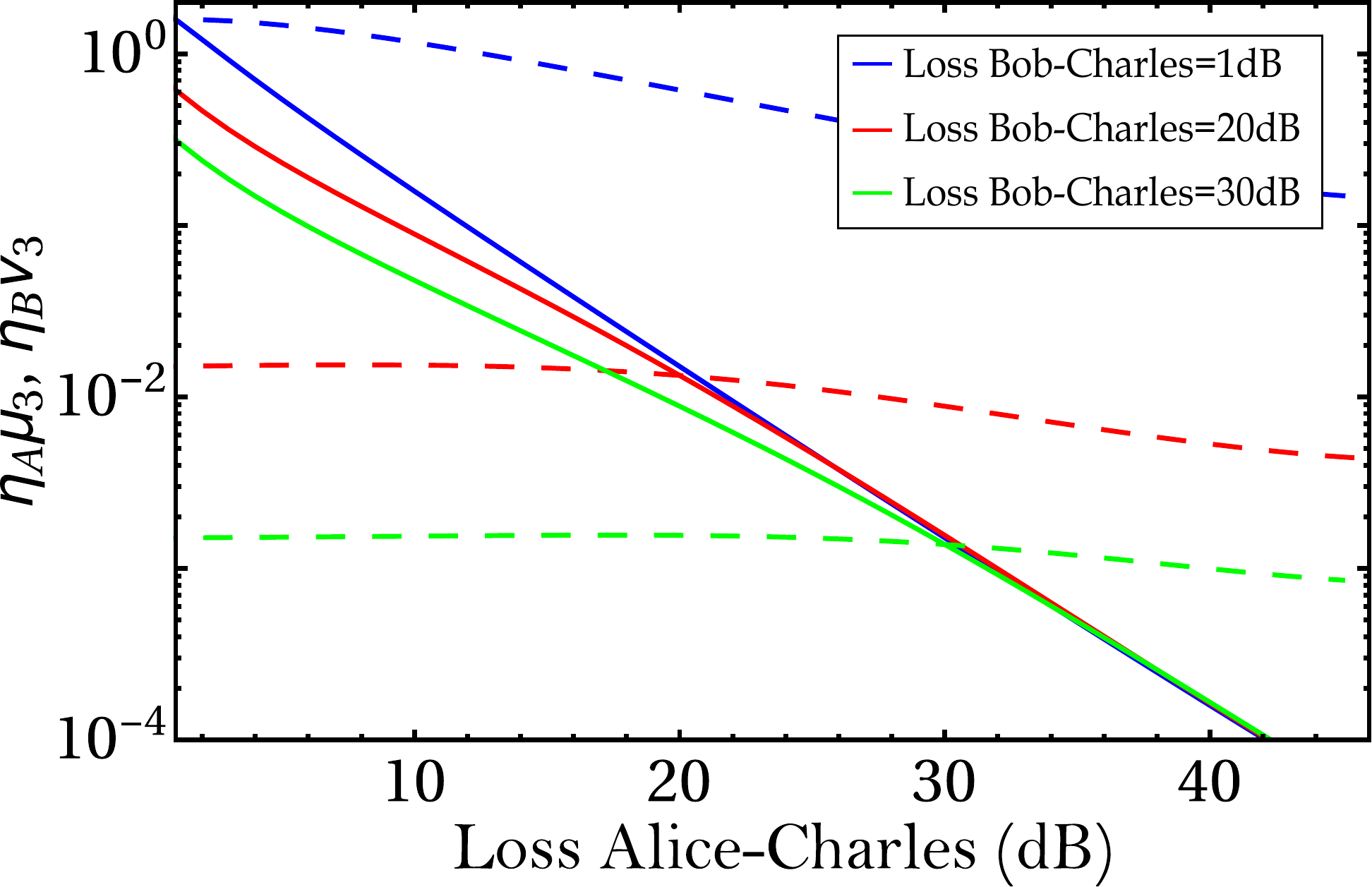}
		\end{center}
	\caption{}
	\end{subfigure}
	\caption{(a) Optimal values of the signal intensities ($\alpha_A^2$ and $\alpha_B^2$) and the arriving signal intensities ($\alpha_A^2\eta_A$ and $\alpha_B^2\eta_B$) both for Alice (solid lines) and Bob (dashed lines). (b) Optimal values of the strongest decoy intensities ($\mu_3$ and $\nu_3$) and the arriving strongest decoy intensities ($\mu_3\eta_A$ and $\nu_3\eta_B$) both for Alice (solid lines) and Bob (dashed lines). All the figures are plotted as a function of the loss in the channel Alice-Charles for three different values of the loss in the channel Bob-Charles. The corresponding optimized key rate is given in Fig~\ref{Fig_AsymDecoyTFQKD4dec}, where each party has four independent decoy intensities. Like in the three-decoy case, we observe that it is optimal for the parties to prepare the intensities of their pulses such that the signals arriving to Charles have similar intensities, especially for the $X$-basis rounds. The experimental parameters used for the simulations are given in Table~\ref{tab:experimental-parameters}.}
	\label{optimal_intensities-4dec}
\end{figure}

In Fig.~\ref{optimal_intensities-4dec} we show, for completeness, the optimal signal and decoy intensities for the four-decoy case.

\section{Upper bounds on the yields with three decoy intensities}
\label{3intensities}

Here we derive the upper bounds on the yields $Y_{00},Y_{11},Y_{02},Y_{20},Y_{22},Y_{13},Y_{31},Y_{04}$ and $Y_{40}$ presented in~\autoref{yields-bounds-2decoys}.

\subsection{Upper bound on $Y_{22}$} \label{bound_on_Y22-3decoys}
We consider the most general combination of the nine constraints (\ref{constr-3decoys}):
\begin{eqnarray}
G_{22}=\sum_{i,j=0}^{2} c_{i,j} \tilde{Q}^{i,j} = \sum_{n,m=0}^{\infty} \frac{Y_{nm}}{n!m!}\left[\sum_{i,j=0}^{2} c_{i,j} \mu_i^n \nu_j^m \right] \,\,, \label{G-22}
\end{eqnarray}
and require that the terms proportional to $Y_{0m},Y_{1m},Y_{n0}$ and $Y_{n1}$ are removed in the combination. We achieve this by imposing proper conditions on the real coefficients $c_{i,j}$:
\begin{eqnarray}
\fl &Y_{n0}\mbox{ removed:}\,\, \sum_{i=0}^{2} \mu_i^n \left(\sum_{j=0}^{2} c_{i,j}\right) =0 \quad\forall\, n \quad\Leftarrow\quad c_{i,0}+c_{i,1}+c_{i,2}=0 
\quad \mbox{for} \,\, i=0,1,2  \label{Yn0removed-22} \\
\fl &Y_{n1}\mbox{ removed:}\,\, \sum_{i=0}^{2} \mu_i^n \left(\sum_{j=0}^{2} \nu_j c_{i,j}\right) =0 \quad\forall\, n \quad\Leftarrow\quad 
\nu_0 c_{i,0}+ \nu_1 c_{i,1}+\nu_2 c_{i,2}=0 \quad \mbox{for} \,\, i=0,1,2  \label{Yn1removed-22} \\
\fl &Y_{0m}\mbox{ removed:}\,\, \sum_{j=0}^{2} \nu_j^m \left(\sum_{i=0}^{2} c_{i,j}\right) =0 \quad\forall\, m \quad\Leftarrow\quad c_{0,j}+c_{1,j}+c_{2,j}=0 
\quad \mbox{for} \,\, j=0,1,2  \label{Y0mremoved-22} \\
\fl &Y_{1m}\mbox{ removed:}\,\, \sum_{j=0}^{2} \nu_j^m \left(\sum_{i=0}^{2} \mu_i c_{i,j}\right) =0 \quad\forall\, m \quad\Leftarrow\quad 
\mu_0 c_{0,j}+\mu_1 c_{1,j}+ \mu_2 c_{2,j}=0 \quad \mbox{for} \,\, j=0,1,2  \label{Y1mremoved-22}  \,.
\end{eqnarray}
The linear system of equations given by (\ref{Yn0removed-22}-\ref{Y1mremoved-22}) has a unique solution in the variables $c_{i,j}$ (up to a global factor that we fix by imposing $c_{0,0}=1$), which reads as follows:
\begin{eqnarray}
&c_{0,0}=1 \nonumber\,\,,\\
&c_{0,1}=\frac{\nu_2-\nu_0}{\nu_1-\nu_2}  \nonumber\,\,,\\
&c_{0,2}=\frac{\nu_0-\nu_1}{\nu_1-\nu_2} \nonumber\,\,,\\
&c_{1,0}=\frac{\mu_2-\mu_0}{\mu_1-\mu_2} \nonumber\,\,,\\
&c_{1,1}= \frac{(\mu_0-\mu_2) (\nu_0-\nu_2)}{(\mu_1-\mu_2) (\nu_1-\nu_2)}  \nonumber\,\,,\\
&c_{1,2}= \frac{(\mu_0-\mu_2) (\nu_0-\nu_1)}{(\mu_2-\mu_1) (\nu_1-\nu_2)} \nonumber\,\,,\\
&c_{2,0}=\frac{\mu_0-\mu_1}{\mu_1-\mu_2} \nonumber\,\,,\\
&c_{2,1}=\frac{(\mu_0-\mu_1) (\nu_0-\nu_2)}{(\mu_1-\mu_2) (\nu_2-\nu_1)} \nonumber\,\,,\\
&c_{2,2}=\frac{(\mu_0-\mu_1) (\nu_0-\nu_1)}{(\mu_1-\mu_2) (\nu_1-\nu_2)}  \label{cij-22}\,\,.
\end{eqnarray}
By substituting the solution for the coefficients $c_{i,j}$ (\ref{cij-22}) back into (\ref{G-22}) one gets:
\begin{eqnarray}
\fl G_{22}= \sum_{n,m=2}^{\infty}\frac{Y_{nm}}{n!m!}\frac{A_{22}(\mu_0,\mu_1,\mu_2,n)A_{22}(\nu_0,\nu_1,\nu_2,m)}{(\mu_1-\mu_2)(\nu_1-\nu_2)}   \,\,, \label{G-22-1}
\end{eqnarray}
where
\begin{eqnarray}
A_{22}(\mu_0,\mu_1,\mu_2,n) \equiv \mu_1^n (\mu_0 - \mu_2) + \mu_2^n (\mu_1 - \mu_0) + \mu_0^n (\mu_2 - \mu_1) \label{A-22}  \,\,,
\end{eqnarray}
is the function defined in \cite{grasselli2019practical} when obtaining the analogous bound on $Y_{22}$ in the symmetric-intensities scenario (i.e. when the decoy intensities of Alice and Bob are drawn from the same set). Thus we can employ the result from \cite{grasselli2019practical} and recast (\ref{A-22}) as follows:
\begin{eqnarray}
\fl	A_{22}(\mu_0,\mu_1,\mu_2,n) =(\mu_0 - \mu_2)(\mu_2 - \mu_1) \sum_{k=0}^{n-1} \mu_2^k (\mu_0^{n-1-k}-\mu_1^{n-1-k})  \label{A-22_1} \,\,.
\end{eqnarray}
Of course we can employ this expression also for $A_{22}(\nu_0,\nu_1,\nu_2,m)$ by making the proper substitutions. We will apply this consideration from now on to similar scenarios.
By employing (\ref{A-22_1}) into (\ref{G-22-1}) one gets:
\begin{eqnarray}
\fl G_{22}= \sum_{n,m=2}^{\infty}\frac{Y_{nm}}{n!m!}(\mu_0 -\mu_2)(\nu_0 -\nu_2)  \sum_{k=0}^{n-1} \mu_2^k (\mu_0^{n-1-k}-\mu_1^{n-1-k}) \sum_{j=0}^{m-1} \nu_2^j (\nu_0^{m-1-j}-\nu_1^{m-1-j}) \,\,. \label{G-22-2}
\end{eqnarray}
From (\ref{G-22-2}) we deduce that the sign of $Y_{nm}$'s coefficient is independent of $n$ and $m$ and it is the same for all terms in the sum. Thus a valid upper bound for $Y_{22}$ is obtained by setting all the other yields to zero in (\ref{G-22-2}), except for $Y_{22}$. By doing this, we obtain~(\ref{Y22-upperbound-3decoys}).

\subsection{Upper bound on $Y_{11}$} \label{bound_on_Y11-3decoys}
We consider the most general combination of the nine equality constraints:
\begin{eqnarray}
G_{11}=\sum_{i,j=0}^{2} c_{i,j} \tilde{Q}^{i,j} = \sum_{n,m=0}^{\infty} \frac{Y_{nm}}{n!m!}\left[\sum_{i,j=0}^{2} c_{i,j} \mu_i^n \nu_j^m \right] \,\,, \label{G-11-3decoys}
\end{eqnarray}
and require that the terms proportional to $Y_{0m},Y_{2m},Y_{n0}$ and $Y_{n2}$ are removed in the combination. We achieve this by imposing proper conditions on the real coefficients $c_{i,j}$:
\begin{eqnarray}
\fl &Y_{n0}\mbox{ removed:}\,\, \sum_{i=0}^{2} \mu_i^n \left(\sum_{j=0}^{2} c_{i,j}\right) =0 \quad\forall\, n \quad\Leftarrow\quad c_{i,0}+c_{i,1}+c_{i,2}=0 
\quad \mbox{for} \,\, i=0,1,2  \label{Yn0removed-11} \\
\fl &Y_{n2}\mbox{ removed:}\,\, \sum_{i=0}^{2} \mu_i^n \left(\sum_{j=0}^{2} \nu^2_j c_{i,j}\right) =0 \quad\forall\, n \quad\Leftarrow\quad 
\nu^2_0 c_{i,0}+ \nu^2_1 c_{i,1}+\nu^2_2 c_{i,2}=0 \quad \mbox{for} \,\, i=0,1,2  \label{Yn2removed-11} \\
\fl &Y_{0m}\mbox{ removed:}\,\, \sum_{j=0}^{2} \nu_j^m \left(\sum_{i=0}^{2} c_{i,j}\right) =0 \quad\forall\, m \quad\Leftarrow\quad c_{0,j}+c_{1,j}+c_{2,j}=0 
\quad \mbox{for} \,\, j=0,1,2  \label{Y0mremoved-11} \\
\fl &Y_{2m}\mbox{ removed:}\,\, \sum_{j=0}^{2} \nu_j^m \left(\sum_{i=0}^{2} \mu^2_i c_{i,j}\right) =0 \quad\forall\, m \quad\Leftarrow\quad 
\mu^2_0 c_{0,j}+\mu^2_1 c_{1,j}+ \mu^2_2 c_{2,j}=0 \quad \mbox{for} \,\, j=0,1,2  \label{Y2mremoved-11}  \,.
\end{eqnarray}
The linear system of equations given by (\ref{Yn0removed-11}-\ref{Y2mremoved-11}) has a unique solution in the variables $c_{i,j}$ (up to a global factor that we fix by imposing $c_{0,0}=1$), which reads as follows:
\begin{eqnarray}
&c_{0,0}=1 \nonumber\,\,,\\
&c_{0,1}=\frac{\nu_2^2-\nu_0^2}{\nu_1^2-\nu_2^2}  \nonumber\,\,,\\
&c_{0,2}=\frac{\nu_0^2-\nu_1^2}{\nu_1^2-\nu_2^2} \nonumber\,\,,\\
&c_{1,0}=\frac{\mu_2^2-\mu_0^2}{\mu_1^2-\mu_2^2} \nonumber\,\,,\\
&c_{1,1}= \frac{\left(\mu_0^2-\mu_2^2\right) \left(\nu_0^2-\nu_2^2\right)}{\left(\mu_1^2-\mu_2^2\right) \left(\nu_1^2-\nu_2^2\right)}  \nonumber\,\,,\\
&c_{1,2}= \frac{\left(\mu_0^2-\mu_2^2\right) \left(\nu_0^2-\nu_1^2\right)}{\left(\mu_2^2-\mu_1^2\right) \left(\nu_1^2-\nu_2^2\right)} \nonumber\,\,,\\
&c_{2,0}=\frac{\mu_0^2-\mu_1^2}{\mu_1^2-\mu_2^2} \nonumber\,\,,\\
&c_{2,1}=\frac{\left(\mu_0^2-\mu_1^2\right) \left(\nu_0^2-\nu_2^2\right)}{\left(\mu_1^2-\mu_2^2\right) \left(\nu_2^2-\nu_1^2\right)} \nonumber\,\,,\\
&c_{2,2}=\frac{\left(\mu_0^2-\mu_1^2\right) \left(\nu_0^2-\nu_1^2\right)}{\left(\mu_1^2-\mu_2^2\right) \left(\nu_1^2-\nu_2^2\right)}  \label{cij-11}\,\,.
\end{eqnarray}
By substituting the solution for the coefficients $c_{i,j}$ (\ref{cij-11}) back into (\ref{G-11-3decoys}) one gets:
\begin{eqnarray}
\fl &G_{11}= Y_{11}\frac{(\mu_0 -\mu_1)(\mu_0 -\mu_2)(\nu_0 -\nu_1)(\nu_0 -\nu_2)}{(\mu_1+\mu_2)(\nu_1+\nu_2)}+ \sum_{m=3}^{\infty} \frac{Y_{1m}}{m!}
\frac{(\mu_0 -\mu_1)(\mu_0 - \mu_2)}{(\mu_1 + \mu_2)(\nu_1^2 - \nu_2^2)} A_{11} (\nu_0,\nu_1,\nu_2,m) \nonumber\\
\fl &+ \sum_{n=3}^{\infty} \frac{Y_{n1}}{n!} \frac{(\nu_0 -\nu_1)(\nu_0 - \nu_2)}{(\nu_1 + \nu_2)(\mu_1^2 - \mu_2^2)} A_{11} (\mu_0,\mu_1,\mu_2,n) +  \sum_{n,m=3}^{\infty}\frac{Y_{nm}}{n!m!}\frac{A_{11}(\mu_0,\mu_1,\mu_2,n)A_{11}(\nu_0,\nu_1,\nu_2,m)}{(\mu^2_1-\mu^2_2)(\nu^2_1-\nu^2_2)}   \,\,. \label{G-11-1}
\end{eqnarray}
The function $A_{11}(\mu_0,\mu_1,\mu_2,n)$ is defined in \cite{grasselli2019practical} when deriving the analogous bound in the symmetric-intensities scenario. It reads:
\begin{equation}
A_{11}(\mu_0,\mu_1,\mu_2,n)\equiv \mu_1^n (\mu^2_0 - \mu^2_2) + \mu_2^n (\mu^2_1 - \mu^2_0) + \mu_0^n (\mu^2_2 - \mu^2_1) \label{A-11} \,\,,
\end{equation}
and can be recast as:
\begin{equation}
A_{11}(\mu_0,\mu_1,\mu_2,n)= (\mu_0 - \mu_2)(\mu_1-\mu_2)(\mu_1-\mu_0) F(\mu_0,\mu_1,\mu_2,n) \quad\mbox{for}\,n\geq 3\,, \label{A-11-1}
\end{equation}
with $F(\mu_0,\mu_1,\mu_2,n)$ being a non-negative quantity independently of the intensities, defined as:
\begin{equation}
F(\mu_0,\mu_1,\mu_2,n) \equiv \sum_{k=0}^{n-3} \mu_2^k \left[(\mu_2+\mu_0)\sum_{j=0}^{n-3-k}\mu_1^{n-2-k-j}\mu_0^j +\mu_2\mu_0^{n-2-k}\right] \label{Fn}\,\,.
\end{equation}
By employing the expression (\ref{A-11-1}) in (\ref{G-11-1}) we obtain:
\begin{eqnarray}
\fl &G_{11}= Y_{11}\frac{(\mu_0 -\mu_1)(\mu_0 -\mu_2)(\nu_0 -\nu_1)(\nu_0 -\nu_2)}{(\mu_1+\mu_2)(\nu_1+\nu_2)} \nonumber\\
\fl &+ \sum_{m=3}^{\infty} \frac{Y_{1m}}{m!}\frac{(\mu_0 -\mu_1)(\mu_0 - \mu_2)}{(\mu_1 + \mu_2)(\nu_1 + \nu_2)} (\nu_0 - \nu_2)(\nu_1 -\nu_0) F(\nu_0,\nu_1,\nu_2,m) \nonumber\\
\fl &+ \sum_{n=3}^{\infty} \frac{Y_{n1}}{n!} \frac{(\nu_0 -\nu_1)(\nu_0 - \nu_2)}{(\nu_1 + \nu_2)(\mu_1 + \mu_2)} (\mu_0 -\mu_2)(\mu_1 - \mu_0) F(\mu_0,\mu_1,\mu_2,n) \nonumber\\
\fl &+  \sum_{n,m=3}^{\infty}\frac{Y_{nm}}{n!m!}\frac{(\mu_0 -\mu_2)(\mu_1 - \mu_0)F(\mu_0,\mu_1,\mu_2,n)(\nu_0 - \nu_2)(\nu_1 -\nu_0) F(\nu_0,\nu_1,\nu_2,m)}{(\mu_1+\mu_2)(\nu_1+\nu_2)}   \,\,. \label{G-11-2}
\end{eqnarray}
By looking at (\ref{G-11-2}), we deduce that a valid upper bound on $Y_{11}$ is obtained by setting the yields $Y_{1m}$ and $Y_{n1}$ to their maximum allowed value and by setting to zero the yields $Y_{nm}$, for $n,m\geq 3$. In particular, we use the upper bounds derived in \ref{bound_on_Y13-3decoys} and \ref{bound_on_Y31-3decoys} to bound $Y_{13}$ and $Y_{31}$, respectively, while we set to 1 all the other yields $Y_{1m}$ and $Y_{n1}$, for $n,m \geq 4$. In so doing, we obtain:
\begin{eqnarray}
\fl &G_{11}= Y^U_{11}\frac{(\mu_0 -\mu_1)(\mu_0 -\mu_2)(\nu_0 -\nu_1)(\nu_0 -\nu_2)}{(\mu_1+\mu_2)(\nu_1+\nu_2)} \nonumber\\
\fl &+ \frac{Y_{13}^U}{6} \frac{(\mu_0 -\mu_1)(\mu_0 - \mu_2)}{(\mu_1 + \mu_2)(\nu_1 + \nu_2)} (\nu_0 - \nu_2)(\nu_1 -\nu_0) (\nu_1 \nu_2 + \nu_0\nu_1+ \nu_2 \nu_0) \nonumber\\
\fl &+ \frac{Y_{31}^U}{6} \frac{(\nu_0 -\nu_1)(\nu_0 - \nu_2)}{(\nu_1 + \nu_2)(\mu_1 + \mu_2)} (\mu_0 -\mu_2)(\mu_1 - \mu_0) (\mu_1 \mu_2 + \mu_0 \mu_1 +\mu_2\mu_0) \nonumber\\
\fl &+ \frac{(\mu_0 -\mu_1)(\mu_0 - \mu_2)}{(\mu_1 + \mu_2)(\nu^2_1 - \nu^2_2)} \sum_{m=4}^{\infty} \left[\frac{\nu_1^m}{m!} (\nu_0^2 - \nu_2^2) +\frac{\nu_2^m}{m!}(\nu_1^2 - \nu_0^2)+\frac{\nu_0^m}{m!}(\nu_2^2 - \nu_1^2) \right] \nonumber\\
\fl &+ \frac{(\nu_0 -\nu_1)(\nu_0 - \nu_2)}{(\nu_1 + \nu_2)(\mu^2_1 - \mu^2_2)} \sum_{n=4}^{\infty} \left[\frac{\mu_1^n}{n!} (\mu_0^2 - \mu_2^2) +\frac{\mu_2^n}{n!}(\mu_1^2 - \mu_0^2)+\frac{\mu_0^n}{n!}(\mu_2^2 - \mu_1^2)\right]  \,\,.   \label{G-11-3}
\end{eqnarray}
By isolating the bound on $Y_{11}$ and summing the series, we obtain \ref{Y11-upperbound-3decoys}.

\subsection{Upper bound on $Y_{02}$ and $Y_{04}$} \label{bound_on_Y02-3decoys}
We consider the most general combination of the nine equality constraints:
\begin{eqnarray}
G_{02}=\sum_{i,j=0}^{2} c_{i,j} \tilde{Q}^{i,j} = \sum_{n,m=0}^{\infty} \frac{Y_{nm}}{n!m!}\left[\sum_{i,j=0}^{2} c_{i,j} \mu_i^n \nu_j^m \right] \,\,, \label{G-02-3decoys}
\end{eqnarray}
and require that the terms proportional to $Y_{1m},Y_{2m},Y_{n0}$ and $Y_{n1}$ are removed in the combination. We achieve this by imposing proper conditions on the real coefficients $c_{i,j}$:
\begin{eqnarray}
\fl &Y_{n0}\mbox{ removed:}\,\, \sum_{i=0}^{2} \mu_i^n \left(\sum_{j=0}^{2} c_{i,j}\right) =0 \quad\forall\, n \quad\Leftarrow\quad c_{i,0}+c_{i,1}+c_{i,2}=0 
\quad \mbox{for} \,\, i=0,1,2  \label{Yn0removed-02} \\
\fl &Y_{n1}\mbox{ removed:}\,\, \sum_{i=0}^{2} \mu_i^n \left(\sum_{j=0}^{2} \nu_j c_{i,j}\right) =0 \quad\forall\, n \quad\Leftarrow\quad 
\nu_0 c_{i,0}+ \nu_1 c_{i,1}+\nu_2 c_{i,2}=0 \quad \mbox{for} \,\, i=0,1,2  \label{Yn1removed-02} \\
\fl &Y_{1m}\mbox{ removed:}\,\, \sum_{j=0}^{2} \nu_j^m \left(\sum_{i=0}^{2} \mu_i c_{i,j}\right) =0 \quad\forall\, m \quad\Leftarrow\quad \mu_0 c_{0,j}+\mu_1 c_{1,j}+\mu_2 c_{2,j}=0 \quad \mbox{for} \,\, j=0,1,2  \label{Y1mremoved-02} \\
\fl &Y_{2m}\mbox{ removed:}\,\, \sum_{j=0}^{2} \nu_j^m \left(\sum_{i=0}^{2} \mu^2_i c_{i,j}\right) =0 \quad\forall\, m \quad\Leftarrow\quad 
\mu^2_0 c_{0,j}+\mu^2_1 c_{1,j}+ \mu^2_2 c_{2,j}=0 \quad \mbox{for} \,\, j=0,1,2  \label{Y2mremoved-02}  \,.
\end{eqnarray}
The linear system of equations given by (\ref{Yn0removed-02}-\ref{Y2mremoved-02}) has a unique solution in the variables $c_{i,j}$ (up to a global factor that we fix by imposing $c_{0,0}=1$), which reads as follows:
\begin{eqnarray}
&c_{0,0}=1 \nonumber\,\,,\\
&c_{0,1}=\frac{\nu_2-\nu_0}{\nu_1-\nu_2}  \nonumber\,\,,\\
&c_{0,2}=\frac{\nu_0-\nu_1}{\nu_1-\nu_2} \nonumber\,\,,\\
&c_{1,0}=\frac{\mu_0 (\mu_2-\mu_0)}{\mu_1 (\mu_1-\mu_2)} \nonumber\,\,,\\
&c_{1,1}=\frac{\mu_0 (\mu_0-\mu_2) (\nu_0-\nu_2)}{\mu_1 (\mu_1-\mu_2) (\nu_1-\nu_2)}  \nonumber\,\,,\\
&c_{1,2}=-\frac{\mu_0 (\mu_0-\mu_2) (\nu_0-\nu_1)}{\mu_1 (\mu_1-\mu_2) (\nu_1-\nu_2)} \nonumber\,\,,\\
&c_{2,0}=\frac{\mu_0 (\mu_0-\mu_1)}{\mu_2 (\mu_1-\mu_2)} \nonumber\,\,,\\
&c_{2,1}=\frac{\mu_0 (\mu_0-\mu_1) (\nu_0-\nu_2)}{\mu_2 (\mu_2-\mu_1) (\nu_1-\nu_2)} \nonumber\,\,,\\
&c_{2,2}=\frac{\mu_0 (\mu_0-\mu_1) (\nu_0-\nu_1)}{\mu_2 (\mu_1-\mu_2) (\nu_1-\nu_2)}  \label{cij-02}\,\,.
\end{eqnarray}
By substituting the solution for the coefficients $c_{i,j}$ (\ref{cij-02}) back into (\ref{G-02-3decoys}) one gets:
\begin{eqnarray}
\fl &G_{02}= \sum_{m=2}^{\infty} \frac{Y_{0m}}{m!} (-1)\frac{(\mu_0-\mu_1)(\mu_0 -\mu_2)A_{22}(\nu_0,\nu_1,\nu_2,m)}{\mu_1 \mu_2 (\nu_1 -\nu_2)} \nonumber\\
\fl &+\sum_{\stackrel[m=2]{n=3}{}}^{\infty} \frac{Y_{nm}}{n!m!} (-1) \frac{B_{02}(\mu_0,\mu_1,\mu_2,n)A_{22}(\nu_0,\nu_1,\nu_2,m)}{\mu_1 \mu_2 (\mu_1 -\mu_2)(\nu_1 -\nu_2)} \,\,, \label{G-02-1}
\end{eqnarray}
where $A_{22}$ is given in (\ref{A-22}) and $B_{02}$ can be written as follows for $n\geq 3$ \cite{grasselli2019practical}: 
\begin{equation}
B_{02}(\mu_0,\mu_1,\mu_2,n) = \mu_0\mu_1\mu_2 (\mu_1-\mu_2)(\mu_0-\mu_2)\sum_{k=0}^{n-2}\mu_2^k (\mu_0^{n-2-k} - \mu_1^{n-2-k}) \,\,. \label{B-02}
\end{equation}
We thus conclude that the sign of $Y_{0m}$ and $Y_{nm}$'s coefficients are always equal in (\ref{G-02-1}), regardless of the values of the intensities. Therefore a valid upper bound on $Y_{0m}$ --for $m=2,4$-- is obtained by setting to zero all the other yields in (\ref{G-02-1}). By doing so, we obtain the upper bounds on $Y_{02}$ and $Y_{04}$ given in Eqs.~(\ref{Y02-upperbound-3decoys}) and~(\ref{Y04-upperbound-3decoys}).

\subsection{Upper bound on $Y_{20}$ and $Y_{40}$} \label{bound_on_Y20-3decoys}
We consider the most general combination of the nine equality constraints:
\begin{eqnarray}
G_{20}=\sum_{i,j=0}^{2} c_{i,j} \tilde{Q}^{i,j} = \sum_{n,m=0}^{\infty} \frac{Y_{nm}}{n!m!}\left[\sum_{i,j=0}^{2} c_{i,j} \mu_i^n \nu_j^m \right] \,\,, \label{G-20-3decoys}
\end{eqnarray}
and require that the terms proportional to $Y_{0m},Y_{1m},Y_{n1}$ and $Y_{n2}$ are removed in the combination. We achieve this by imposing proper conditions on the real coefficients $c_{i,j}$:
\begin{eqnarray}
\fl &Y_{n1}\mbox{ removed:}\,\, \sum_{i=0}^{2} \mu_i^n \left(\sum_{j=0}^{2} \nu_j c_{i,j}\right) =0 \quad\forall\, n \quad\Leftarrow\quad \nu_0 c_{i,0}+ \nu_1 c_{i,1}+\nu_2 c_{i,2}=0 \quad \mbox{for} \,\, i=0,1,2  \label{Yn1removed-20} \\
\fl &Y_{n2}\mbox{ removed:}\,\, \sum_{i=0}^{2} \mu_i^n \left(\sum_{j=0}^{2} \nu^2_j c_{i,j}\right) =0 \quad\forall\, n \quad\Leftarrow\quad 
\nu^2_0 c_{i,0}+ \nu^2_1 c_{i,1}+\nu^2_2 c_{i,2}=0 \quad \mbox{for} \,\, i=0,1,2  \label{Yn2removed-20} \\
\fl &Y_{0m}\mbox{ removed:}\,\, \sum_{j=0}^{2} \nu_j^m \left(\sum_{i=0}^{2} c_{i,j}\right) =0 \quad\forall\, m \quad\Leftarrow\quad c_{0,j}+ c_{1,j}+ c_{2,j}=0 \quad \mbox{for} \,\, j=0,1,2  \label{Y0mremoved-20} \\
\fl &Y_{1m}\mbox{ removed:}\,\, \sum_{j=0}^{2} \nu_j^m \left(\sum_{i=0}^{2} \mu_i c_{i,j}\right) =0 \quad\forall\, m \quad\Leftarrow\quad 
\mu_0 c_{0,j}+\mu_1 c_{1,j}+ \mu_2 c_{2,j}=0 \quad \mbox{for} \,\, j=0,1,2  \label{Y1mremoved-20}  \,.
\end{eqnarray}
The linear system of equations given by (\ref{Yn1removed-20}-\ref{Y1mremoved-20}) has a unique solution in the variables $c_{i,j}$ (up to a global factor that we fix by imposing $c_{0,0}=1$), which reads as follows:
\begin{eqnarray}
&c_{0,0}=1 \nonumber\,\,,\\
&c_{0,1}=\frac{\nu_0 (\nu_2-\nu_0)}{\nu_1 (\nu_1-\nu_2)}  \nonumber\,\,,\\
&c_{0,2}=\frac{\nu_0 (\nu_0-\nu_1)}{\nu_2 (\nu_1-\nu_2)} \nonumber\,\,,\\
&c_{1,0}=\frac{\mu_2-\mu_0}{\mu_1-\mu_2} \nonumber\,\,,\\
&c_{1,1}=\frac{\nu_0 (\mu_0-\mu_2) (\nu_0-\nu_2)}{\nu_1 (\mu_1-\mu_2) (\nu_1-\nu_2)}  \nonumber\,\,,\\
&c_{1,2}=\frac{\nu_0 (\mu_0-\mu_2) (\nu_0-\nu_1)}{\nu_2 (\mu_2-\mu_1) (\nu_1-\nu_2)} \nonumber\,\,,\\
&c_{2,0}=\frac{\mu_0-\mu_1}{\mu_1-\mu_2} \nonumber\,\,,\\
&c_{2,1}=-\frac{\nu_0 (\mu_0-\mu_1) (\nu_0-\nu_2)}{\nu_1 (\mu_1-\mu_2) (\nu_1-\nu_2)} \nonumber\,\,,\\
&c_{2,2}=\frac{\nu_0 (\mu_0-\mu_1) (\nu_0-\nu_1)}{\nu_2 (\mu_1-\mu_2) (\nu_1-\nu_2)}  \label{cij-20}\,\,.
\end{eqnarray}
By substituting the solution for the coefficients $c_{i,j}$ (\ref{cij-20}) back into (\ref{G-20-3decoys}) one gets:
\begin{eqnarray}
\fl &G_{20}= \sum_{n=2}^{\infty} \frac{Y_{n0}}{n!} (-1)\frac{(\nu_0-\nu_1)(\nu_0 -\nu_2)A_{22}(\mu_0,\mu_1,\mu_2,n)}{\nu_1 \nu_2 (\mu_1 -\mu_2)} \nonumber\\
\fl &+\sum_{\stackrel[m=3]{n=2}{}}^{\infty} \frac{Y_{nm}}{n!m!} (-1) \frac{A_{22}(\mu_0,\mu_1,\mu_2,n)B_{02}(\nu_0,\nu_1,\nu_2,m)}{\nu_1 \nu_2 (\mu_1 -\mu_2)(\nu_1 -\nu_2)} \,\,, \label{G-20-1}
\end{eqnarray}
where $A_{22}$ is given in (\ref{A-22}) and $B_{02}$ in (\ref{B-02}).
From (\ref{G-20-1}) we observe that the sign of $Y_{n0}$ and $Y_{nm}$'s coefficients are always the same, regardless of the values of the intensities. Therefore a valid upper bound on $Y_{n0}$ --for $n=2,4$-- is obtained by setting to zero all the other yields in (\ref{G-20-1}). By doing so, we obtain the upper bounds on $Y_{20}$ and $Y_{40}$ given in Eqs.~(\ref{Y20-upperbound-3decoys}) and~(\ref{Y40-upperbound-3decoys}).

\subsection{Upper bound on $Y_{13}$} \label{bound_on_Y13-3decoys}
We consider the most general combination of the nine equality constraints:
\begin{eqnarray}
G_{13}=\sum_{i,j=0}^{2} c_{i,j} \tilde{Q}^{i,j} = \sum_{n,m=0}^{\infty} \frac{Y_{nm}}{n!m!}\left[\sum_{i,j=0}^{2} c_{i,j} \mu_i^n \nu_j^m \right] \,\,, \label{G-13-3decoys}
\end{eqnarray}
and require that the terms proportional to $Y_{0m},Y_{2m},Y_{n0}$ and $Y_{n1}$ are removed in the combination. We achieve this by imposing proper conditions on the real coefficients $c_{i,j}$:
\begin{eqnarray}
\fl &Y_{n0}\mbox{ removed:}\,\, \sum_{i=0}^{2} \mu_i^n \left(\sum_{j=0}^{2} c_{i,j}\right) =0 \quad\forall\, n \quad\Leftarrow\quad c_{i,0}+c_{i,1}+c_{i,2}=0 
\quad \mbox{for} \,\, i=0,1,2  \label{Yn0removed-13} \\
\fl &Y_{n1}\mbox{ removed:}\,\, \sum_{i=0}^{2} \mu_i^n \left(\sum_{j=0}^{2} \nu_j c_{i,j}\right) =0 \quad\forall\, n \quad\Leftarrow\quad 
\nu_0 c_{i,0}+ \nu_1 c_{i,1}+\nu_2 c_{i,2}=0 \quad \mbox{for} \,\, i=0,1,2  \label{Yn1removed-13} \\
\fl &Y_{0m}\mbox{ removed:}\,\, \sum_{j=0}^{2} \nu_j^m \left(\sum_{i=0}^{2} c_{i,j}\right) =0 \quad\forall\, m \quad\Leftarrow\quad  c_{0,j}+ c_{1,j}+ c_{2,j}=0 \quad \mbox{for} \,\, j=0,1,2  \label{Y0mremoved-13} \\
\fl &Y_{2m}\mbox{ removed:}\,\, \sum_{j=0}^{2} \nu_j^m \left(\sum_{i=0}^{2} \mu^2_i c_{i,j}\right) =0 \quad\forall\, m \quad\Leftarrow\quad 
\mu^2_0 c_{0,j}+\mu^2_1 c_{1,j}+ \mu^2_2 c_{2,j}=0 \quad \mbox{for} \,\, j=0,1,2  \label{Y2mremoved-13}  \,.
\end{eqnarray}
The linear system of equations given by (\ref{Yn0removed-13}-\ref{Y2mremoved-13}) has a unique solution in the variables $c_{i,j}$ (up to a global factor that we fix by imposing $c_{0,0}=1$), which reads as follows:
\begin{eqnarray}
&c_{0,0}=1 \nonumber\,\,,\\
&c_{0,1}=\frac{\nu_2-\nu_0}{\nu_1-\nu_2}  \nonumber\,\,,\\
&c_{0,2}=\frac{\nu_0-\nu_1}{\nu_1-\nu_2} \nonumber\,\,,\\
&c_{1,0}=\frac{\mu_2^2-\mu_0^2}{\mu_1^2-\mu_2^2} \nonumber\,\,,\\
&c_{1,1}=\frac{\left(\mu_0^2-\mu_2^2\right) (\nu_0-\nu_2)}{\left(\mu_1^2-\mu_2^2\right) (\nu_1-\nu_2)}  \nonumber\,\,,\\
&c_{1,2}=\frac{\left(\mu_0^2-\mu_2^2\right) (\nu_0-\nu_1)}{\left(\mu_2^2-\mu_1^2\right) (\nu_1-\nu_2)} \nonumber\,\,,\\
&c_{2,0}=\frac{\mu_0^2-\mu_1^2}{\mu_1^2-\mu_2^2} \nonumber\,\,,\\
&c_{2,1}=\frac{\left(\mu_0^2-\mu_1^2\right) (\nu_0-\nu_2)}{\left(\mu_1^2-\mu_2^2\right) (\nu_2-\nu_1)} \nonumber\,\,,\\
&c_{2,2}=\frac{\left(\mu_0^2-\mu_1^2\right) (\nu_0-\nu_1)}{\left(\mu_1^2-\mu_2^2\right) (\nu_1-\nu_2)}  \label{cij-13}\,\,.
\end{eqnarray}
By substituting the solution for the coefficients $c_{i,j}$ (\ref{cij-13}) back into (\ref{G-13-3decoys}) one gets:
\begin{eqnarray}
\fl &G_{13}= \sum_{m=2}^{\infty} \frac{Y_{1m}}{m!} \frac{(\mu_0-\mu_1)(\mu_0 -\mu_2)A_{22}(\nu_0,\nu_1,\nu_2,m)}{(\mu_1 +\mu_2) (\nu_1 -\nu_2)} \nonumber\\
\fl &+\sum_{\stackrel[m=2]{n=3}{}}^{\infty} \frac{Y_{nm}}{n!m!} \frac{A_{11}(\mu_0,\mu_1,\mu_2,n)A_{22}(\nu_0,\nu_1,\nu_2,m)}{(\mu^2_1 -\mu^2_2)(\nu_1 -\nu_2)} \,\,, \label{G-13-1}
\end{eqnarray}
where $A_{22}$ is given in (\ref{A-22_1}) and $A_{11}$ is given in (\ref{A-11-1}). We thus conclude that $Y_{1m}$ and $Y_{nm}$'s coefficients have always opposite sign in (\ref{G-13-1}), regardless of the values of the intensities. Therefore a valid upper bound on $Y_{13}$ is obtained by setting to zero all the yields of the form $Y_{1m}$ for $m\neq 3$ and by setting to 1 all the other yields of the form $Y_{nm}$ with $n\geq 3$ and $m\geq 2$. In so doing, we obtain the following expression:
\begin{eqnarray}
\fl &G_{13}=  \frac{Y^U_{13}}{6} \frac{(\mu_0-\mu_1)(\mu_0 -\mu_2)(\nu_0 -\nu_2)(\nu_2 -\nu_1)[\nu_0^2-\nu_1^2+\nu_2 (\nu_0 -\nu_1)]}{(\mu_1 +\mu_2) (\nu_1 -\nu_2)} \nonumber\\
\fl &+\sum_{\stackrel[m=2]{n=3}{}}^{\infty} \frac{[\mu_1^n (\mu^2_0 - \mu^2_2) + \mu_2^n (\mu^2_1 - \mu^2_0) + \mu_0^n (\mu^2_2 - \mu^2_1)][\nu_1^m (\nu_0 - \nu_2) + \nu_2^m (\nu_1 - \nu_0) + \nu_0^m (\nu_2 - \nu_1)]}{n! m! (\mu^2_1 -\mu^2_2)(\nu_1 -\nu_2)} \,\,, \label{G-13-2}
\end{eqnarray}
where we used in the series the original expressions of $A_{22}$ and $A_{11}$ that are given in (\ref{A-22}) and (\ref{A-11}), respectively.
By summing and rearranging the terms, we obtain the upper bound on $Y_{13}$ given in~(\ref{Y13-upperbound-3decoys}).

\subsection{Upper bound on $Y_{31}$} \label{bound_on_Y31-3decoys}
In a similar fashion to $Y_{13}$'s bound, one first removes the terms proportional to $Y_{0m},Y_{1m},Y_{n0}$ and $Y_{n2}$ from the general combination of the nine gains:
\begin{eqnarray}
G_{31}=\sum_{i,j=0}^{2} c_{i,j} \tilde{Q}^{i,j} = \sum_{n,m=0}^{\infty} \frac{Y_{nm}}{n!m!}\left[\sum_{i,j=0}^{2} c_{i,j} \mu_i^n \nu_j^m \right] \,\,, \label{G-31-3decoys}
\end{eqnarray}
by properly fixing the coefficients $c_{i,j}$ as follows:
\begin{eqnarray}
&c_{0,0}=1 \nonumber\,\,,\\
&c_{0,1}=\frac{\nu_2^2-\nu_0^2}{\nu_1^2-\nu_2^2}  \nonumber\,\,,\\
&c_{0,2}=\frac{\nu_0^2-\nu_1^2}{\nu_1^2-\nu_2^2} \nonumber\,\,,\\
&c_{1,0}=\frac{\mu_2-\mu_0}{\mu_1-\mu_2} \nonumber\,\,,\\
&c_{1,1}=\frac{(\mu_0-\mu_2) \left(\nu_0^2-\nu_2^2\right)}{(\mu_1-\mu_2) \left(\nu_1^2-\nu_2^2\right)}  \nonumber\,\,,\\
&c_{1,2}=\frac{(\mu_0-\mu_2) \left(\nu_0^2-\nu_1^2\right)}{(\mu_2-\mu_1) \left(\nu_1^2-\nu_2^2\right)} \nonumber\,\,,\\
&c_{2,0}=\frac{\mu_0-\mu_1}{\mu_1-\mu_2} \nonumber\,\,,\\
&c_{2,1}=\frac{(\mu_0-\mu_1) \left(\nu_0^2-\nu_2^2\right)}{(\mu_1-\mu_2) \left(\nu_2^2-\nu_1^2\right)} \nonumber\,\,,\\
&c_{2,2}=\frac{(\mu_0-\mu_1) \left(\nu_0^2-\nu_1^2\right)}{(\mu_1-\mu_2) \left(\nu_1^2-\nu_2^2\right)}  \label{cij-31}\,\,.
\end{eqnarray}
Then one substitutes the solution (\ref{cij-31}) back into (\ref{G-31-3decoys}) and gets:
\begin{eqnarray}
\fl &G_{31}= \sum_{n=2}^{\infty} \frac{Y_{n1}}{n!} \frac{(\nu_0-\nu_1)(\nu_0 -\nu_2)A_{22}(\mu_0,\mu_1,\mu_2,n)}{(\nu_1 +\nu_2) (\mu_1 -\mu_2)} \nonumber\\
\fl &+\sum_{\stackrel[m=3]{n=2}{}}^{\infty} \frac{Y_{nm}}{n!m!} \frac{A_{22}(\mu_0,\mu_1,\mu_2,n)A_{11}(\nu_0,\nu_1,\nu_2,m)}{(\nu^2_1 -\nu^2_2)(\mu_1 -\mu_2)} \,\,, \label{G-31-1}
\end{eqnarray}
where $A_{22}$ and $A_{11}$ are given in (\ref{A-22_1}) and (\ref{A-11-1}), respectively. By noting that the coefficients of the $Y_{n1}$ terms have opposite sign to those of the $Y_{nm}$ terms, we derive an upper bound on $Y_{31}$ by setting to zero all the $Y_{n1}$ yields (for $n \neq 3$) and to 1 all the other ones. The upper bound on $Y_{31}$ is given in~(\ref{Y31-upperbound-3decoys}).

\subsection{Upper bound on $Y_{00}$} \label{bound_on_Y00-3decoys}
We consider the most general combination of the nine equality constraints:
\begin{eqnarray}
G_{00}=\sum_{i,j=0}^{2} c_{i,j} \tilde{Q}^{i,j} = \sum_{n,m=0}^{\infty} \frac{Y_{nm}}{n!m!}\left[\sum_{i,j=0}^{2} c_{i,j} \mu_i^n \nu_j^m \right] \,\,, \label{G-00-3decoys}
\end{eqnarray}
and require that the terms proportional to $Y_{1m},Y_{2m},Y_{n1}$ and $Y_{n2}$ are removed in the combination. We achieve this by imposing proper conditions on the real coefficients $c_{i,j}$:
\begin{eqnarray}
\fl &Y_{n1}\mbox{ removed:}\,\, \sum_{i=0}^{2} \mu_i^n \left(\sum_{j=0}^{2} \nu_j c_{i,j}\right) =0 \quad\forall\, n \quad\Leftarrow\quad \nu_0 c_{i,0}+ \nu_1 c_{i,1}+\nu_2 c_{i,2}=0 \quad \mbox{for} \,\, i=0,1,2  \label{Yn1removed-00} \\
\fl &Y_{n2}\mbox{ removed:}\,\, \sum_{i=0}^{2} \mu_i^n \left(\sum_{j=0}^{2} \nu^2_j c_{i,j}\right) =0 \quad\forall\, n \quad\Leftarrow\quad 
\nu^2_0 c_{i,0}+ \nu^2_1 c_{i,1}+\nu^2_2 c_{i,2}=0 \quad \mbox{for} \,\, i=0,1,2  \label{Yn2removed-00} \\
\fl &Y_{1m}\mbox{ removed:}\,\, \sum_{j=0}^{2} \nu_j^m \left(\sum_{i=0}^{2} \mu_i c_{i,j}\right) =0 \quad\forall\, m \quad\Leftarrow\quad  \mu_0 c_{0,j}+\mu_1 c_{1,j}+ \mu_2 c_{2,j}=0 \quad \mbox{for} \,\, j=0,1,2  \label{Y1mremoved-00} \\
\fl &Y_{2m}\mbox{ removed:}\,\, \sum_{j=0}^{2} \nu_j^m \left(\sum_{i=0}^{2} \mu^2_i c_{i,j}\right) =0 \quad\forall\, m \quad\Leftarrow\quad 
\mu^2_0 c_{0,j}+\mu^2_1 c_{1,j}+ \mu^2_2 c_{2,j}=0 \quad \mbox{for} \,\, j=0,1,2  \label{Y2mremoved-00}  \,.
\end{eqnarray}
The linear system of equations given by (\ref{Yn1removed-00}-\ref{Y2mremoved-00}) has a unique solution in the variables $c_{i,j}$ (up to a global factor that we fix by imposing $c_{0,0}=1$), which reads as follows:
\begin{eqnarray}
&c_{0,0}=1 \nonumber\,\,,\\
&c_{0,1}=\frac{\nu_0 (\nu_2-\nu_0)}{\nu_1 (\nu_1-\nu_2)}  \nonumber\,\,,\\
&c_{0,2}=\frac{\nu_0 (\nu_0-\nu_1)}{\nu_2 (\nu_1-\nu_2)} \nonumber\,\,,\\
&c_{1,0}=\frac{\mu_0 (\mu_2-\mu_0)}{\mu_1 (\mu_1-\mu_2)} \nonumber\,\,,\\
&c_{1,1}=\frac{\mu_0 \nu_0 (\mu_0-\mu_2) (\nu_0-\nu_2)}{\mu_1 \nu_1 (\mu_1-\mu_2) (\nu_1-\nu_2)}  \nonumber\,\,,\\
&c_{1,2}=-\frac{\mu_0 \nu_0 (\mu_0-\mu_2) (\nu_0-\nu_1)}{\mu_1 \nu_2 (\mu_1-\mu_2) (\nu_1-\nu_2)} \nonumber\,\,,\\
&c_{2,0}=\frac{\mu_0 (\mu_0-\mu_1)}{\mu_2 (\mu_1-\mu_2)} \nonumber\,\,,\\
&c_{2,1}=\frac{\mu_0 \nu_0 (\mu_0-\mu_1) (\nu_0-\nu_2)}{\mu_2 \nu_1 (\mu_2-\mu_1) (\nu_1-\nu_2)} \nonumber\,\,,\\
&c_{2,2}=\frac{\mu_0 \nu_0 (\mu_0-\mu_1) (\nu_0-\nu_1)}{\mu_2 \nu_2 (\mu_1-\mu_2) (\nu_1-\nu_2)}  \label{cij-00}\,\,.
\end{eqnarray}
By substituting the solution for the coefficients $c_{i,j}$ (\ref{cij-13}) back into (\ref{G-13-3decoys}) one gets:
\begin{eqnarray}
\fl &G_{00}= \frac{Y_{00}(\mu_0 -\mu_1)(\mu_0 -\mu_2)(\nu_0 -\nu_1)(\nu_0 -\nu_2)}{\mu_1 \mu_2 \nu_1 \nu_2} + \sum_{m=3}^{\infty} \frac{Y_{0m}}{m!} \frac{(\mu_0-\mu_1)(\mu_0 -\mu_2)A_{00}(\nu_0,\nu_1,\nu_2,m)}{\mu_1\mu_2 \nu_1\nu_2 (\nu_1 -\nu_2)} \nonumber\\
\fl &+\sum_{n=3}^{\infty} \frac{Y_{n0}}{n!} \frac{(\nu_0-\nu_1)(\nu_0 -\nu_2)A_{00}(\mu_0,\mu_1,\mu_2,n)}{\mu_1\mu_2 \nu_1\nu_2 (\mu_1 -\mu_2)} + \sum_{\stackrel[m=3]{n=3}{}}^{\infty} \frac{Y_{nm}}{m!} \frac{A_{00}(\mu_0,\mu_1,\mu_2,n)A_{00}(\nu_0,\nu_1,\nu_2,m)}{\mu_1\mu_2 \nu_1\nu_2 (\mu_1 -\mu_2)(\nu_1 -\nu_2)} \,\,, \label{G-00-1}
\end{eqnarray}
where $A_{00}$ is defined as \cite{grasselli2019practical}:
\begin{equation}
A_{00}(\mu_0,\mu_1,\mu_2,n) \equiv \mu_1^n (\mu^2_2\mu_0 - \mu_2\mu^2_0) + \mu_2^n (\mu^2_0\mu_1 - \mu_0\mu^2_1) + \mu_0^n (\mu^2_1\mu_2 - \mu_1\mu_2^2) \label{A-00}  \,\,.
\end{equation}
Using the result in \cite{grasselli2019practical}, one can recast the function $A_{22}$ as follows:
\begin{eqnarray}
A_{00}(\mu_0,\mu_1,\mu_2,n)=  \mu_0\mu_1 \mu_2 (\mu_0 - \mu_2)(\mu_1 - \mu_2)\sum_{k=0}^{n-2} \mu_2^k (\mu_0^{n-2-k}-\mu_1^{n-2-k})  \,\,, \label{A-00-1}
\end{eqnarray}
and notice that all the yields in (\ref{G-00-1}) have coefficients with equal sign, regardless of the intensities' values. Hence a valid upper bound on $Y_{00}$ is obtained by setting all the other yields to zero (except for the yield to be bounded) in (\ref{G-00-1}). The upper bound on $Y_{00}$ is given in~(\ref{Y00-upperbound-3decoys}).



\section{Upper bounds on the yields with four decoy intensities}
\label{4intensities}

In this case each party prepares phase-randomized coherent states with four possible intensities, namely $\{\mu_0,\mu_1,\mu_2,\mu_3\}$ for Alice and $\{\nu_0,\nu_1,\nu_2,\nu_3\}$ for Bob. The yields are then subjected to the following sixteen equality constraints:
\begin{equation}
\tilde{Q}^{k,l} \equiv e^{\mu_k + \nu_l} Q^{k,l} =\sum_{n,m=0}^{\infty} \frac{Y_{nm}}{n!m!} {\mu_k}^n {\nu_l}^m \quad k,l \in \{0,1,2,3\} \,\,, \label{constr-4decoys}
\end{equation}
and to the inequality constraints given in (\ref{ineq-constr}).\\
Below we derive tighter upper bounds on the yields $Y_{04},Y_{40},Y_{13}$ and $Y_{31}$, since the bounds derived on the yields $Y_{00}, Y_{11}, Y_{02}, Y_{20}$ and $Y_{22}$ in \ref{3intensities} are already good enough, i.e bounding them with one additional decoy intensity would not result in a significant improvement of the performance of the protocol. Note that the bounds presented here are not valid when two decoy intensities of the same party have the same value. This case would then reduce to the three decoy intensity case. Thus, without loss of generality, we assume the following ordering within each set of intensities: $\mu_3>\mu_0>\mu_1>\mu_2$ and $\nu_3>\nu_0>\nu_1>\nu_2$.

\subsection{Upper bound on $Y_{04}$} \label{bound_on_Y04-4decoys}
We consider the combination of gains (\ref{G-02-1}) that leads to the bound on $Y_{04}$ in the case of three decoy intensity settings:
\begin{equation}
G^{0,1,2}_{02}= \sum_{n,m}^{\infty} \frac{Y_{nm}}{n!m!} C^{0,1,2}_{02} (n,m)  \label{G-02-4decoys-012} \,\,,
\end{equation}
where the function $C^{0,1,2}_{02} (n,m)$ is defined by the r.h.s. of (\ref{G-02-1}), while $G^{0,1,2}_{02}$ is the combination of gains given by (\ref{G-02-3decoys}), with the coefficients $c_{i,j}$ of the combination given in (\ref{cij-02}). The subscript indicates the combination of gains to which it refers, while the superscript indicates the decoy intensities that are involved, namely $\{\mu_0,\mu_1,\mu_2\}$ for Alice and $\{\nu_0,\nu_1,\nu_2\}$ for Bob. From \ref{bound_on_Y02-3decoys} we know that the terms $Y_{n0},Y_{n1},Y_{1m}$ and $Y_{2m}$ are removed in (\ref{G-02-4decoys-012}), i.e. $ C^{0,1,2}_{02} (n,0)= C^{0,1,2}_{02} (n,1)= C^{0,1,2}_{02} (1,m)= C^{0,1,2}_{02} (2,m)=0$, for any $n,m$. Now that the parties have at their disposal the fourth decoy intensity ($\mu_3$ for Alice and $\nu_3$ for Bob), one can derive three additional combinations like (\ref{G-02-4decoys-012}) by simply replacing one of the first three intensities with the fourth one:
\begin{eqnarray}
G^{0,1,3}_{02}&= \sum_{n,m}^{\infty} \frac{Y_{nm}}{n!m!} C^{0,1,3}_{02} (n,m)  \label{G-02-4decoys-013} \,\,, \\
G^{0,2,3}_{02}&= \sum_{n,m}^{\infty} \frac{Y_{nm}}{n!m!} C^{0,2,3}_{02} (n,m)  \label{G-02-4decoys-023} \,\,, \\
G^{1,2,3}_{02}&= \sum_{n,m}^{\infty} \frac{Y_{nm}}{n!m!} C^{1,2,3}_{02} (n,m)  \label{G-02-4decoys-123} \,\,.
\end{eqnarray}
For instance, the combination (\ref{G-02-4decoys-013}) is obtained by replacing $\mu_2\rightarrow\mu_3$ and $\nu_2\rightarrow\nu_3$ in the function $C^{0,1,2}_{02} (n,m)$, thus obtaining $C^{0,1,3}_{02} (n,m)$. Regarding the r.h.s, $G^{0,1,3}_{02}$ is obtained by replacing $\mu_2\rightarrow\mu_3$ and $\nu_2\rightarrow\nu_3$ in the coefficients $c_{i,j}$ appearing in the combination $G^{0,1,2}_{02}$, and by making the substitution $\tilde{Q}^{2,l}\rightarrow \tilde{Q}^{3,l}$ and $\tilde{Q}^{k,2}\rightarrow \tilde{Q}^{k,3}$ on the gains in $G^{0,1,2}_{02}$. In so doing, we obtain three more combinations of gains (\ref{G-02-4decoys-013}-\ref{G-02-4decoys-123}) in which the terms $Y_{n0},Y_{n1},Y_{1m}$ and $Y_{2m}$ are removed.\\
At this point, we further combine the expressions (\ref{G-02-4decoys-012},\ref{G-02-4decoys-013},\ref{G-02-4decoys-023},\ref{G-02-4decoys-123}) with arbitrary real coefficients $d_{i,j,k}$\footnote{Note that we identify such a combination as $H_{04}$ since it appears in bounding $Y_{04}$. However the elements in the combination, namely $G^{i,j,k}_{02}$, have a different subscript since they are borrowed from the bounds on $Y_{02}$ and $Y_{04}$ with three decoy intensity settings.}:
\begin{eqnarray}
\fl &H_{04} \equiv d_{0,1,2} G^{0,1,2}_{02} +d_{0,1,3} G^{0,1,3}_{02}+d_{0,2,3} G^{0,2,3}_{02}+d_{1,2,3} G^{1,2,3}_{02} =  \nonumber \\
\fl &\sum_{n,m}^{\infty} \frac{Y_{nm}}{n!m!}\left[d_{0,1,2} C^{0,1,2}_{02} (n,m) +d_{0,1,3} C^{0,1,3}_{02} (n,m) + d_{0,2,3} C^{0,2,3}_{02} (n,m)+ d_{1,2,3} C^{1,2,3}_{02} (n,m)\right] \label{G-02-4decoys-new} \,\,,
\end{eqnarray}
and impose that even the terms $Y_{n2}$ and $Y_{3m}$ are removed:
\begin{equation}
\fl \left\{
{\begin{array}{rcl}
	d_{0,1,2} C^{0,1,2}_{02} (n,2) +d_{0,1,3} C^{0,1,3}_{02} (n,2) + d_{0,2,3} C^{0,2,3}_{02} (n,2)+ d_{1,2,3} C^{1,2,3}_{02} (n,2)	 & = & 0 \quad\forall\, n \\[0.8ex]
	d_{0,1,2} C^{0,1,2}_{02} (3,m) +d_{0,1,3} C^{0,1,3}_{02} (3,m) + d_{0,2,3} C^{0,2,3}_{02} (3,m)+ d_{1,2,3} C^{1,2,3}_{02} (3,m)	 & = & 0 \quad\forall\, m \\[0.8ex]
	d_{0,1,2} & = & 1 \,\,,
	\end{array}}
\right.   \label{system-Y-04}
\end{equation}
where we fixed the remaining degree of freedom (global factor on all the $d_{i,j,k}$) by requiring that $d_{0,1,2}=1$. The solution of the linear system (\ref{system-Y-04}) reads:
\begin{eqnarray}
\fl d_{0,1,2} &=1 \nonumber\\
\fl d_{0,1,3} &= \frac{(\mu_0-\mu_2) (\nu_0-\nu_2) \left[\mu_0 (\nu_1-\nu_3)+\mu_1 (\nu_3-\nu_0)+\mu_3 (\nu_0-\nu_1)\right]}{(\mu_0-\mu_3) (\nu_0-\nu_3) \left[\mu_0 (\nu_2-\nu_1)+\mu_1 (\nu_0-\nu_2)+\mu_2 (\nu_1-\nu_0)\right]} \nonumber\\
\fl d_{0,2,3} &= \frac{(\mu_0-\mu_1) (\nu_0-\nu_1) \left[\mu_0 (\nu_2-\nu_3)+\mu_2 (\nu_3-\nu_0)+\mu_3 (\nu_0-\nu_2)\right]}{(\mu_0-\mu_3) (\nu_0-\nu_3) \left[\mu_0 (\nu_1-\nu_2)+\mu_1 (\nu_2-\nu_0)+\mu_2 (\nu_0-\nu_1)\right]} \nonumber\\
\fl d_{1,2,3} &= \frac{\mu_0 (\mu_0-\mu_1) (\mu_0-\mu_2) (\nu_0-\nu_1) (\nu_0-\nu_2) \left[\mu_1 (\nu_3-\nu_2)+\mu_2 (\nu_1-\nu_3)+\mu_3 (\nu_2-\nu_1)\right]}{\mu_1 (\mu_1-\mu_2) (\mu_1-\mu_3)
	(\nu_1-\nu_2) (\nu_1-\nu_3) \left[\mu_0 (\nu_1-\nu_2)+\mu_1 (\nu_2-\nu_0)+\mu_2 (\nu_0-\nu_1)\right]} \,\,. \label{d_ijk}
\end{eqnarray}
By substituting the solution (\ref{d_ijk}) back into (\ref{G-02-4decoys-new}) and by rearranging the r.h.s, one gets a combination of gains where all the terms $Y_{n0},Y_{n1},Y_{n2},Y_{1m},Y_{2m}$ and $Y_{3m}$ are removed:
\begin{eqnarray}
\fl H_{04} = \sum_{m=3}^{\infty} \frac{Y_{0m}}{m!} A_{04}(\mu_0,\mu_1,\mu_2,\mu_3,\nu_0,\nu_1,\nu_2,\nu_3,m) + \sum_{\stackrel[m=3]{n=4}{}}^{\infty} \frac{Y_{nm}}{n!m!} B_{04}(\mu_0,\mu_1,\mu_2,\mu_3,\nu_0,\nu_1,\nu_2,\nu_3,n,m) \,\,,\nonumber\\ \fl \label{G-02-4decoys-new-1}
\end{eqnarray}
where:
\begin{eqnarray}
\fl &A_{04}(\mu_0,\mu_1,\mu_2,\mu_3,\nu_0,\nu_1,\nu_2,\nu_3,m) = -\frac{(\mu_0 -\mu_1)(\mu_0 -\mu_2)(\nu_0 -\nu_1)(\nu_0 -\nu_2)p_{04}(\mu_0,\mu_1,\mu_2,\mu_3,\nu_0,\nu_1,\nu_2,\nu_3)}{\mu_1\mu_2\mu_3\left[\nu_0(\mu_1 -\mu_2) -\nu_1(\mu_0 -\mu_2)+\nu_2(\mu_0 -\mu_1)\right]} \nonumber\\
\fl &\times \left(\sum_{i_1 \leq i_2 \leq \dots \leq i_{m-3}} \nu_{i_1}\nu_{i_2} \cdot \dots \cdot \nu_{i_{m-3}}\right) \,\,; \label{A04} \\
\fl &p_{04}(\mu_0,\mu_1,\mu_2,\mu_3,\nu_0,\nu_1,\nu_2,\nu_3) = \mu_0 [\mu_1 (\nu_0-\nu_1) (\nu_2-\nu_3)-\mu_2 (\nu_0-\nu_2) (\nu_1-\nu_3)+\mu_3(\nu_0-\nu_3)  \nonumber\\
\fl &\times(\nu_1-\nu_2)]+\mu_1 \left[\mu_2 (\nu_0-\nu_3) (\nu_1-\nu_2)-\mu_3 (\nu_0-\nu_2) (\nu_1-\nu_3)\right]+\mu_2 \mu_3 (\nu_0-\nu_1) (\nu_2-\nu_3)  \label{p04}
\end{eqnarray}
and
\begin{eqnarray}
\fl &B_{04}(\mu_0,\mu_1,\mu_2,\mu_3,\nu_0,\nu_1,\nu_2,\nu_3,n,m)= -\mu_0\mu_1 \mu_2 \mu_3 \,A_{04}(\mu_0,\mu_1,\mu_2,\mu_3,\nu_0,\nu_1,\nu_2,\nu_3,m) \nonumber\\
\fl &\times \left(\sum_{i_1 \leq i_2 \leq \dots \leq i_{n-4}} \mu_{i_1}\mu_{i_2} \cdot \dots \cdot \mu_{i_{n-4}}\right) \label{B04}  \,\,.
\end{eqnarray}
We assume that the indexes in the sums run over the set $\{0,1,2,3\}$ and we define \mbox{$\sum_{i_1 \leq i_2 \leq \dots \leq i_{m-3}} \nu_{i_1}\nu_{i_2} \cdot \dots \cdot \nu_{i_{m-3}}|_{m=3} =1$}.
From (\ref{A04}) and (\ref{B04}) we deduce that the coefficients of $Y_{0m}$ and $Y_{nm}$ have always opposite sign, hence the upper bound on $Y_{04}$ is obtained from (\ref{G-02-4decoys-new-1}) by setting all the yields $Y_{0m}$ (with $m\neq 4$) to zero and the yields $Y_{nm}$ (with $n\geq 4,\,m \geq 3$) to 1. After rearranging the terms, we get the following expression for the upper bound on $Y_{04}$:
\begin{eqnarray}
\fl Y^U_{04}=\frac{24}{A_{04}(\mu_0,\mu_1,\mu_2,\mu_3,\nu_0,\nu_1,\nu_2,\nu_3,4)}\left[H_{04} - \sum_{\stackrel[m=3]{n=4}{}}^{\infty} \frac{B_{04}(\mu_0,\mu_1,\mu_2,\mu_3,\nu_0,\nu_1,\nu_2,\nu_3,n,m)}{n!m!} \right]  \label{Y04-upperbound-4decoys} \,\,,
\end{eqnarray}
where $H_{04}$ is given in the first line of (\ref{G-02-4decoys-new}), the function $A_{04}$ evaluated for $m=4$ reads:
\begin{eqnarray}
\fl &A_{04}(\mu_0,\mu_1,\mu_2,\mu_3,\nu_0,\nu_1,\nu_2,\nu_3,4)= \nonumber\\
\fl &-\frac{(\mu_0 -\mu_1)(\mu_0 -\mu_2)(\nu_0 -\nu_1)(\nu_0 -\nu_2)(\nu_0+\nu_1+\nu_2+\nu_3)p_{04}(\mu_0,\mu_1,\mu_2,\mu_3,\nu_0,\nu_1,\nu_2,\nu_3)}{\mu_1\mu_2\mu_3\left[\nu_0(\mu_1 -\mu_2) -\nu_1(\mu_0 -\mu_2)+\nu_2(\mu_0 -\mu_1)\right]} \,\,, \label{A04-evaluated}
\end{eqnarray} 
and the series of $B_{04}$ sums to:
{\small\begin{eqnarray}\footnotesize 
\fl &\sum_{\stackrel[m=3]{n=4}{}}^{\infty} \frac{B_{04}(\mu_0,\mu_1,\mu_2,\mu_3,\nu_0,\nu_1,\nu_2,\nu_3,n,m)}{n!m!}= \nonumber\\
\fl &\frac{\mu_0 \, p_{04}(\mu_0,\mu_1,\mu_2,\mu_3,\nu_0,\nu_1,\nu_2,\nu_3) \left[\nu_2 (\mu_0-\mu_1)-\nu_1 (\mu_0-\mu_2)+\nu_0 (\mu_1-\mu_2)\right]^{-1} }{(\mu_0-\mu_3) (\mu_1-\mu_2) (\mu_1-\mu_3) (\mu_2-\mu_3) (\nu_0-\nu_3) (\nu_1-\nu_2) (\nu_1-\nu_3) (\nu_2-\nu_3)} \nonumber\\
\fl &\times  \left[\left(e^{\mu_0}-\frac{\mu_0^2}{2}-\mu_0-1\right) (\mu_1-\mu_2) (\mu_1-\mu_3) (\mu_2-\mu_3)-\left(e^{\mu_1}-\frac{\mu_1^2}{2}-\mu_1-1\right) (\mu_0-\mu_2) (\mu_0-\mu_3) (\mu_2-\mu_3) \right.\nonumber\\
\fl  &\left. +\left(e^{\mu_2}-\frac{\mu_2^2}{2}-\mu_2-1\right) (\mu_0-\mu_1) (\mu_0-\mu_3) (\mu_1-\mu_3)-\left(e^{\mu_3}-\frac{\mu_3^2}{2}-\mu_3-1\right) (\mu_0-\mu_1) (\mu_0-\mu_2) (\mu_1-\mu_2)\right] \nonumber\\
\fl &\times  \left[\left(e^{\nu_0}-\frac{\nu_0^2}{2}-\nu_0-1\right) (\nu_1-\nu_2) (\nu_1-\nu_3) (\nu_2-\nu_3)-\left(e^{\nu_1}-\frac{\nu_1^2}{2}-\nu_1-1\right) (\nu_0-\nu_2) (\nu_0-\nu_3) (\nu_2-\nu_3) \right.\nonumber\\
\fl &\left. +\left(e^{\nu_2}-\frac{\nu_2^2}{2}-\nu_2-1\right) (\nu_0-\nu_1) (\nu_0-\nu_3) (\nu_1-\nu_3)-\left(e^{\nu_3}-\frac{\nu_3^2}{2}-\nu_3-1\right) (\nu_0-\nu_1) (\nu_0-\nu_2) (\nu_1-\nu_2)\right] \,. \nonumber
\end{eqnarray}}
\begin{equation}
\label{B04-summed}\medskip\\
\end{equation}
We remark that in deriving the bound (\ref{Y04-upperbound-4decoys}) we implicitly assumed that at least one of the following equalities does not hold: $\nu_0=\mu_0$, $\nu_1=\mu_1$ and $\nu_2=\mu_2$. Indeed, when all three equalities hold (i.e. when Alice and Bob are using the same intensities settings for three out of four decoy pulses) one gets a ``$\frac{0}{0}$ form'' in the bound expression (\ref{Y04-upperbound-4decoys}). In order to overcome this issue (which is not likely to happen in practice due to intensity fluctuations), we derive an additional upper bound on $Y_{04}$ which is valid in the particular case of: $\nu_0=\mu_0$, $\nu_1=\mu_1$ and $\nu_2=\mu_2$.\\
The procedure resembles that used in deriving (\ref{Y04-upperbound-4decoys}). We start by considering the four combinations of gains (\ref{G-02-4decoys-012}), (\ref{G-02-4decoys-013}), (\ref{G-02-4decoys-023}) and (\ref{G-02-4decoys-123}) and we impose the conditions: $\nu_0=\mu_0$, $\nu_1=\mu_1$ and $\nu_2=\mu_2$. Let's indicate the resulting gains combinations as follows:
\begin{eqnarray}
{\tilde{G}}^{0,1,2}_{02}&= \sum_{n,m}^{\infty} \frac{Y_{nm}}{n!m!} {\tilde{C}}^{0,1,2}_{02} (n,m)  \label{G-02-4decoys-012-tilde} \,\,, \\
{\tilde{G}}^{0,1,3}_{02}&= \sum_{n,m}^{\infty} \frac{Y_{nm}}{n!m!} {\tilde{C}}^{0,1,3}_{02} (n,m)  \label{G-02-4decoys-013-tilde} \,\,, \\
{\tilde{G}}^{0,2,3}_{02}&= \sum_{n,m}^{\infty} \frac{Y_{nm}}{n!m!} {\tilde{C}}^{0,2,3}_{02} (n,m)  \label{G-02-4decoys-023-tilde} \,\,, \\
{\tilde{G}}^{1,2,3}_{02}&= \sum_{n,m}^{\infty} \frac{Y_{nm}}{n!m!} {\tilde{C}}^{1,2,3}_{02} (n,m)  \label{G-02-4decoys-123-tilde} \,\,.
\end{eqnarray}
The tilde symbol above the gains combinations $G_{02}$ and the corresponding yields coefficients $C_{02}$ indicates that we operated the substitutions $\nu_0\longrightarrow\mu_0$, $\nu_1\longrightarrow\mu_1$ and $\nu_2\longrightarrow\mu_2$ in their original expressions.\\
We further combine the expressions (\ref{G-02-4decoys-012-tilde}), (\ref{G-02-4decoys-013-tilde}), (\ref{G-02-4decoys-023-tilde}) and (\ref{G-02-4decoys-123-tilde}) with arbitrary real coefficients ${\tilde{d}}_{i,j,k}$:
\begin{eqnarray}
\fl &{\tilde{H}}_{04} \equiv {\tilde{d}}_{0,1,2} {\tilde{G}}^{0,1,2}_{02} +{\tilde{d}}_{0,1,3} {\tilde{G}}^{0,1,3}_{02}+{\tilde{d}}_{0,2,3} {\tilde{G}}^{0,2,3}_{02}+{\tilde{d}}_{1,2,3} {\tilde{G}}^{1,2,3}_{02} =  \nonumber \\
\fl &\sum_{n,m}^{\infty} \frac{Y_{nm}}{n!m!}\left[{\tilde{d}}_{0,1,2} {\tilde{G}}^{0,1,2}_{02} (n,m) +{\tilde{d}}_{0,1,3} {\tilde{G}}^{0,1,3}_{02} (n,m) + {\tilde{d}}_{0,2,3} {\tilde{G}}^{0,2,3}_{02} (n,m)+ {\tilde{d}}_{1,2,3} {\tilde{G}}^{1,2,3}_{02} (n,m)\right] \label{G-02-4decoys-new-tilde} \,\,,
\end{eqnarray}
and impose that even the terms $Y_{n2}$ and $Y_{3m}$ are removed. The solution for the coefficients ${\tilde{d}}_{i,j,k}$ reads:
\begin{eqnarray}
\fl {\tilde{d}}_{0,1,2} &=0 \nonumber\\
\fl {\tilde{d}}_{0,1,3} &= \mu_3 \nonumber\\
\fl {\tilde{d}}_{0,2,3} &= -\frac{(\mu_0 -\mu_1)\mu_3}{\mu_0 -\mu_2} \nonumber\\
\fl {\tilde{d}}_{1,2,3} &= \frac{\mu_0 \mu_3 (\mu_0 -\mu_1)(\mu_0 -\mu_3)(\mu_0 -\nu_3)}{\mu_1 (\mu_1 -\mu_2)(\mu_1 - \mu_3)(\mu_1 - \nu_3)} \,\,. \label{d_ijk-tilde}
\end{eqnarray}
By substituting the solution (\ref{d_ijk}) back into (\ref{G-02-4decoys-new}) and by rearranging the r.h.s, one gets a combination of gains where all the terms $Y_{n0},Y_{n1},Y_{n2},Y_{1m},Y_{2m}$ and $Y_{3m}$ are removed:
\begin{eqnarray}
\fl {\tilde{H}}_{04} = \sum_{m=3}^{\infty} \frac{Y_{0m}}{m!} {\tilde{A}}_{04}(\mu_0,\mu_1,\mu_2,\mu_3,\nu_3,m) + \sum_{\stackrel[m=3]{n=4}{}}^{\infty} \frac{Y_{nm}}{n!m!} {\tilde{B}}_{04}(\mu_0,\mu_1,\mu_2,\mu_3,\nu_3,n,m) \,\,,\nonumber\\ \fl \label{G-02-4decoys-new-1-tilde}
\end{eqnarray}
where:
\begin{eqnarray}
\fl {\tilde{A}}_{04}(\mu_0,\mu_1,\mu_2,\mu_3,\nu_3,m) &= -\frac{(\mu_0 -\mu_1)^2(\mu_0 -\mu_2)(\mu_1 -\mu_2)(\mu_0 -\mu_3)(\mu_0 -\nu_3)}{\mu_1 \mu_2} \nonumber\\
\fl &\times\left. \left(\sum_{i_1 \leq i_2 \leq \dots \leq i_{m-3}} \mu_{i_1}\mu_{i_2} \cdot \dots \cdot \mu_{i_{m-3}}\right)\right\vert_{\mu_3\longrightarrow \nu_3} \label{A04-tilde} 
\end{eqnarray}
and
\begin{eqnarray}
\fl {\tilde{B}}_{04}(\mu_0,\mu_1,\mu_2,\mu_3,\nu_3,n,m) &= -\mu_0\mu_1 \mu_2 \mu_3 \,{\tilde{A}}_{04}(\mu_0,\mu_1,\mu_2,\mu_3,\nu_3,m) \nonumber\\
\fl &\times \left(\sum_{i_1 \leq i_2 \leq \dots \leq i_{n-4}} \mu_{i_1}\mu_{i_2} \cdot \dots \cdot \mu_{i_{n-4}}\right) \label{B04-tilde}  \,\,.
\end{eqnarray}
We assume that the indexes in the sums run over the set $\{0,1,2,3\}$, we define \\\mbox{$\sum_{i_1 \leq i_2 \leq \dots \leq i_{m-3}} \nu_{i_1}\nu_{i_2} \cdot \dots \cdot \nu_{i_{m-3}}|_{m=3} =1$} and with $\mu_3\longrightarrow \nu_3$ in (\ref{A04-tilde}) we intend that every $\mu_3$ contained in the sum must be replaced with a $\nu_3$.\\
From (\ref{A04-tilde}) and (\ref{B04-tilde}) we deduce that the coefficients of $Y_{0m}$ and $Y_{nm}$ have always opposite sign, hence the upper bound on $Y_{04}$ is obtained from (\ref{G-02-4decoys-new-1-tilde}) by setting all the yields $Y_{0m}$ (with $m\neq 4$) to zero and the yields $Y_{nm}$ (with $n\geq 4,\,m \geq 3$) to 1. After rearranging the terms, we get the following expression for the upper bound on $Y_{04}$ under the conditions $\nu_0=\mu_0$, $\nu_1=\mu_1$ and $\nu_2=\mu_2$:
\begin{eqnarray}
\fl {\tilde{Y}}^U_{04}=\frac{24}{{\tilde{A}}_{04}(\mu_0,\mu_1,\mu_2,\mu_3,\nu_3,4)}\left[{\tilde{H}}_{04} - \sum_{\stackrel[m=3]{n=4}{}}^{\infty} \frac{{\tilde{B}}_{04}(\mu_0,\mu_1,\mu_2,\mu_3,\nu_3,n,m)}{n!m!} \right]  \label{Y04-upperbound-4decoys-tilde} \,\,,
\end{eqnarray}
where ${\tilde{H}}_{04}$ is given in the first line of (\ref{G-02-4decoys-new-tilde}), the function ${\tilde{A}}_{04}$ evaluated for $m=4$ reads:
\begin{eqnarray}
\fl &{\tilde{A}}_{04}(\mu_0,\mu_1,\mu_2,\mu_3,\nu_3,4)= -\frac{(\mu_0 -\mu_1)^2(\mu_0 -\mu_2)(\mu_1 -\mu_2)(\mu_0 -\mu_3)(\mu_0 -\nu_3)}{\mu_1 \mu_2} (\mu_0 +\mu_1 +\mu_2 + \nu_3) \,\,, \label{A04-evaluated-tilde}
\end{eqnarray} 
and the series of ${\tilde{B}}_{04}$ sums to:
{\small\begin{eqnarray}
\fl &\sum_{\stackrel[m=3]{n=4}{}}^{\infty} \frac{{\tilde{B}}_{04}(\mu_0,\mu_1,\mu_2,\mu_3,\nu_3,n,m)}{n!m!}= \frac{\mu_0 \mu_3}{(\mu_0 -\mu_2)(\mu_1 -\mu_2)(\mu_1 -\mu_3)(\mu_2 -\mu_3)(\mu_1 -\nu_3)(\mu_2 -\nu_3)} \nonumber\\
\fl &\times  \left[\left(e^{\mu_0}-\frac{\mu_0^2}{2}-\mu_0-1\right) (\mu_1-\mu_2) (\mu_1-\mu_3) (\mu_2-\mu_3)-\left(e^{\mu_1}-\frac{\mu_1^2}{2}-\mu_1-1\right) (\mu_0-\mu_2) (\mu_0-\mu_3) (\mu_2-\mu_3) \right.\nonumber\\
\fl  &\left. +\left(e^{\mu_2}-\frac{\mu_2^2}{2}-\mu_2-1\right) (\mu_0-\mu_1) (\mu_0-\mu_3) (\mu_1-\mu_3)-\left(e^{\mu_3}-\frac{\mu_3^2}{2}-\mu_3-1\right) (\mu_0-\mu_1) (\mu_0-\mu_2) (\mu_1-\mu_2)\right] \nonumber\\
\fl &\times  \left[\left(e^{\mu_0}-\frac{\mu_0^2}{2}-\mu_0-1\right) (\mu_1-\mu_2) (\mu_1-\nu_3) (\mu_2-\nu_3)-\left(e^{\mu_1}-\frac{\mu_1^2}{2}-\mu_1-1\right) (\mu_0-\mu_2) (\mu_0-\nu_3) (\mu_2-\nu_3) \right.\nonumber\\
\fl &\left. +\left(e^{\mu_2}-\frac{\mu_2^2}{2}-\mu_2-1\right) (\mu_0-\mu_1) (\mu_0-\nu_3) (\mu_1-\nu_3)-\left(e^{\nu_3}-\frac{\nu_3^2}{2}-\nu_3-1\right) (\mu_0-\mu_1) (\mu_0-\mu_2) (\mu_1-\mu_2)\right] \,. \nonumber
\end{eqnarray}}
\begin{equation}
\label{B04-summed-tilde}
\end{equation}
\subsection{Upper bound on $Y_{40}$} \label{bound_on_Y40-4decoys}
Similarly to the bound on $Y_{04}$, we consider the combination of gains (\ref{G-20-1}) that leads to the bound on $Y_{40}$ in the case of three decoy intensity settings:
\begin{equation}
G^{0,1,2}_{20}= \sum_{n,m}^{\infty} \frac{Y_{nm}}{n!m!} C^{0,1,2}_{20} (n,m)  \label{G-20-4decoys-012} \,\,,
\end{equation}
where the function $C^{0,1,2}_{20} (n,m)$ is defined by the r.h.s. of (\ref{G-20-1}), while $G^{0,1,2}_{20}$ is the combination of gains given by (\ref{G-20-3decoys}), with the coefficients $c_{i,j}$ of the combination given in (\ref{cij-20}). From \ref{bound_on_Y20-3decoys} we know that the terms $Y_{n1},Y_{n2},Y_{0m}$ and $Y_{1m}$ are removed in (\ref{G-20-4decoys-012}). Following the same procedure described in \ref{bound_on_Y04-4decoys}, we derive three additional combinations of gains in which the terms $Y_{n1},Y_{n2},Y_{0m}$ and $Y_{1m}$ are removed:
\begin{eqnarray}
G^{0,1,3}_{20}&= \sum_{n,m}^{\infty} \frac{Y_{nm}}{n!m!} C^{0,1,3}_{20} (n,m)  \label{G-20-4decoys-013} \,\,, \\
G^{0,2,3}_{20}&= \sum_{n,m}^{\infty} \frac{Y_{nm}}{n!m!} C^{0,2,3}_{20} (n,m)  \label{G-20-4decoys-023} \,\,, \\
G^{1,2,3}_{20}&= \sum_{n,m}^{\infty} \frac{Y_{nm}}{n!m!} C^{1,2,3}_{20} (n,m)  \label{G-20-4decoys-123} \,\,.
\end{eqnarray}
Now we further combine these expressions with arbitrary real coefficients $d_{i,j,k}$:
\begin{eqnarray}
\fl &H_{40} \equiv d_{0,1,2} G^{0,1,2}_{20} +d_{0,1,3} G^{0,1,3}_{20}+d_{0,2,3} G^{0,2,3}_{20}+d_{1,2,3} G^{1,2,3}_{20} =  \nonumber \\
\fl &\sum_{n,m}^{\infty} \frac{Y_{nm}}{n!m!}\left[d_{0,1,2} C^{0,1,2}_{20} (n,m) +d_{0,1,3} C^{0,1,3}_{20} (n,m) + d_{0,2,3} C^{0,2,3}_{20} (n,m)+ d_{1,2,3} C^{1,2,3}_{20} (n,m)\right] \label{G-20-4decoys-new} \,\,,
\end{eqnarray}
and impose that even the terms $Y_{n3}$ and $Y_{2m}$ are removed from the r.h.s. of (\ref{G-20-4decoys-new}). This yields a linear system of equations in the variables $d_{i,j,k}$, whose unique solution (up to a global rescaling) reads as follows:
\begin{eqnarray}
\fl d_{0,1,2} &=1 \nonumber\\
\fl d_{0,1,3} &= \frac{(\mu_0-\mu_2) (\nu_0-\nu_2) \left[\mu_0 (\nu_1-\nu_3)+\mu_1 (\nu_3-\nu_0)+\mu_3 (\nu_0-\nu_1)\right]}{(\mu_0-\mu_3) (\nu_0-\nu_3) \left[\mu_0 (\nu_2-\nu_1)+\mu_1 (\nu_0-\nu_2)+\mu_2 (\nu_1-\nu_0)\right]} \nonumber\\
\fl d_{0,2,3} &= \frac{(\mu_0-\mu_1) (\nu_0-\nu_1) \left[\mu_0 (\nu_2-\nu_3)+\mu_2 (\nu_3-\nu_0)+\mu_3 (\nu_0-\nu_2)\right]}{(\mu_0-\mu_3) (\nu_0-\nu_3) \left[\mu_0 (\nu_1-\nu_2)+\mu_1 (\nu_2-\nu_0)+\mu_2 (\nu_0-\nu_1)\right]} \nonumber\\
\fl d_{1,2,3} &=\frac{\nu_0 (\mu_0-\mu_1) (\mu_0-\mu_2) (\nu_0-\nu_1) (\nu_0-\nu_2) \left[\mu_1 (\nu_3-\nu_2)+\mu_2 (\nu_1-\nu_3)+\mu_3 (\nu_2-\nu_1)\right]}{\nu_1 (\mu_1-\mu_2) (\mu_1-\mu_3)
	(\nu_1-\nu_2) (\nu_1-\nu_3) \left[\mu_0 (\nu_1-\nu_2)+\mu_1 (\nu_2-\nu_0)+\mu_2 (\nu_0-\nu_1)\right]} \,\,. \label{d_ijk-40}
\end{eqnarray}
By substituting the solution (\ref{d_ijk-40}) back into (\ref{G-20-4decoys-new}) and by rearranging the r.h.s, one gets a combination of gains where all the terms $Y_{n1},Y_{n2},Y_{n3},Y_{0m},Y_{1m}$ and $Y_{2m}$ are removed:
\begin{eqnarray}
\fl H_{40} = \sum_{n=3}^{\infty} \frac{Y_{n0}}{n!} A_{04}(\nu_0,\nu_1,\nu_2,\nu_3,\mu_0,\mu_1,\mu_2,\mu_3,n) + \sum_{\stackrel[m=4]{n=3}{}}^{\infty} \frac{Y_{nm}}{n!m!} B_{04}(\nu_0,\nu_1,\nu_2,\nu_3,\mu_0,\mu_1,\mu_2,\mu_3,m,n) \,\,,\nonumber\\ \fl \label{G-20-4decoys-new-1}
\end{eqnarray}
where the functions $A_{04}$ and $B_{04}$ are the same found in bounding $Y_{04}$ with four decoys and are given by (\ref{A04}) and (\ref{B04}), respectively. Note that in this case the roles of the intensities $\mu_i$ and $\nu_i$ are exchanged with respect to the bound on $Y_{04}$ (see \ref{G-02-4decoys-new-1}), as well as the roles of $n$ and $m$. Following the same reasoning of \ref{bound_on_Y04-4decoys}, we can conclude that the coefficients of $Y_{n0}$ and $Y_{nm}$ have always opposite sign. Hence the upper bound on $Y_{40}$ is obtained from (\ref{G-20-4decoys-new-1}) by setting all the yields $Y_{n0}$ (with $n\neq 4$) to zero and the yields $Y_{nm}$ (with $n\geq 3,\,m \geq 4$) to 1. After rearranging the terms, we get the following expression for the upper bound on $Y_{40}$:
\begin{eqnarray}
\fl Y^U_{40}=\frac{24}{A_{04}(\nu_0,\nu_1,\nu_2,\nu_3,\mu_0,\mu_1,\mu_2,\mu_3,4)}\left[H_{40} - \sum_{\stackrel[m=4]{n=3}{}}^{\infty} \frac{B_{04}(\nu_0,\nu_1,\nu_2,\nu_3,\mu_0,\mu_1,\mu_2,\mu_3,m,n)}{n!m!} \right]  \label{Y40-upperbound-4decoys} \,\,,
\end{eqnarray}
where $H_{40}$ is given in the first line of (\ref{G-20-4decoys-new}), while $A_{04}(\nu_0,\nu_1,\nu_2,\nu_3,\mu_0,\mu_1,\mu_2,\mu_3,4)$ and the sum of the series are given in (\ref{A04-evaluated}) and (\ref{B04-summed}), respectively, under the replacement $\mu_i \leftrightarrow \nu_i$ for $i=0,1,2,3$.\medskip\\
We remark that in deriving the bound (\ref{Y40-upperbound-4decoys}) we implicitly assumed --as in the $Y_{04}$ case-- that at least one of the following equalities does not hold: $\nu_0=\mu_0$, $\nu_1=\mu_1$ and $\nu_2=\mu_2$. Indeed, when all three equalities hold (i.e. when Alice and Bob are using the same intensities settings for three out of four decoy pulses) one gets a ``$\frac{0}{0}$ form'' in the bound expression (\ref{Y40-upperbound-4decoys}). In order to overcome this issue, one can follow an analogous procedure to that performed for the same issue affecting the bound on $Y_{04}$ (see last paragraph in \ref{bound_on_Y04-4decoys}), and obtain an additional upper bound on $Y_{40}$ which is valid in the particular case of: $\nu_0=\mu_0$, $\nu_1=\mu_1$ and $\nu_2=\mu_2$. The new bound on $Y_{40}$ reads:
\begin{eqnarray}
\fl {\tilde{Y}}^U_{40}=\frac{24}{{\tilde{A}}_{04}(\mu_0,\mu_1,\mu_2,\nu_3,\mu_3,4)}\left[{\tilde{H}}_{40} - \sum_{\stackrel[m=4]{n=3}{}}^{\infty} \frac{{\tilde{B}}_{04}(\mu_0,\mu_1,\mu_2,\nu_3,\mu_3,n,m)}{n!m!} \right]  \label{Y40-upperbound-4decoys-tilde} \,\,,
\end{eqnarray}
where ${\tilde{H}}_{40}$ is given by:
\begin{equation}
{\tilde{H}}_{40}= {\tilde{d}}_{0,1,2} {\tilde{G}}^{0,1,2}_{20} +{\tilde{d}}_{0,1,3} {\tilde{G}}^{0,1,3}_{20}+{\tilde{d}}_{0,2,3} {\tilde{G}}^{0,2,3}_{20}+{\tilde{d}}_{1,2,3} {\tilde{G}}^{1,2,3}_{20} \,\,,  \label{H40-tilde}
\end{equation}
where:
\begin{eqnarray}
\fl {\tilde{d}}_{0,1,2} &=0 \nonumber\\
\fl {\tilde{d}}_{0,1,3} &= \nu_3 \nonumber\\
\fl {\tilde{d}}_{0,2,3} &= -\frac{(\mu_0 -\mu_1)\nu_3}{\mu_0 -\mu_2} \nonumber\\
\fl {\tilde{d}}_{1,2,3} &= \frac{\mu_0 \nu_3 (\mu_0 -\mu_1)(\mu_0 -\mu_3)(\mu_0 -\nu_3)}{\mu_1 (\mu_1 -\mu_2)(\mu_1 - \mu_3)(\mu_1 - \nu_3)} \,\,, \label{d_ijk-40-tilde}
\end{eqnarray}
and ${\tilde{G}}_{20}$ are the same gains combinations (\ref{G-20-4decoys-012}), (\ref{G-20-4decoys-013}), (\ref{G-20-4decoys-023}) and (\ref{G-20-4decoys-123}) derived at the beginning of this Subsection, under the replacements: $\nu_0\longrightarrow\mu_0$, $\nu_1\longrightarrow\mu_1$ and $\nu_2\longrightarrow\mu_2$.
The quantity ${\tilde{A}}_{04}$ and the sum of the series are instead given in (\ref{A04-evaluated-tilde}) and (\ref{B04-summed-tilde}), respectively, under the replacement $\mu_3 \leftrightarrow \nu_3$.

\subsection{Upper bound on $Y_{13}$} \label{bound_on_Y13-4decoys}
We follow the same procedure used in bounding the other yields in the case of four decoy intensity settings. We start by considering the four combination of gains in which the terms $Y_{n0},Y_{n1},Y_{0m}$ and $Y_{2m}$ are removed:
\begin{eqnarray}
G^{0,1,2}_{13}&= \sum_{n,m}^{\infty} \frac{Y_{nm}}{n!m!} C^{0,1,2}_{13} (n,m)  \label{G-13-4decoys-012} \,\,, \\
G^{0,1,3}_{13}&= \sum_{n,m}^{\infty} \frac{Y_{nm}}{n!m!} C^{0,1,3}_{13} (n,m)  \label{G-13-4decoys-013} \,\,, \\
G^{0,2,3}_{13}&= \sum_{n,m}^{\infty} \frac{Y_{nm}}{n!m!} C^{0,2,3}_{13} (n,m)  \label{G-13-4decoys-023} \,\,, \\
G^{1,2,3}_{13}&= \sum_{n,m}^{\infty} \frac{Y_{nm}}{n!m!} C^{1,2,3}_{13} (n,m)  \label{G-13-4decoys-123} \,\,.
\end{eqnarray}
where the last three combinations are derived from the first one as described in \ref{bound_on_Y04-4decoys}, while the first combination is given by (\ref{G-13-1}).
Now we further combine these expressions with arbitrary real coefficients $d_{i,j,k}$:
\begin{eqnarray}
\fl &H_{13} \equiv d_{0,1,2} G^{0,1,2}_{13} +d_{0,1,3} G^{0,1,3}_{13}+d_{0,2,3} G^{0,2,3}_{13}+d_{1,2,3} G^{1,2,3}_{13} =  \nonumber \\
\fl &\sum_{n,m}^{\infty} \frac{Y_{nm}}{n!m!}\left[d_{0,1,2} C^{0,1,2}_{13} (n,m) +d_{0,1,3} C^{0,1,3}_{13} (n,m) + d_{0,2,3} C^{0,2,3}_{13} (n,m)+ d_{1,2,3} C^{1,2,3}_{13} (n,m)\right] \label{G-13-4decoys-new} \,\,,
\end{eqnarray}
and impose that even the terms $Y_{n2}$ and $Y_{3m}$ are removed from the r.h.s. of (\ref{G-13-4decoys-new}). This yields a linear system of equations in the variables $d_{i,j,k}$, whose unique solution (up to a global rescaling) reads as follows:
\begin{eqnarray}
\fl d_{0,1,2} &=1 \nonumber\\
\fl d_{0,1,3} &= \frac{(\mu_0-\mu_2)(\mu_1+\mu_3)(\nu_0-\nu_2)}{(\mu_0-\mu_3)(\mu_1+\mu_2)(\nu_0-\nu_3)} \nonumber\\ 
\fl &\times \left\lbrace\mu_0^2 (\mu_1+\mu_2) (\mu_2+\mu_3)(\nu_1-\nu_3)+(\mu_0+\mu_2) \left[\mu_1^2 (\mu_2+\mu_3) (\nu_3-\nu_0)+ \mu_3^2 (\mu_1+\mu_2)(\nu_0-\nu_1)\right]\right\rbrace \nonumber\\
\fl &/\left\lbrace -\mu_0^2 (\mu_1+\mu_3) (\mu_2+\mu_3) (\nu_1-\nu_2)+(\mu_0+\mu_3) \left[\mu_1^2 (\mu_2+\mu_3) (\nu_0-\nu_2)+ \mu_2^2 (\mu_1+\mu_3)(\nu_1-\nu_0)\right]\right\rbrace \nonumber\\
\fl d_{0,2,3} &= -\frac{(\mu_0-\mu_1)(\mu_2+\mu_3)(\nu_0-\nu_1)}{(\mu_0-\mu_3)(\mu_1+\mu_2)(\nu_0-\nu_3)} \nonumber\\
\fl &\times \left\lbrace\mu_0^2 (\mu_1+\mu_2)(\mu_1+\mu_3)(\nu_2-\nu_3)+(\mu_0+\mu_1) \left[\mu_2^2 (\mu_1+\mu_3)(\nu_3-\nu_0)+\mu_3^2 (\mu_1+\mu_2)(\nu_0-\nu_2)\right]\right\rbrace \nonumber\\
\fl &/\left\lbrace -\mu_0^2 (\mu_1+\mu_3)(\mu_2+\mu_3)(\nu_1-\nu_2)+(\mu_0+\mu_3) \left[\mu_1^2 (\mu_2+\mu_3) (\nu_0-\nu_2)+\mu_2^2 (\mu_1+\mu_3)(\nu_1-\nu_0)\right]\right\rbrace \nonumber\\
\fl d_{1,2,3} &= \frac{(\mu_0-\mu_1) (\mu_0-\mu_2) (\mu_2+\mu_3) (\nu_0-\nu_1) (\nu_0-\nu_2)}{\left(\mu_1^2-\mu_2^2\right) (\mu_1-\mu_3) (\nu_1-\nu_2) (\nu_1-\nu_3)} \left\lbrace \mu_0^2 \left[\mu_1^2 (\nu_3-\nu_2)+\mu_2^2 (\nu_1-\nu_3)+\mu_3^2 (\nu_2-\nu_1)\right] \right. \nonumber\\
\fl &\left.+ (\mu_0\mu_1\mu_2+ \mu_0\mu_1\mu_3 + \mu_0\mu_2 \mu_3+ \mu_1 \mu_2 \mu_3) [\mu_1 (\nu_3-\nu_2)+\mu_2(\nu_1-\nu_3)+\mu_3 (\nu_2-\nu_1)]\right\rbrace \nonumber\\
\fl &/ \left\lbrace \mu_0^2 (\mu_1+\mu_3) (\mu_2+\mu_3) (\nu_1-\nu_2)+(\mu_0+\mu_3) \left[-\mu_1^2 (\mu_2+\mu_3) (\nu_0-\nu_2)+ \mu_2^2 (\mu_1+\mu_3)(\nu_0-\nu_1)\right]\right\rbrace  \,\,. \nonumber\\
\fl  \label{d_ijk-13}
\end{eqnarray}
By substituting the solution (\ref{d_ijk-13}) back into (\ref{G-13-4decoys-new}) and by rearranging the r.h.s, one gets a combination of gains where all the terms $Y_{n0},Y_{n1},Y_{n2},Y_{0m},Y_{2m}$ and $Y_{3m}$ are removed:
\begin{eqnarray}
\fl H_{13} = &\sum_{m=3}^{\infty} \frac{Y_{1m}}{m!} A_{13}(\mu_0,\mu_1,\mu_2,\mu_3,\nu_0,\nu_1,\nu_2,\nu_3,m)  \nonumber\\
\fl &+ \sum_{\stackrel[m=3]{n=4}{}}^{\infty} \frac{Y_{nm}}{n!m!} A_{13}(\mu_0,\mu_1,\mu_2,\mu_3,\nu_0,\nu_1,\nu_2,\nu_3,m) D_n(\mu_0,\mu_1,\mu_2,\mu_3) \,\,, \label{G-13-4decoys-new-1}
\end{eqnarray}
where:
\begin{eqnarray}
\fl &A_{13}(\mu_0,\mu_1,\mu_2,\mu_3,\nu_0,\nu_1,\nu_2,\nu_3,m) =  -\frac{(\mu_0 -\mu_1)(\mu_0 -\mu_2)(\nu_0 -\nu_1)(\nu_0 -\nu_2)}{(\mu_1+\mu_2)} \nonumber\\
\fl &\times\frac{p_{13}(\mu_0,\mu_1,\mu_2,\mu_3,\nu_0,\nu_1,\nu_2,\nu_3)}{q_{13}(\mu_0,\mu_1,\mu_2,\mu_3,\nu_0,\nu_1,\nu_2,\nu_3)} \left(\sum_{i_1 \leq i_2 \leq \dots \leq i_{m-3}} \nu_{i_1}\nu_{i_2} \cdot \dots \cdot \nu_{i_{m-3}}\right) \,\,; \label{A13}
\end{eqnarray}
\begin{eqnarray}
\fl &p_{13}(\mu_0,\mu_1,\mu_2,\mu_3,\nu_0,\nu_1,\nu_2,\nu_3) = \mu_1^2 \left[\mu_2^2 (\nu_0-\nu_3) (\nu_1-\nu_2)-\mu_3^2 (\nu_0-\nu_2) (\nu_1-\nu_3)\right] \nonumber\\
\fl &+\mu_0^2 \left[\mu_1^2 (\nu_0-\nu_1) (\nu_2-\nu_3)-\mu_2^2 (\nu_0-\nu_2) (\nu_1-\nu_3)+\mu_3^2 (\nu_0-\nu_3) (\nu_1-\nu_2)\right]\nonumber \\
\fl &+\mu_2^2 \mu_3^2 (\nu_0-\nu_1) (\nu_2-\nu_3) \,\,;\\
\fl &q_{13}(\mu_0,\mu_1,\mu_2,\mu_3,\nu_0,\nu_1,\nu_2,\nu_3) = \mu_0^2 (\mu_1+\mu_3) (\mu_2+\mu_3) (\nu_1-\nu_2)-\mu_1^2 (\mu_0+\mu_3) (\mu_2+\mu_3)(\nu_0-\nu_2) \nonumber\\
\fl &+\mu_2^2 (\mu_0+\mu_3) (\mu_1+\mu_3) (\nu_0-\nu_1) \,\,,
\end{eqnarray}
and $D_n$ is defined recursively as \cite{grasselli2019practical}:
\begin{eqnarray}
\fl  \left\{
{\begin{array}{lcl}
	D_{n}(\mu_0,\mu_1,\mu_2,\mu_3) &=& \frac{\sum_{j=1}^{n-4} (\mu^j_0+\mu^j_1+\mu^j_2+\mu^j_3) D_{n-j}(\mu_0,\mu_1,\mu_2,\mu_3) -\mu_0 \mu_1 \mu_2 \mu_3 \left(\sum_{i_1 \leq i_2 \leq \dots \leq i_{n-5}} \mu_{i_1}\mu_{i_2} \cdot \dots \cdot
		\mu_{i_{n-5}}\right)}{n-4}   \nonumber \\
	D_4(\mu_0,\mu_1,\mu_2,\mu_3) &=& \mu_0 \mu_1 \mu_2 +\mu_0 \mu_1 \mu_3 +\mu_0 \mu_2 \mu_3 + \mu_1 \mu_2 \mu_3  \,\,.
	\end{array}}
\right.  \nonumber\\ \fl  \label{Dn}  
\end{eqnarray}
We assume that the indexes in the sums run over the set $\{0,1,2,3\}$ and we define \mbox{$\sum_{i_1 \leq i_2 \leq \dots \leq i_{m-3}} \nu_{i_1}\nu_{i_2} \cdot \dots \cdot \nu_{i_{m-3}}|_{m=3} =1$}. Since $D_n \geq 0$ for every $n\geq 4$, we deduce that the coefficients of $Y_{1m}$ and $Y_{nm}$ in (\ref{G-13-4decoys-new-1}) have always equal sign.
Hence the upper bound on $Y_{13}$ is obtained from (\ref{G-13-4decoys-new-1}) by setting all the other yields to zero. After rearranging the terms, we get the following expression for the upper bound on $Y_{13}$:
\begin{eqnarray}
Y^U_{13}=\frac{6\,H_{13}}{A_{13}(\mu_0,\mu_1,\mu_2,\mu_3,\nu_0,\nu_1,\nu_2,\nu_3,3)}  \label{Y13-upperbound-4decoys} \,\,,
\end{eqnarray}
where $H_{13}$ is given in the first line of (\ref{G-13-4decoys-new}), while $A_{13}(\mu_0,\mu_1,\mu_2,\mu_3,\nu_0,\nu_1,\nu_2,\nu_3,3)$ is given by:
\begin{eqnarray}
\fl  A_{13}(\mu_0,\mu_1,\mu_2,\mu_3,\nu_0,\nu_1,\nu_2,\nu_3,3)=& -\frac{(\mu_0 -\mu_1)(\mu_0 -\mu_2)(\nu_0 -\nu_1)(\nu_0 -\nu_2)}{(\mu_1+\mu_2)}  \,\,. \nonumber\\ 
\fl &\times \frac{p_{13}(\mu_0,\mu_1,\mu_2,\mu_3,\nu_0,\nu_1,\nu_2,\nu_3)}{q_{13}(\mu_0,\mu_1,\mu_2,\mu_3,\nu_0,\nu_1,\nu_2,\nu_3)}  \label{A13-evaluated}
\end{eqnarray}

\subsection{Upper bound on $Y_{31}$} \label{bound_on_Y31-4decoys}
We follow the same procedure used in bounding the other yields in the case of four decoy intensity settings. We start by considering the four combination of gains in which the terms $Y_{n0},Y_{n2},Y_{0m}$ and $Y_{1m}$ are removed:
\begin{eqnarray}
G^{0,1,2}_{31}&= \sum_{n,m}^{\infty} \frac{Y_{nm}}{n!m!} C^{0,1,2}_{31} (n,m)  \label{G-31-4decoys-012} \,\,, \\
G^{0,1,3}_{31}&= \sum_{n,m}^{\infty} \frac{Y_{nm}}{n!m!} C^{0,1,3}_{31} (n,m)  \label{G-31-4decoys-013} \,\,, \\
G^{0,2,3}_{31}&= \sum_{n,m}^{\infty} \frac{Y_{nm}}{n!m!} C^{0,2,3}_{31} (n,m)  \label{G-31-4decoys-023} \,\,, \\
G^{1,2,3}_{31}&= \sum_{n,m}^{\infty} \frac{Y_{nm}}{n!m!} C^{1,2,3}_{31} (n,m)  \label{G-31-4decoys-123} \,\,,
\end{eqnarray}
where the last three combinations are derived from the first one as described in \ref{bound_on_Y04-4decoys}, while the first combination is given by (\ref{G-31-1}).
Now we further combine these expressions with arbitrary real coefficients $d_{i,j,k}$:
\begin{eqnarray}
\fl &H_{31} \equiv d_{0,1,2} G^{0,1,2}_{31} +d_{0,1,3} G^{0,1,3}_{31}+d_{0,2,3} G^{0,2,3}_{31}+d_{1,2,3} G^{1,2,3}_{31} =  \nonumber \\
\fl &\sum_{n,m}^{\infty} \frac{Y_{nm}}{n!m!}\left[d_{0,1,2} C^{0,1,2}_{31} (n,m) +d_{0,1,3} C^{0,1,3}_{31} (n,m) + d_{0,2,3} C^{0,2,3}_{31} (n,m)+ d_{1,2,3} C^{1,2,3}_{31} (n,m)\right] \label{G-31-4decoys-new} \,\,,
\end{eqnarray}
and impose that even the terms $Y_{n3}$ and $Y_{2m}$ are removed from the r.h.s. of (\ref{G-31-4decoys-new}). This yields a linear system of equations in the variables $d_{i,j,k}$, whose unique solution (up to a global rescaling) is given in (\ref{d_ijk-13}), under the replacement: $\mu_i \leftrightarrow \nu_i$ for $i=0,1,2,3$.
By substituting the solution back into (\ref{G-31-4decoys-new}) and by rearranging the r.h.s, one gets a combination of gains where all the terms $Y_{n0},Y_{n2},Y_{n3},Y_{0m},Y_{1m}$ and $Y_{2m}$ are removed:
\begin{eqnarray}
\fl H_{31} = &\sum_{n=3}^{\infty} \frac{Y_{n1}}{n!} A_{13}(\nu_0,\nu_1,\nu_2,\nu_3,\mu_0,\mu_1,\mu_2,\mu_3,n) \nonumber\\
\fl &+ \sum_{\stackrel[m=4]{n=3}{}}^{\infty} \frac{Y_{nm}}{n!m!} A_{13}(\nu_0,\nu_1,\nu_2,\nu_3,\mu_0,\mu_1,\mu_2,\mu_3,n) D_m(\nu_0,\nu_1,\nu_2,\nu_3) \,\,, \label{G-31-4decoys-new-1}
\end{eqnarray}
where the functions $A_{13}$ and $D_m$ are defined in (\ref{A13}) and (\ref{Dn}), respectively. Since $D_m \geq 0$ for every $m\geq 4$, we deduce that the coefficients of $Y_{n1}$ and $Y_{nm}$ in (\ref{G-31-4decoys-new-1}) have always equal sign.
Hence the upper bound on $Y_{31}$ is obtained from (\ref{G-31-4decoys-new-1}) by setting all the other yields to zero. After rearranging the terms, we get the following expression for the upper bound on $Y_{31}$:
\begin{eqnarray}
Y^U_{31}=\frac{6\,H_{31}}{A_{13}(\nu_0,\nu_1,\nu_2,\nu_3,\mu_0,\mu_1,\mu_2,\mu_3,3)}  \label{Y31-upperbound-4decoys} \,\,,
\end{eqnarray}
where $H_{31}$ is given in the first line of (\ref{G-31-4decoys-new}), while $A_{13}(\nu_0,\nu_1,\nu_2,\nu_3,\mu_0,\mu_1,\mu_2,\mu_3,3)$ is given by (\ref{A13-evaluated}) under the substitution: $\mu_i \leftrightarrow \nu_i$ for $i=0,1,2,3$.


\section*{References}

\bibliographystyle{iopart-num}
\bibliography{library}

\end{document}